\documentclass{aa}  

\usepackage{graphicx}
\usepackage[flushleft]{threeparttable}
\usepackage{longtable}
\usepackage{txfonts}
\usepackage{hyperref}
\usepackage{lscape}

\newcommand{\modifLE}[1]{#1}
\begin{document} 

\title{A population of transition disks around evolved stars: Fingerprints of planets\thanks{Table\,\ref{Table:list} is only available in electronic form
at the CDS via anonymous ftp to cdsarc.u-strasbg.fr (130.79.128.5)
or via http://cdsweb.u-strasbg.fr/cgi-bin/qcat?J/A+A/}}
\titlerunning{A population of transition disks around evolved stars}

\subtitle{Catalog of disks surrounding Galactic post-AGB binaries}



   \author{
          J. Kluska
          \inst{1}
          \and
          H. Van Winckel\inst{1}
          \and
          Q. Coppée\inst{1}
            \and
            G.-M. Oomen\inst{1}
            \and
            K. Dsilva\inst{1}
               \and
            D. Kamath\inst{2,3}
            \and
            V. Bujarrabal\inst{4}
            \and
            M. Min\inst{5}
          }

   \institute{Institute of Astronomy, KU Leuven, Celestijnenlaan 200D, 3001 Leuven, Belgium\\
              \email{jacques.kluska@kuleuven.be}
    \and
     Department of Physics and Astronomy, Macquarie University, Sydney, NSW 2109, Australia
      \and
     Astronomy, Astrophysics and Astrophotonics Research Centre, Macquarie University, Sydney, NSW 2109, Australia
     \and
    Observatorio Astronómico Nacional (OAN-IGN), Apartado 112, E-28803 Alcalá de Henares, Spain
     \and
    SRON Netherlands Institute for Space Research, Sorbonnelaan 2, 3584 CA Utrecht, The Netherlands}
   \date{}

 
  \abstract
  {Post-asymptotic giant branch (post-AGB) binaries are surrounded by massive disks of gas and dust that are similar to the protoplanetary disks that are known to surround young stars.}
   {We assembled a catalog of all known Galactic post-AGB binaries featuring disks. We  explore the correlations between the different observables with the aim of learning more about potential disk-binary interactions.}
   {We compiled spectral energy distributions of 85 Galactic post-AGB binary systems. 
   We built a color-color diagram to differentiate between the different disk morphologies traced by the characteristics of the infrared excess. 
   We categorized the different disk types and searched for correlations with other observational characteristics of these systems.}
   {Between 8 and 12\% of our targets are surrounded by transition disks, that is, disks having no or low near-infrared excess.
   We find a strong link between these transition disks and the depletion of refractory elements seen on the surface of \modifLE{the} post-AGB star. 
   We interpret this correlation as evidence of the presence of a mechanism that stimulates the dust and gas separation within the disk and that also produces the transition disk structure.
   We propose that such a mechanism \modifLE{is likely to be due to} a giant planet carving a hole in the disk, effectively trapping the dust in the outer disk parts. 
   We propose two disk evolutionary scenarios, depending on the actual presence of such a giant planet in the disk.  }
   {We advocate that giant planets can successfully explain the correlation between the transition disks and the depletion of refractory materials observed in post-AGB binaries. 
   If the planetary scenario is confirmed, disks around post-AGB binaries could be a unique laboratory for testing planet-disk interactions and their influence on the late evolution of binary stars. The question of whether such planets are first- or second-generation bodies also remains to be considered. 
   We argue that these disks are ideal for studying planet formation scenarios in an unprecedented parameter space. 
   }


   \maketitle
%

\section{Introduction}

Historically, binaries located among post-asymptotic giant branch stars (post-AGB) have been  serendipitously detected, however, they have also turned out to have a certain property in common, namely: a characteristic spectral energy distribution (SED) with a near- and mid-infrared excess pointing to the presence of hot dust in the system and a long-wavelength tail exhibiting a Rayleigh-Jeans slope. By now, it has been well established that these SED characteristics point to the presence of a stable  circumbinary disk of gas and dust in Keplerian rotation. The observational characteristics of these objects were recently reviewed by \cite{VanWinckel2018}.

Subsequent systematic radial velocity monitoring campaigns focusing on stars \modifLE{surrounded} by a disk showed that these systems were indeed mainly binaries \citep[and references therein]{Oomen2018}. 
Their typical orbital periods are between a hundred and a few thousand days with, at times, surprisingly high eccentricities (up to $e\sim0.6$).

The global picture that emerges from the observed orbital elements is that a star evolved in a system which is too short to accommodate a full-grown red giant branch (RGB) or AGB star\footnote{Although the targets of the study can be either post-AGB or post-RGB binaries, we refer to them as post-AGB binaries in the rest of the paper.}.  
During a poorly understood phase of strong interaction, the system did not suffer the dramatic spiral-in phase that is predicted for a common envelope evolution. 
The observed orbits show that the common envelope was either very rapidly expelled or somehow avoided.  The product of this interaction is a disk of gas and dust with Keplerian rotation and exhibiting similar properties to protoplanetary disks around young stars. 
This Keplerian rotation was resolved in interferometric CO observations in several systems \citep[e.g.,][and references therein]{Gallardo2021}.
Therefore, these plausible protoplanetary disks form as the result of an unconstrained interaction process that takes place during the final evolution of the evolved component of the binary. 


The interactions between the central system and the circumbinary disk are varied \modifLE{and} lead to distinct observational signatures. 
One of them is a photospheric chemical anomaly called ``depletion'' \citep[and references therein]{VanWinckel2003,Maas2005,Maas2007,Giridhar2010,Venn2014, Gezer2015, Kamath2019}.
The abundance trend in such atmospheres resembles the gas phase of the interstellar medium: refractory elements are underabundant, while volatiles retain their original abundances. 
There is general acknowledgment that this is caused by a chemical process in which gas that was subject to dust formation in the circumstellar environment, becomes separated from the dust and accreted onto the star.
Good tracers of depletion are, for example, [Fe/H] or [Zn/Ti] abundance ratios. 
The $s$-process elements are also refractory \citep{Lodders2003} but the eventual enrichment of AGB nucleosynthesis products is negligible compared to the depletion process. 
In a recent quantified study, the observed depletion patterns were explained by the accretion of already depleted matter from the circumbinary disk, provided this process is efficient and long-lived \citep{Oomen2019, Oomen2020}.
It was shown that the extremely low envelope mass of the post-AGB star makes it so that its composition is very quickly dominated by accreted matter. Abundance studies provide an efficient way to probe the composition of the accreted gas.
However, the origin of the separation between volatiles and refractory elements is not well constrained.
The origin of this depletion pattern is debated.
\citet{Waters1992} suggested that the dust is separated from the gas at the inner rim of the disk, because of the high radiation pressure exerted on the dust grains that are not coupled to the gas (at fairly low accretion rates; M$_\mathrm{acc}$<10$^{-6}$M$_\odot$/yr).
Finally, it seems that the accretion of the matter from the circumbinary disk can prolong the life-time of the evolved star on the post-AGB branch by a factor of between 2 and 5 \citep{Oomen2019}.

Another interaction mechanism has been revealed by phase resolved line-profile studies,  showing that at a superior conjunction, a fast outflow originating around the companion is seen in the absorption of the background luminous primary \citep{Gorlova2015, Bollen2017, Bollen2019}. 
The jets with wide opening angles are found in almost all cases where there is enough spectroscopic monitoring data available. The jets are proven to have been launched by an accretion disk around the companion. This disk was also directly detected by infrared interferometry \citep{Hillen2016}. 
The magnitude of accretion is such that the likely accretion source is indeed the circumbinary disk \citep{Bollen2020}, strengthening the link between accretion and depletion patterns.


The presence of stable circumbinary disks in the evolved systems has been well established thus far, their role in the (orbital) evolution of the central stars is far from clear. 
The disks and the disk-binary interactions do appear to be key ingredients in improving our understanding of the late evolution of a very significant binary population.

Despite the evidence for dust grain growth \citep{Sahai2011,Gielen2008,Gielen2011,Scicluna2020}, the evolution of disks around post-AGB binaries is currently not well studied.
On the contrary, the evolution of disks around young stellar objects was extensively investigated both theoretically and via observations. 
As we partly use the framework developed for the study of disks around young stars here (see Sect.\,\ref{sec:nomenclature}), we briefly describe what is known on the evolution of disks around young stars.
While protoplanetary disks around young low-mass stars were first discovered through the presence of a continuous infrared excess from the near-infrared \modifLE{(NIR)} to the millimeter  \citep[as in ][for ex.]{Dalessio1999}, there was also evidence that not all disks extend from the dust sublimation radius outward.
It was later discovered that around 10\% of disks around young stars share a lack of near-infrared excess -- which was interpreted as the disks having large inner dust cavities. These disks are called transition disks, as they are thought to be in a transition phase between full disks and a later stage of gas-less debris disks \citep[e.g.,][]{Strom1989,Calvet2002,Calvet2005}.
Some disks with a small mid-infrared \modifLE{(MIR)} excess combine this with an excess in the near-infrared as well as at longer wavelengths, thereby forming the class of pre-transitional disks \citep{Esapaillat2010}.

Similar disk classes were also found around intermediate-mass young stars, namely, Herbig stars \citep{Herbig1960}, where group~I sources displayed an excess that could be fitted by a blackbody and a power law and group~II by a power-law only \citep{Meeus2001}.
These groups were first interpreted as group~I disks being younger and still significantly flared, while group~II were older flat disks.
However, the evolutionary sequence changed with accumulating evidence for gaps in group~I disks having produc\modifLE{e}d a puffed-up inner rim of the outer disk that increases the surface \modifLE{that emits} in the mid-infrared \citep{Maaskant2013,Menu2015}.
The advent of high-spatial resolution observations of disks with the Atacama large (sub-)millimeter array (ALMA) and adaptive optics direct imaging instruments have confirmed the cavities of transition disks \citep{Andrews2011} and have also shown a broad range of substructures such as gaps, rings, spirals, misaligned disks, and vortices \citep[e.g.,][]{vanderMarel2013,Menu2015,Benisty2017,Benisty2018,Andrews2018}, as recently reviewed in \citet{Andrews2020}.

It is interesting to note that depletion is also observed in young stellar objects or young main sequence stars, although that is often at a much lower amplitude \citep[e.g.,][]{Kama2015,Booth2020}. For instance, the chemically peculiar $\lambda$\,B\"oo stars \citep[e.g.,][]{Murphy2017} are also linked to the depletion of refractory elements.

Despite this great observational and theoretical effort, the evolutionary scenario for disks around young stars is still not entirely clear.
The disks are expected to evolve viscously \citep{LyndenBell1974,Hartmann1998} and dissipate quickly via photo-evaporation \citep[][and references therein]{Alexander2014}.
However, all disks probably do not go through a transition disk phase. 
It was shown that the origin of a transition disk is more likely related to the presence of an additional (stellar or sub-stellar) component creating the cavity \citep[e.g.,][]{Biller2012,Lacour2016,Norfolk2021}, although some of the transition disks are still consistent with photo-evaporation due to high-energy photons from the central star \citep[e.g.][]{Owen2011,Ercolano2017}.
It seems, however, that massive disks produce massive planets and evolve through the transition disk phase \citep{vanderMarel2018,Cieza2021}, while less massive disk may evolve through viscous disk dissipation and photo-evaporation, possibly avoiding the transition disk phase.

In this paper, we focus on disks around post-AGB binaries. 
First, we present some nomenclature that we use in this paper to describe the targets in Sect.\,\ref{sec:nomenclature}.
In Sect.\,\ref{Section:Catalog} we present what is, to our knowledge, a complete catalog of all the known Galactic optically bright post-AGB disk sources. 
Subsequently, we focus on the associated SEDs and investigate observational evidence for disk evolution within this full sample. 
We display the population of SEDs in Sect.\,\ref{Section:SEDs} and characterize them in an infrared color-color diagram with the help of a synthetic disk population in Sect.\,\ref{sec:carac}. 
Finally, we discuss the results and include a comparison between the SED and other system characteristics, to isolate the evidence supporting disk evolution in Sect.\,\ref{sec:discussion}. We present our conclusions in Sect.\,\ref{sec:conclusion}.

\section{Nomenclature}
\label{sec:nomenclature}

As this paper is at the boundary between binary evolution and protoplanetary disks studies, we define several terms that we use in the paper. 
The goal is to remove any ambiguities associated with the use of these terms as they can have different meaning in different communities.
\textit{Post-AGB binary} refers here to the evolved star of the binary. It can be a post-AGB or post-RGB star as it is believed that the binary interaction may stop the giant branch evolution of the evolved star \citep{Kamath2015,Kamath2016}.
\textit{Depletion}: small photospheric abundances of refractory elements compared to volatiles. The depletion correlates with condensation temperature.
    \textit{Protoplanetary disk}: a disk of dust and gas that will produce planets whatever the evolutionary stage of the central star(s).
    \textit{Full disk}: a disk that has dust that extends continuously from the dust sublimation radius to the outer radius.
    \textit{Transition disk}: a disk that has the inner dust rim significantly larger than the dust sublimation radius. These disks are also characterized by the lack of near-infrared emission from very hot dust. This description is independent on the origin of the transition disk morphology.
    \textit{Cavity}: lack of dust within a certain radius from the central star(s). This radius is significantly larger than the theoretical dust sublimation radius.




\section{Complete Galactic catalog}
\label{Section:Catalog}

We limited ourselves to optically bright Galactic stars classified as post-AGB or post-RGB objects and we used three criteria in the selection process. The complete list was assembled throughout the years based on both systematic research as well as findings on individual objects.  Firstly, we selected evolved objects with spectral types warmer than late K, M, or C. These are typical for AGB stars. Secondly, the SED needs to points to the presence of hot dust in the system. This hot dust creates an infrared excess that is already starting to appear in the near-infrared. 
We complemented this selection with evolved objects whose photospheres are affected by the chemical anomaly called depletion.  Finally, we added spectroscopic binaries from the sample of optically bright post-AGB stars for which the dust excess was not detected.
In Table~\ref{Table:list} we list the objects and the main reference we used to select the object for this catalog. In what follows, we report on the reconstruction of the full list.

We build on the systematic survey by \citet[and references therein]{DeRuyter2006} who selected post-AGB stars with hot dust. 
We retained 49 of the 51 sources from this catalog.
We omitted the M-type stars IRAS09400$-$4733 and IRAS\,10174$-$5704, as these are too cool to be considered to be post-AGB objects.
We complemented the list with the dusty objects labeled as "disk objects" from \cite{Gezer2015}. 
The latter was a systematic study focusing on RV\,Tauri pulsators, which are the most luminous population II Cepheids.  
One of the main conclusions of this study was that dusty RV\,Tauri stars are mainly disk sources. 
Hence, we added another 16 objects from this list. 
More recent additions to this list of optically bright objects with a strong and hot dust excess was HD\,101584 \citep{Olofsson2015,Kluska2020b} and objects with a characteristic infrared (IR) excess that have not yet been published (assigned as "this work") but have been selected in the framework of the long-term radial velocity monitoring program we run at our Mercator telescope \citep{VanWinckel2010, Oomen2018}.

Additionally, we also searched in the literature for examples of objects which display depleted photospheric abundance patterns. 
Clearly depleted objects complementing the references above came from \citet[one object]{Gezer2019}, and \citet[one object]{Giridhar2005a}, \citet[one object]{Klochkova1996}, \citet[one object]{Maas2007}, \citet[two objects]{Giridhar2010}, and \citet[one object]{Manick2021}. 
Many of these systems turned out to display an infrared excess as well, although it is typically much less strong.
Lastly, we included in the list those objects labeled as post-AGB stars, for which their binary nature has been well established via radial-velocity programs \citep{Oomen2018}. 

Our list consists of 85 Galactic sources and to our knowledge, this is the list of all such Galactic objects known to date. 
The observational bias lies toward the presence of an infrared excess and the presence of depleted photospheres. 
Given the fact that central stars are proven or suspected binaries, we find that the Gaia distances are still too unreliable to aim at a volume-limited sample or to quantify the completeness.
Given that there are 209 likely Galactic post-AGB stars and 112 Galactic RV\,Tauri pulsators \citep{Szczerba2007} we highlight that the established figure of 85 post-AGB stars is quite large compared with other, similar categories of stars.


\section{SEDs and fitting of the photosphere}
\label{Section:SEDs}

\begin{figure}[t]
    \centering
    \includegraphics[width=9.5cm]{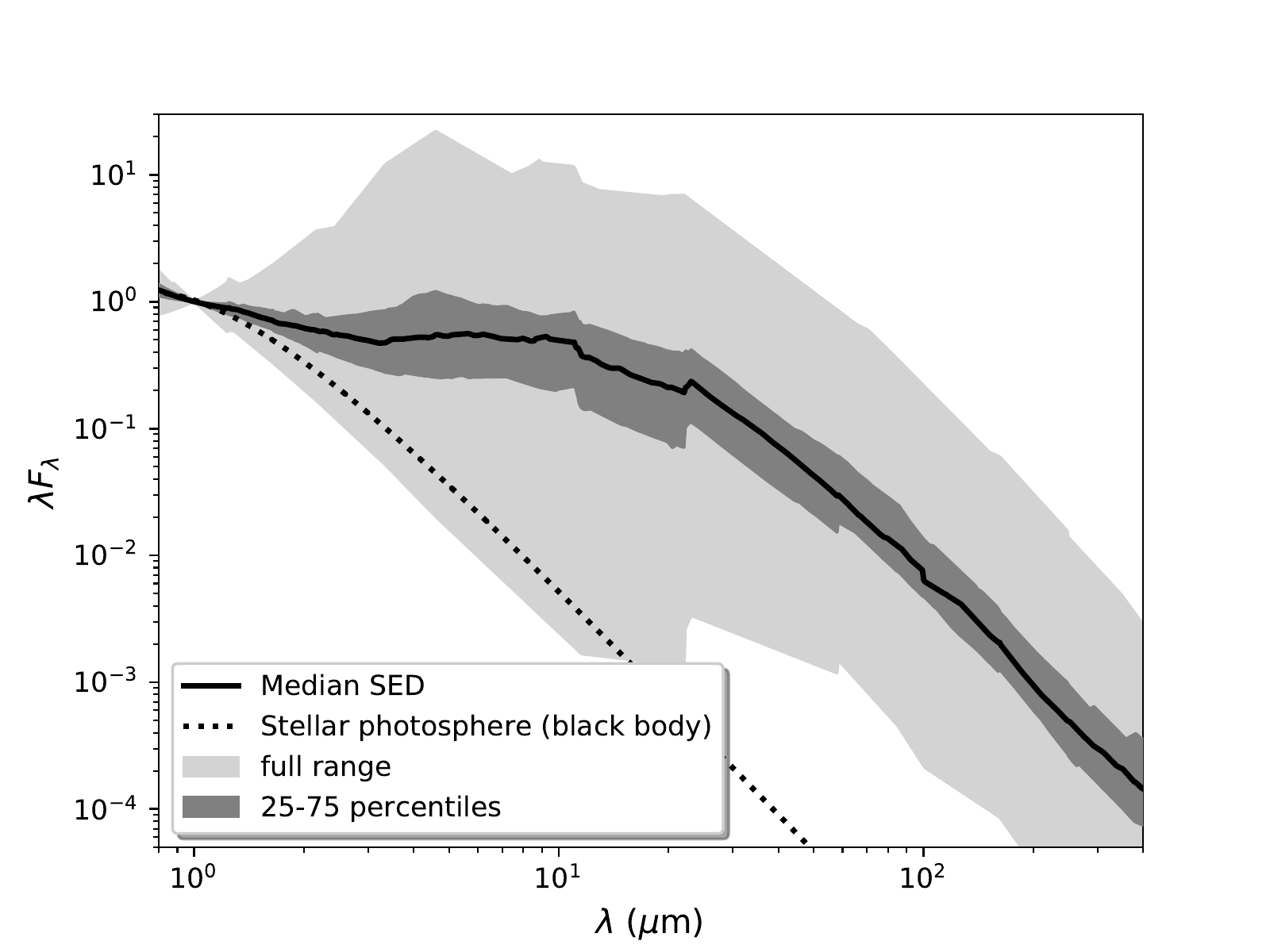}
    \caption{Median SED built by linearly interpolating between the photometric points of all SEDs normalized to unity at 1\,$\mu$m. The dark gray area represents the SED between the 25th and 75th percentiles while the light gray are represents the full range of SEDs. A black-body with a temperature of 5500\,K is represented for comparison.}
    \label{fig:medianSEDall}
\end{figure}

We used the same method as \cite{Oomen2018} to construct all the SEDs for this study. 
In brief, we started by determining the total line-of-sight reddening by minimizing the difference between the optical fluxes and the reddened photospheric models, the parameters of which were found in the spectroscopic analyses found the literature. 
We interpolate in the $\chi^2$ landscape between the models centered around the spectroscopically determined parameters and applied ranges of $\Delta T_\mathrm{eff}$ $\pm$ 500K, $\Delta\log g$ $\pm$ 0.5, $\Delta [\mathrm{Fe/H}]$ $\pm$ 0.5. 
We used the Kurucz photospheric models, as determined in \citet{Castelli2003}\footnote{ \href{https://wwwuser.oats.inaf.it/castelli/grids.html}{https://wwwuser.oats.inaf.it/castelli/grids.html} }. 

The photometric data points are automatically retrieved from the Vizier database \citep{Ochsenbein2000}. 
We refer to the Appendix A of \cite{Oomen2018} for a list of the most common catalogs. 
We expanded the wavelength coverage by including the data obtained by Spectral and Photometric Imaging REceiver \citep[SPIRE;][]{Griffin2010}, Photodetector Array Camera and Spectrometer \citep[PACS;][]{Poglitsch2010} for which we applied aperture photometry in the Herschel Interactive Processing Environment (HIPE) environment to extract the fluxes \citep{Dsilva2019}. 
For some sources we obtained 
Sub-millimeter Common-User Bolometer Array \citep[SCUBA-1 and 2][]{Holland1999, Holland2013} sub-millimeter fluxes, which are in part previously unpublished; while others are from \citet{DeRuyter2005}. 
In the SED plots, we color-coded the different photometric systems and sources. 
The full list of SEDs is given in Appendix A.
To complement the SEDs we also included the infrared spectra as obtained by 
Infrared Space Observatory \citep[ISO;][and references therein]{Molster2002, VanWinckel2003}, as well as Spitzer \citep{Gielen2008, Gielen2011, Gielen2011b, Kemper2010}, when available.

Many of the targets are photometric variables and, especially, the RV\,Tauri pulsators show often high pulsation amplitudes that can go up to several magnitudes in V. 
The SEDs of these pulsators are hence intrinsically uncertain because we do not have a full coverage of all photometric bands in the same pulsation phase. 
Reddening determinations are therefore notoriously difficult. 
For this sample study, we determined the SED in a strictly homogeneous way for all objects.
We expect that the pulsations will affect a small fraction of the targets and that a group study of the infrared excesses is still possible.
We therefore neglect this effect in the rest of the paper.

\section{Characterizing the disks through infrared colors}
\label{sec:carac}

In this section, we investigate what the common disk features are for the whole population of disks in the Galaxy using the SEDs.
The first disks that were observed with infrared interferometry and modeled with a radiative transfer code have either their inner dust rim at the theoretical dust sublimation radius \citep[89\,Her and IRAS08544-4431;][]{Hillen2014,Kluska2018} or one that is several times larger \citep[AC\,Her;][]{Hillen2015}, similarly to transition disks found around young stars. 
Inspired by these targets, we performed our investigation using a color-color diagram in the infrared to probe the inner disk parts and check whether the aforementioned targets are representative of the larger sample.

\subsection{Color-color diagram}


We use the color-color diagram with near-infrared colors, $H$-$K$s from the 2MASS photometry, to probe the presence of emission close to dust sublimation radius and the Wide-field Infrared Survey Explorer (WISE) mid-infrared colors, $W1$-$W3$ for emission further away.
For IRAS08544-4431, the WISE band W1 is saturated and was replaced by 
South African Astronomical Observatory (SAAO) magnitude in the $L$ band. 
For 89\,Her, we used the Cosmic Background Explorer/Diffuse Infrared Background Experiment (COBE/DIRBE) measurement in the F3.5 filter instead of the WISE W1 filter.
As no such combination of colors exist for IRAS10456-5712, this target is not present in our diagram and was therefore not included in further analysis.

\begin{figure}[t]
    \centering
    \includegraphics[width=10cm]{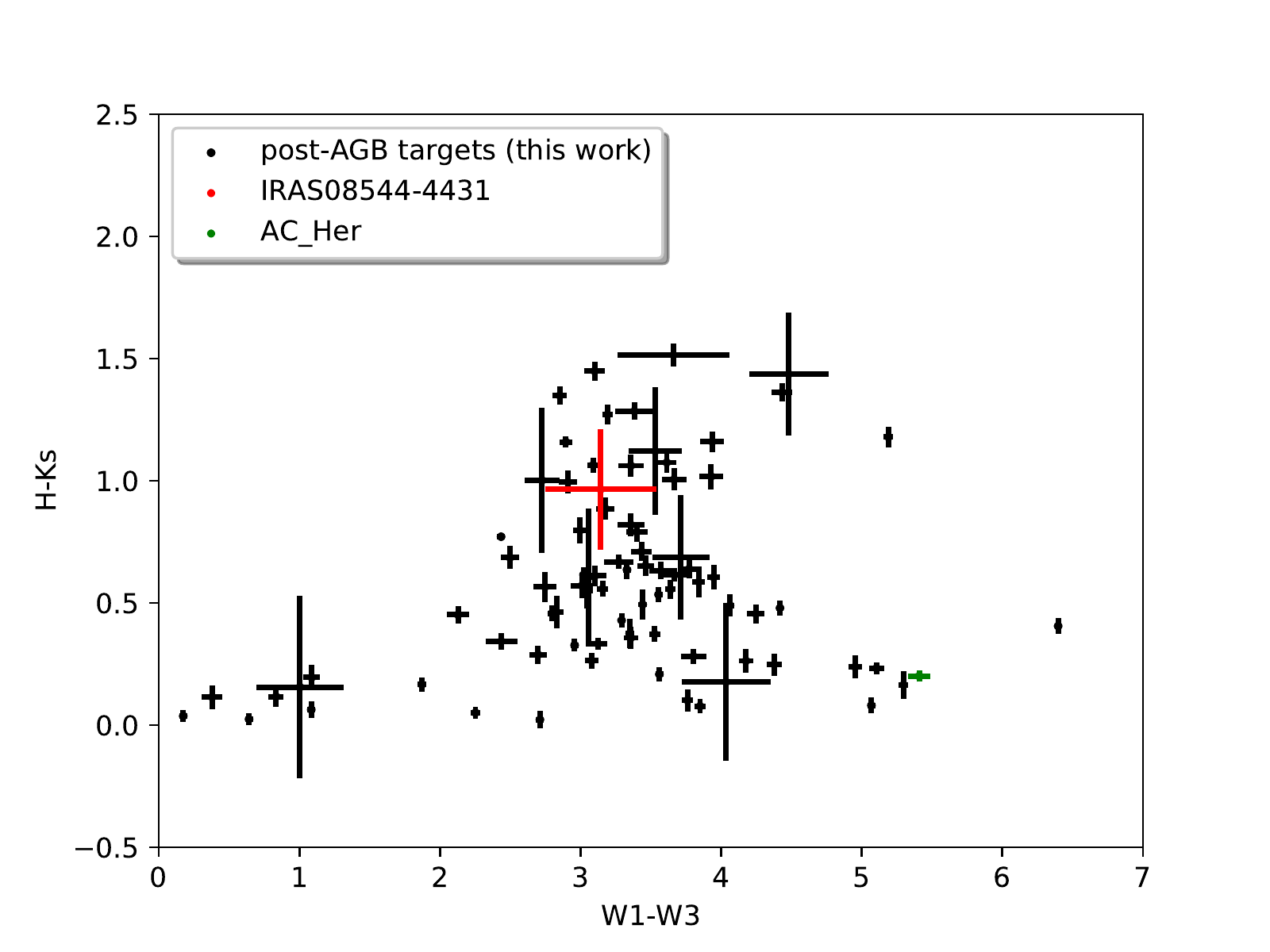}
    \caption{Infrared color-color diagram for galactic post-AGB targets. Arbitrarily selected representative targets are indicated with different colors.}
    \label{fig:ccdata}
\end{figure}

The color-color diagram with our observed disk population is displayed in Fig\,\ref{fig:ccdata} and is significantly structured.
Most of the targets are located in a narrow range of mid-infrared colors (3 < $W1$-$W3$ < 4) and within a larger range of near-infrared colors (0 < $H$-$K$s < 2).
We can find in this location of the color-color diagram IRAS08544-4431 and 89\,Her that have hot dust emission with an inner disk rim at the dust sublimation radius.
There is also a strip of targets, including AC\,Her, with low values of near-infrared colors (0 < $H$-$K$s < 0.5) but with a large range of mid-infrared colors (0 < $W1$-$W3$ < 7).

\subsection{Synthetic disk population}
In order to interpret the observed distribution of targets in the color-color diagram, we used the radiative transfer code MCMax \citep{Min2009} to produce a synthetic population of disk models.
Exploring all the possible disk parameters is outside the scope of this paper.
Instead, we want to see whether variations in the best-known models for both a full disk and a disk with a large cavity are sufficient to reproduce the observed infrared color-color diagram.
In this study, we used it to populate the color-color diagram with models and compare their locations to the observed population.
The near and mid infrared emission is dominated by the inner disk regions. If the disk inner rim is located at the dust sublimation radius, then the infrared excess will start in the near-infrared, where the thermal emission peaks at the typical dust sublimation temperature ($\sim\,1500$\,K).
If the disk has a larger cavity, the infrared excess will start at longer wavelengths, depending on the temperature of dust located at the disk inner rim.
Our models are inspired by the  ones used for the full disk of IRAS\,08544-4431 \citep{Kluska2018} and the transition disk of AC\,Her \citep{Hillen2015}.

In particular, MCMax is a radiative transfer code that computes the disk scale height as resulting from hydrostatic equilibrium.
The details of the code and the disk parametrization can be found in the above-cited papers, but we recall here the main properties of the disk models.
The dusty disk extends from the inner disk radius, $R_\mathrm{in}$, to the outer disk radius, $R_\mathrm{out}$.
We used a double power-law with the radius to define the disk surface density, as in \citet{Hillen2015} for example.
The indices are $p_\mathrm{in}$ for the inner disk, which extends from $R_\mathrm{in}$ to the radius at which the surface density power-law changes, $R_\mathrm{mid}$ and $p_\mathrm{out}$, for the outer disk between $R_\mathrm{mid}$ and $R_\mathrm{out}$.
It results in the following surface density law at a given radius $r$: 
\begin{eqnarray}
     \Sigma(r) \propto 
     \begin{cases} 
    r^{p_\mathrm{in}} \text{ for $R_\mathrm{in} < r < R_\mathrm{mid}$ } \\
      r^{p_\mathrm{out}} \text{for $R_\mathrm{mid} < r < R_\mathrm{out}$}
     \end{cases}
,\end{eqnarray}
such that the value of surface density is continuous at $R_\mathrm{mid}$ and the total disk mass equals $100\times M_\mathrm{dust}$.

The dust composition is assumed to be a mixture of silicates in the distribution of hollow spheres approximation and a size distribution between $a_\mathrm{min}$ and $a_\mathrm{max}$, following a power-law with an index $-q$ such that the number density of the dust grains ($n_\mathrm{\text{$a_\mathrm{d}$}}$) is defined as:
\begin{equation}
    n_\mathrm{\text{$a_\mathrm{d}$}} \propto a_\mathrm{d}^{-q}.
\end{equation}

A larger  value of $q$ indicates a grain distribution dominated by smaller grains. Dust settling is included self-consistently using the $\alpha_\mathrm{turb}$ parameter \citep{Mulders2012} following the \citet{Shakura1973} prescription.
We chose to use a dust composition with a mixture of astronomical silicates \citep{Min2007}.
We did not explore other dust mixtures as there are only six targets that show carbonaceous dust in a very peculiar form as these are polycyclic aromatic hydrocarbons (PAHs) and Fullerenes \citep{Gielen2011b} and, more importantly, because the bulk composition of these disks remains oxygen rich \citep{Gielen2011}. 
We also do not expect the dust mixture to have a crucial effect on the analysis we perform here. 

\begin{figure}[t]
\begin{center}
\includegraphics[scale=0.45]{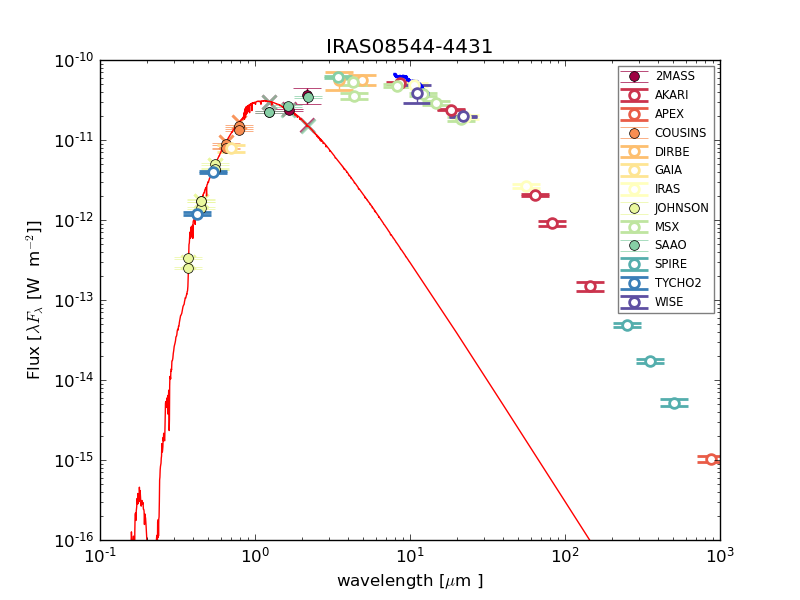}
\includegraphics[scale=0.45]{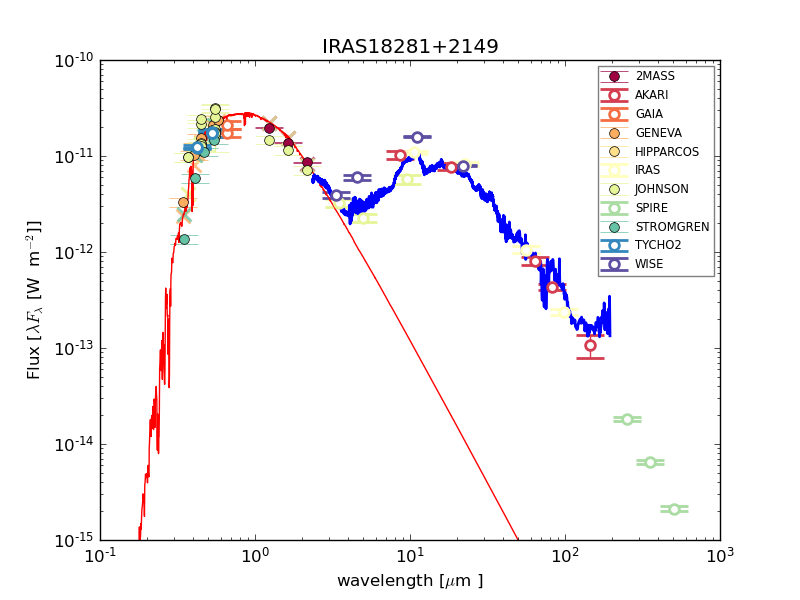}
\caption{\modifLE{Examples of SEDs of post-AGB binaries.} Top panel: IRAS08544-4431. Lower panel:\ IRAS18281+2149 (AC\,Her). Labels show the sources of photometric data points (see also Section~\ref{Section:SEDs}). We used these well-studied objects as prototypes for objects where the inner rim of the disk starts at sublimation radius (top panel) and objects where this inner rim is further out (bottom panel). The wavelength where the dust excess starts to appear is used as observational signature. }\label{Fig:SED2} 
\end{center}
\end{figure}


\begin{table*}[t]
    \centering
        \caption{Parameters of the radiative transfer model grid used to generate a synthetic population of post-AGB disks.  }
    \begin{tabular}{c c c   c c}
    \hline
    Parameters     &  \multicolumn{2}{c}{IRAS08544-4431} & \multicolumn{2}{c}{AC\,Her}\\
    \hline
     \multicolumn{5}{c}{Stellar properties}\\
    \hline
    $T_\mathrm{eff}$ (K) & \multicolumn{2}{c}{7250 $\pm$ 250} & \multicolumn{2}{c}{5800 $\pm$ 250} \\ 
    $\log g$ (dex) & \multicolumn{2}{c}{1.5$\pm$0.5} & \multicolumn{2}{c}{1.0$\pm$0.5} \\
    $Z$ (dex) & \multicolumn{2}{c}{-0.3$\pm$0.3} & \multicolumn{2}{c}{-1.4$\pm$0.3} \\
    E(B-V) (mag) & \multicolumn{2}{c}{1.32} & \multicolumn{2}{c}{0.46} \\
    L$_*$ (L$_\odot$) & \multicolumn{2}{c}{(1.4$^{+0.4}_{-0.3}) \times$ $10^4$}& \multicolumn{2}{c}{(2.4$\pm0.2) \times$ $10^3$}  \vspace{0.1cm} \\
    \hline
    \multicolumn{5}{c}{Disk parameters}\\
    \hline
    & Best fit & Grid & Best fit & Grid \\
    \hline
    $M_\mathrm{dust}$ ($M_\odot$) & 2$\times10^{-3}$  & [2$\times10^{-4}$, 2$\times10^{-3}$, 2$\times10^{-2}$] & 2.5$\times10^{-3}$ & [2.5$\times10^{-4}$, 2.5$\times10^{-3}$, 2.5$\times10^{-2}$]\\
    $R_\mathrm{in}$ (au) & 8.25 & [6.60, 7.43, 8.25, 9.08, 9.90] & 34 & [27.2, 34.0, 40.8]\\
    $\frac{R_\mathrm{mid}}{R_\mathrm{in}}$ & 3 & [2, 2.5, 3] & 2.5 & [2, 2.5, 3]\\
    $R_\mathrm{out}$ (au) & 200 & \ldots & 200 & \ldots \\
    $a_\mathrm{min}$ ($\mu$m) & 0.1 & \ldots & 0.01 & \ldots \\
    $a_\mathrm{max}$ (mm) & 1 & [1, 10] & 1 & [1, 10]  \\
    $p_\mathrm{in}$ & $-$1.5 & [$-$2.5, $-$1.5, $-$0.5] & $-$3 & [$-$4, $-$3, $-$2]\\
    $i$ ($^\circ$) & 19 & [0, 35, 50, 64, 75] & 50 & [0, 35, 50, 64, 75] \\
    $q$ & 2.75 & [2.2, 2.75, 3.3] & 3.25  & [2.6, 3.25, 3.9]\\
    $\alpha_\mathrm{turb}$ & 0.01 & \ldots & 0.01 & \ldots
    \end{tabular}
    \tablefoot{The models are based on the best fit models of IRAS08544-4431 \citep{Kluska2018}, representing full disks, and AC\,Her \citep{Hillen2015}, representative of transition disks.}
    \label{tab:models}
\end{table*}

\textbf{Full disk models based on IRAS08544-4431}: a full disk model with the inner rim starting at the sublimation radius. The surface density power-law inflection happens at 24.75\,au. This model was fitted to the SED as well as to long-baseline interferometric data in the near-infrared. However, an ad-hoc extended component with a black-body temperature of 1000\,K and 10\% of the total $H$-band flux has been added to reproduce the data.

\textbf{Transition disk models based on AC\,Her}: a transition disk model with a cavity of 34\,au, around seven times larger than the theoretical dust sublimation radius. It was fitted to SED data together with mid-infrared interferometric data. 

Starting with these two types of models for full and transition disks, we explored a range of parameters with those that have a potential influence on both the near-infrared and mid-infrared colors \citep[e.g.,][]{Woitke2016}.  
Our goal here is not to explore all the possible models, nor to carry out a full-parameter study but, rather, to see whether slight changes in some key parameters can populate similar areas of the color-color diagram as the observed targets do.
Thus, we  made a grid by changing the following parameters exhibiting an influence on the inner disk morphology: the inner disk radius, $R_\mathrm{in}$; the radius at which the surface density power-law changes ,$R_\mathrm{mid}$; and the inner surface density power law index, $p_\mathrm{in}$.
We also changed several global disk parameters, such as the disk dust mass, $M_\mathrm{dust}$; the size of large grains, $a_\mathrm{max}$; the dust size distribution power-law, $q;$ and the disk inclination, $i,$ equally spaced in $\cos{i}$.
We varied each parameter in ranges around the best-fit values for each model (see Table\,\ref{tab:models}).
We therefore generated 6480 models composed of 4050 full disk models and 2430 transition disk models.
The effect on the color-color diagram of each of the parameters is displayed in Fig.\,\ref{fig:CCeffect}.

\subsection{Results}


The synthetic population we created by varying the different disk parameters is sufficient to populate the color-color-diagram (Fig.\,\ref{fig:CCmodels}, top) in a similar way to the observations (Fig.\,\ref{fig:CCmodels}, bottom).
There is a clear distinction between full disk models that cover a narrow range of mid-infrared colors between 2.5 and 4.5, and transition disk models that have either a near-infrared color between 0.0 and 0.3 or a mid-infrared color larger than 4.5.
We have also tested the influence of the ad-hoc extended component on the position of the full disks in the color-color diagram. 
Without the addition of such an extended component, the full disk models have their near-infrared color reduced by at most $\sim$0.3 and their mid-infrared color increased by $\sim$0.3 (Fig.\,\ref{fig:OReffect}).
While it is necessary to include an extended component of 10\% of the flux to model IRAS08544-4431, it is unknown whether such a component is needed for other targets as well.
It was shown, however, that 50\% of the 23 targets of the near-infrared interferometric sample require more than 5\% of such an extended emission to reproduce the data \citep{Kluska2019}.
There is no extended component needed to model AC\,Her, therefore, we did not incorporate it into the transition disk models.


\section{Categories and correlation analysis}
\label{sec:analysis}

In this section, we analyze the properties of our sample by defining a categorization based on the infrared colors of both our synthetic model population and the observed one (Sect.\,\ref{sec:categories}).
Subsequently, we look for correlations between the categories and several other observables, such as the infrared excess luminosity (Sect.\,\ref{sec:IRlum}), the RVb phenomenon (Sect.\,\ref{sec:RVb}), stellar and orbital parameters (Sect.\,\ref{sec:nocor}), and depletion tracers (Sect.\,\ref{sec:depletion}).

\subsection{SED categories}
\label{sec:categories}

\begin{figure}[t]
    \centering
    \includegraphics[scale=0.55]{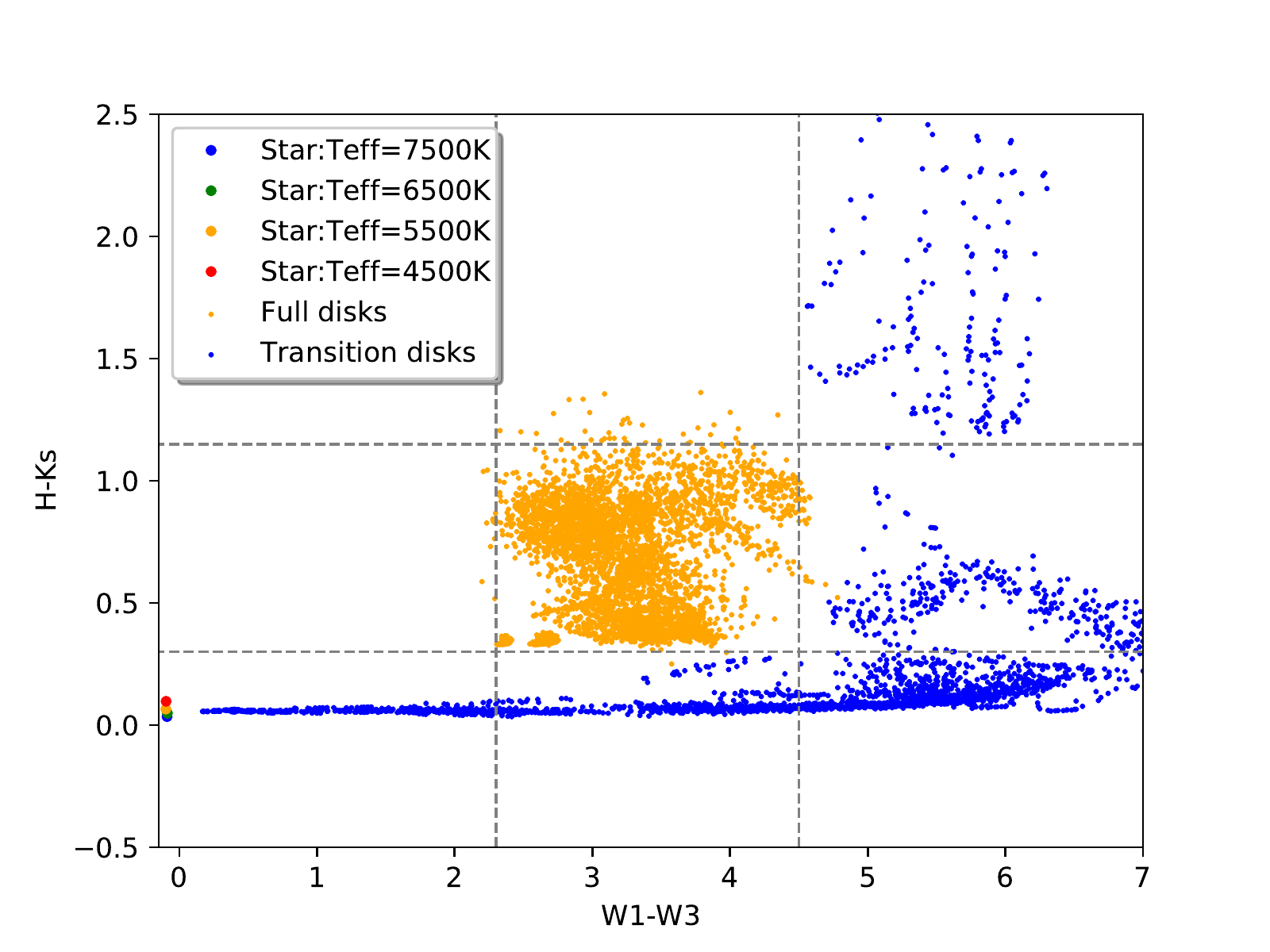}
    \includegraphics[scale=0.55]{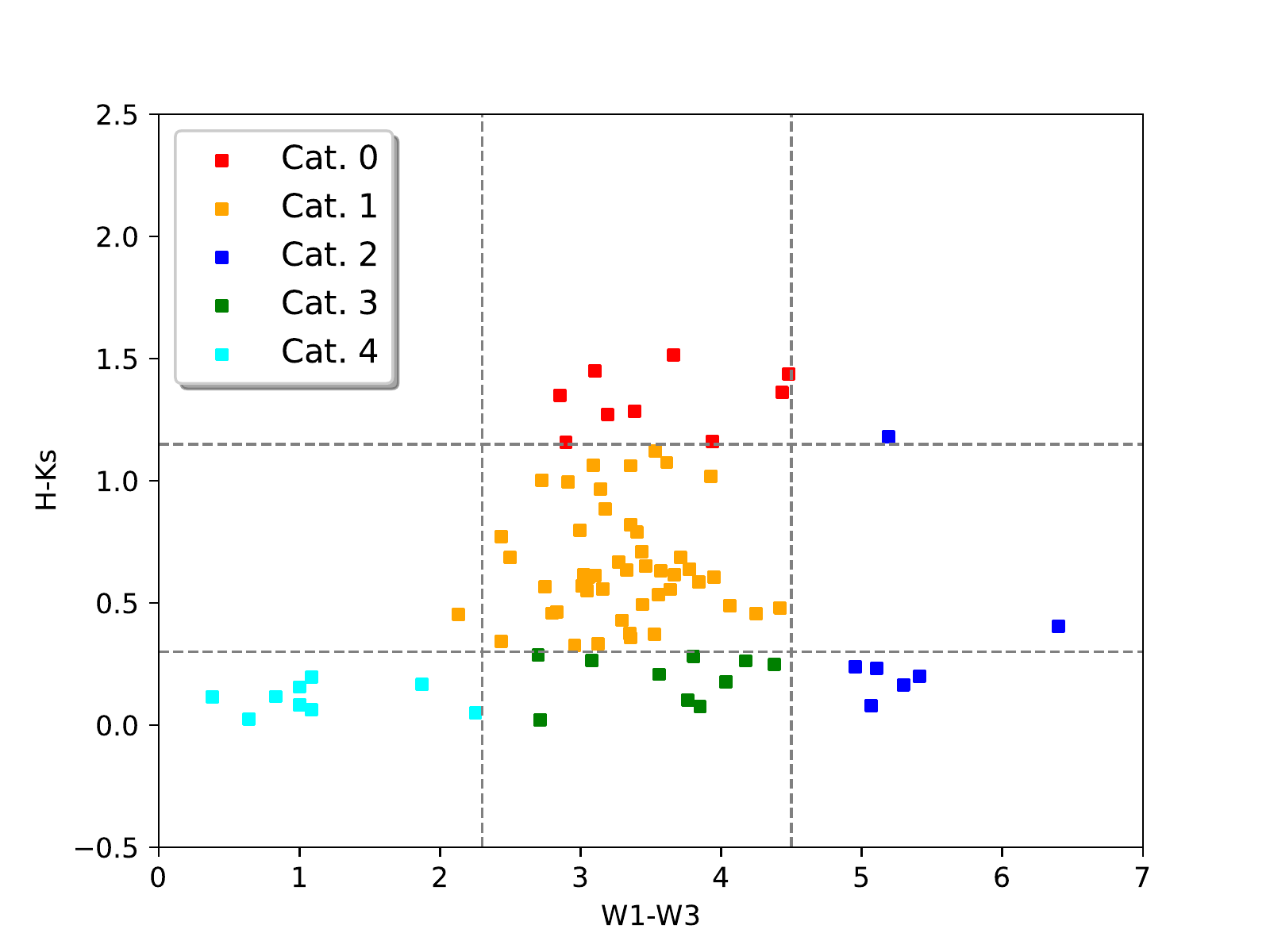}
    \caption{\modifLE{Color-color diagrams for models and targets.} Top: Positions of the models of IRAS08544-4431-like (orange) and AC\,Her-like (blue) disks in the color-color diagram. The dashed gray lines represent the boundaries between the different categories defined in the text. Bottom: Locations of the observed disks color-coded with their assigned categories. }
    \label{fig:CCmodels}
\end{figure}

Given the structure in the color-color diagram, we defined five categories to classify the targets and help us interpret the results.
We used the positions in the color-color diagram of the population of synthetic models to delineate the categories.
The goal of these categories is not to unambiguously classify every target but to compare the group properties of each category.
We define the categories in Table\,\ref{tab:categ} (see Fig\,\ref{fig:CCmodels}).
In the following subsections, we explore and compare the properties of targets in each category.

\begin{table*}[]
   \caption{Categories defined in the study.}
    \centering
    \begin{tabular}{c|c|c|l}
        Category & NIR & MIR & Description  \\
        \hline 
        \rule{0pt}{1.0\normalbaselineskip} 0 & $H$-$K$s > 1.15 & W1-W3 < 4.5 & This region of the color-color diagram is populated by  \\
        &&&several observed targets and is on top of the full disk models.\\
        1 & 0.3 < $H$-$K$s < 1.15 & W1-W3 < 4.5 & It encompasses the area occupied by models of full disks. \\
        2 & \ldots & W1-W3>4.5 & It includes all the models that have high mid-infrared colors \\
        &&&and that are populated by variations of models of AC\,Her, namely, \\
        &&&of disks with large cavities. \\
        3 & $H$-$K$s < 0.3 & 2.3 < W1-W3 < 4.5 & It covers the targets with low near-infrared colors and medium \\
        &&&mid-infrared colors. This area is populated by transition disk models but \\
        &&& include can also full disk models without the ad-hoc extended component. \\
        4 & $H$-$K$s < 0.3 & W1-W3 < 2.3 & It includes targets with low near-infrared and mid-infrared excesses. \\
        &&&It is occupied by models of transition disks. \\
    \end{tabular}
 
    \label{tab:categ}
\end{table*}

\subsection{Infrared luminosity}
\label{sec:IRlum}

\begin{figure}[t]
    \centering
    \includegraphics[scale=0.55]{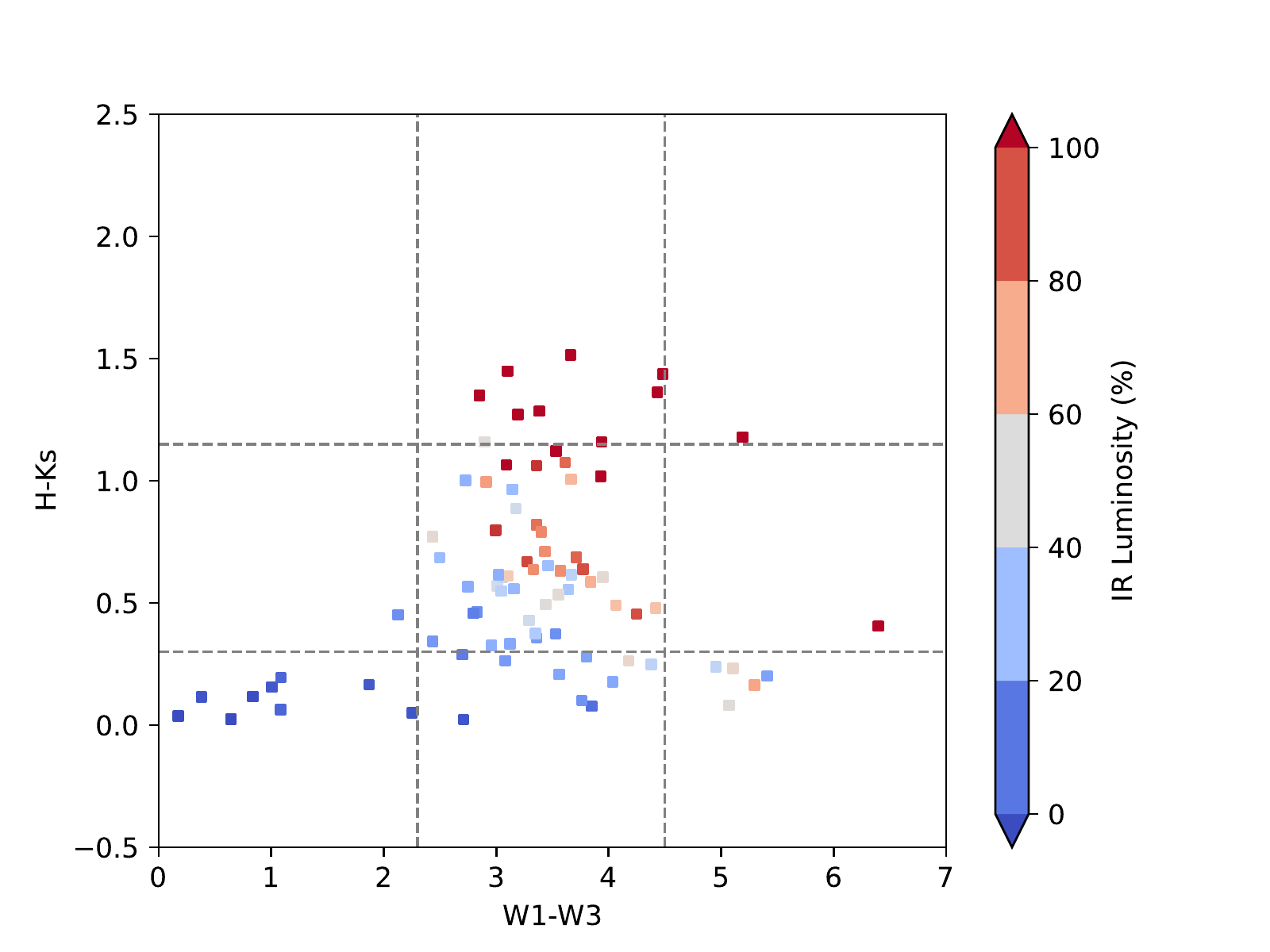}
    \caption{Distribution of the targets with respect to the infrared to stellar luminosity ratio that is color-coded. The dashed gray lines represent the boundaries of the object categories. }
    \label{fig:IRLum}
\end{figure}

The infrared excess luminosity ratios is computed by integrating under the photometric points and subtracting the stellar luminosity using the best-fit photospheric fit (described in Sect.\,\ref{Section:SEDs}).
We computed the excess-to-stellar luminosity ratio that we express in percentage.
The color-color diagram clearly discriminates between targets with high and low infrared excess luminosities (Fig.\,\ref{fig:IRLum}).
Indeed, targets with high infrared-to-stellar luminosity ratios (>100\%) belong to category 0, which populates the top of the color-color diagram with high near-infrared colors (H-Ks>1.15).
We can also see (in Fig.\,\ref{fig:IRLum}) a gradient in the infrared-to-star luminosity ratio across the color-color diagram reaching 0\% for low near-infrared and mid-infrared colors.
This is expected,  as low infrared colors indicate no or low infrared excess at wavelengths that are shorter than $\sim$20$\mu$m.

\begin{figure}[t]
    \centering
    \includegraphics[scale=0.55]{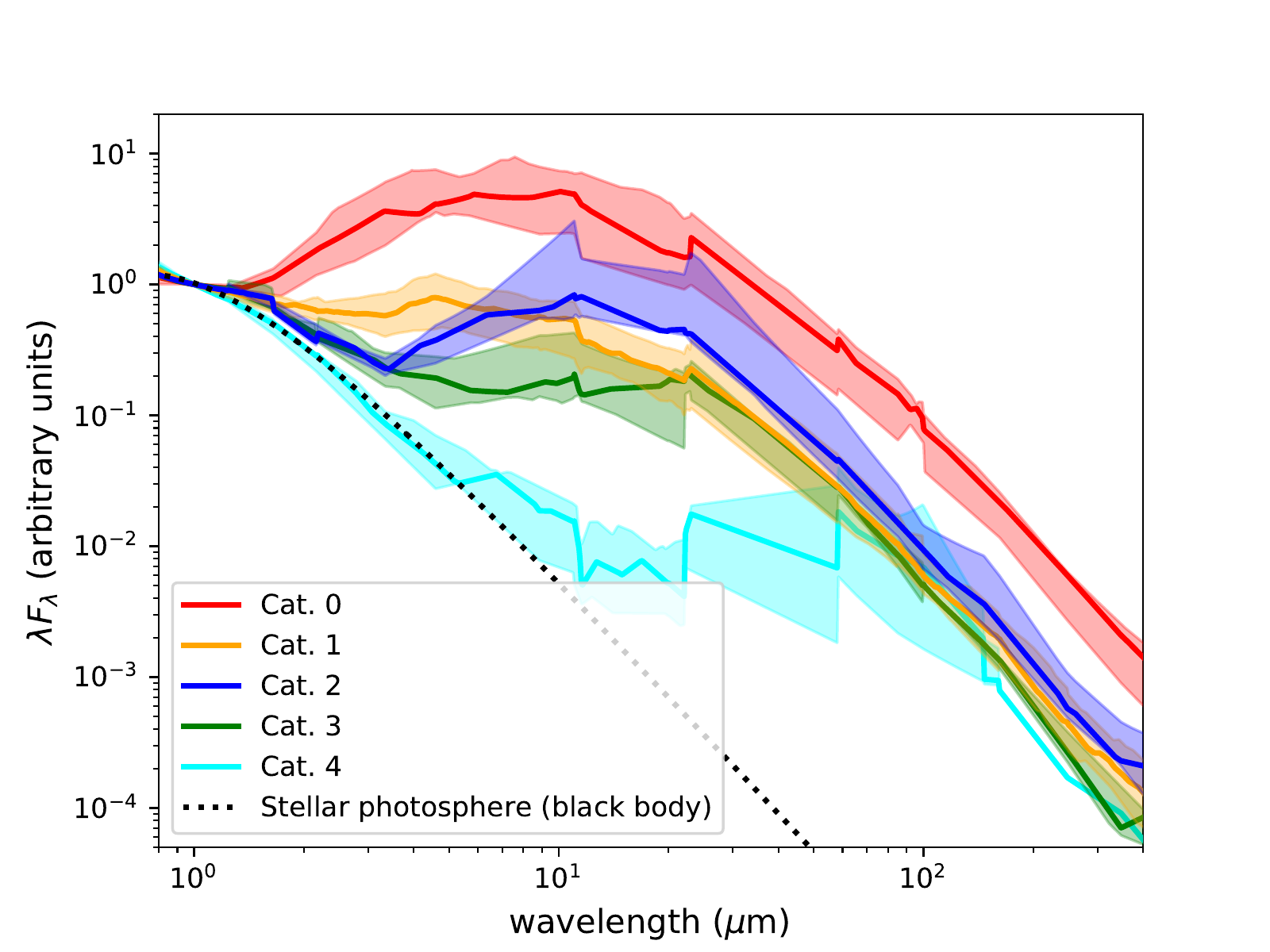}
    \caption{\modifLE{Median infrared excess in SEDs.} Solid lines represent the median SED for each category normalized at 1$\mu$m. The SEDs are enveloped by an area representing the 25th-to-75th percentiles SED of each category. A black body of T=5500\,K is plotted for illustration to mimic a stellar photosphere.}
    \label{fig:IRexcesses}
\end{figure}

In Fig.\,\ref{fig:IRexcesses}, we have plotted the median and 25th to 75th percentiles SED for each category.
Category 0 has the largest excess, in line with the fact they have the largest infrared luminosity.
Then, category 1 objects have an infrared excess starting at near-infrared wavelengths and then slowly decreasing toward longer wavelengths.
The long tail of the infrared excess in category 1 objects is similar to objects in category 3.
Nevertheless, the latter have a significantly lower near-infrared excess.
Category 2 objects have an infrared excess starting at longer wavelengths than objects of category 1. 
The excess is getting higher at mid-infrared wavelengths.
Lastly, category 4 objects have an infrared excess starting at wavelengths equal or larger than $\sim$20$\mu$m.

\subsection{RVb phenomenon}
\label{sec:RVb}

\begin{figure}[t]
    \centering
    \includegraphics[scale=0.55]{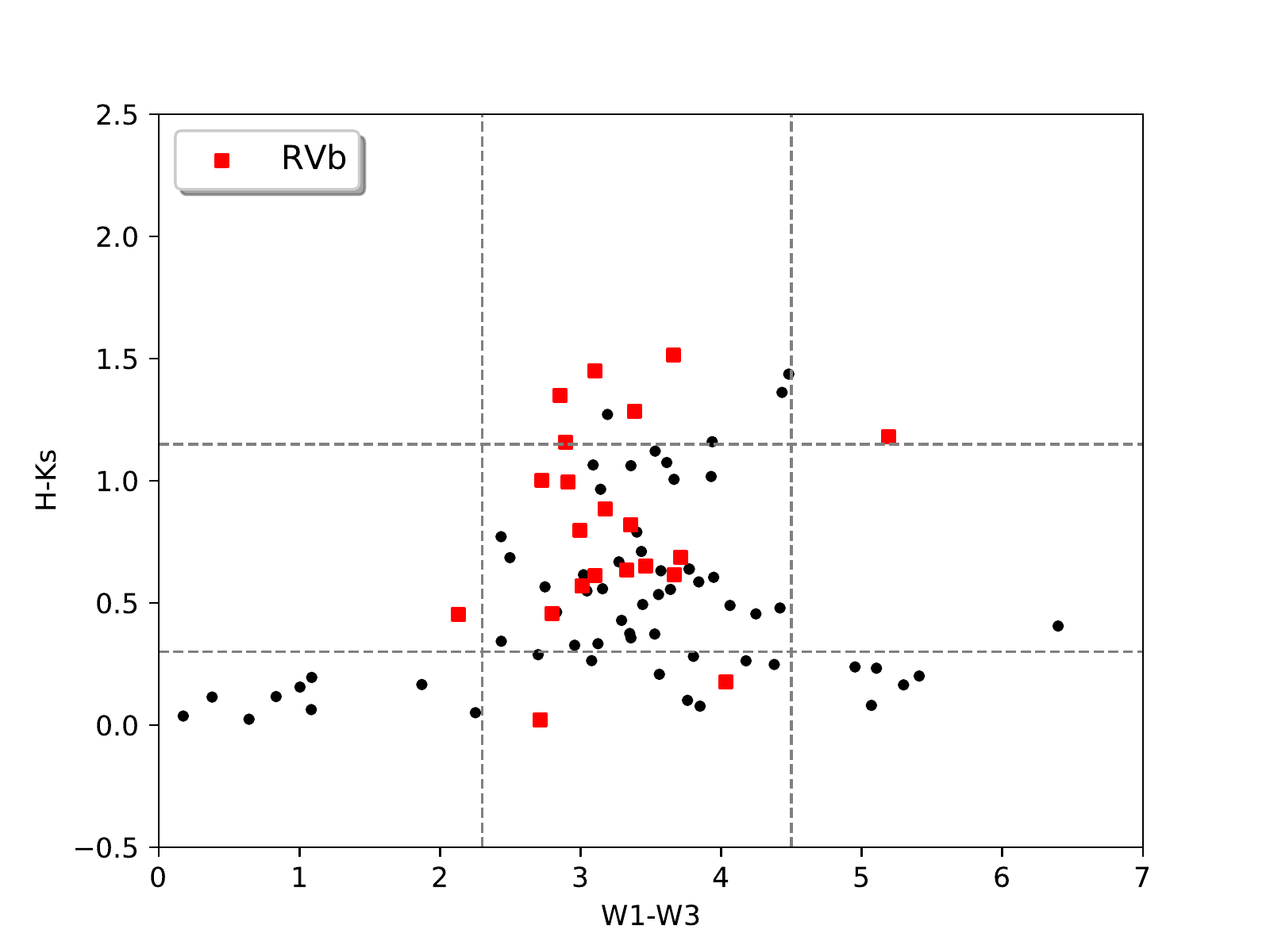}
    \caption{Distribution of the RVb targets in the color-color diagram. The dashed gray lines represent the boundaries of the object categories.}
    \label{fig:rvb}
\end{figure}

The RVb phenomenon consists of a strong and periodic variation of brightness at optical wavelengths with the same period as the orbital one \citep[e.g.,][]{Kiss2007,KissBodi2017,BodiKiss2019,Manick2019}.
While first seen around RV\,Tauri stars, the RVb phenomenon is also observed in non RV\,Tauri, post-AGB binaries.

The proportion of targets \modifLE{displaying} the RVb phenomenon is different in each category (see Fig.\,\ref{fig:rvb}).
The largest number of targets including the RVb phenomenon are in category 0 (5/9; >50\%).
On the contrary, categories 2 and 3 have 20\% or fewer of their objects displaying the RVb phenomenon, while there are no such objects in the eight targets composing category 4.
Finally, in category 1, there are $\sim$25\% (13/49) targets displaying the RVb phenomenon.
We conclude that there are clear differences in the distribution of the RVb phenomenon among the different categories.

\subsection{No straightforward correlations with orbital properties or effective temperature}
\label{sec:nocor}

The effective temperature might be a proxy for age as the post-AGB star is expected to contract and, thus, to heat up with time.
However, we do not find any correlation or preferred category for stars of a given effective temperature (T$_\mathrm{eff}$, see Fig.\,\ref{fig:Teff}).
For full disks in category 2, we can see that the hotter stars have H-Ks colors above $\sim$0.6 while colder stars have lower colors.
To test the impact of a different effective temperature but a constant luminosity, we have changed the spectrum of the central star of our full disk models. We focused on the effect on the position in the color-color diagram.
Indeed, hotter stars lie higher up in the diagram than colder ones (see Fig.\,\ref{fig:TeffTest}). 
The disk emission or temperature structure is, however, only marginally changed.

We also investigated whether the categories correlate with given orbital properties as it can be expected that disk-binary interactions can modify the orbital properties through Lindblad resonances \citep{Artymowicz1994} or angular momentum exchange through mass accretion from the disk onto the stars \citep{Munoz2019}.
But both for the orbital period or the inclination there is no obvious differences between the different categories (Fig.\,\ref{fig:orbits}).
It only seems that there is a greater number of short-period orbits for category 3 than in others.
But given the low number of targets, this needs to be further confirmed.
We conclude that the disk structure in the dust does not significantly influence the orbital properties of the binary and vice-versa.

\subsection{Depletion tracers}
\label{sec:depletion}

\begin{figure}[t]
    \centering
    \includegraphics[scale=0.55]{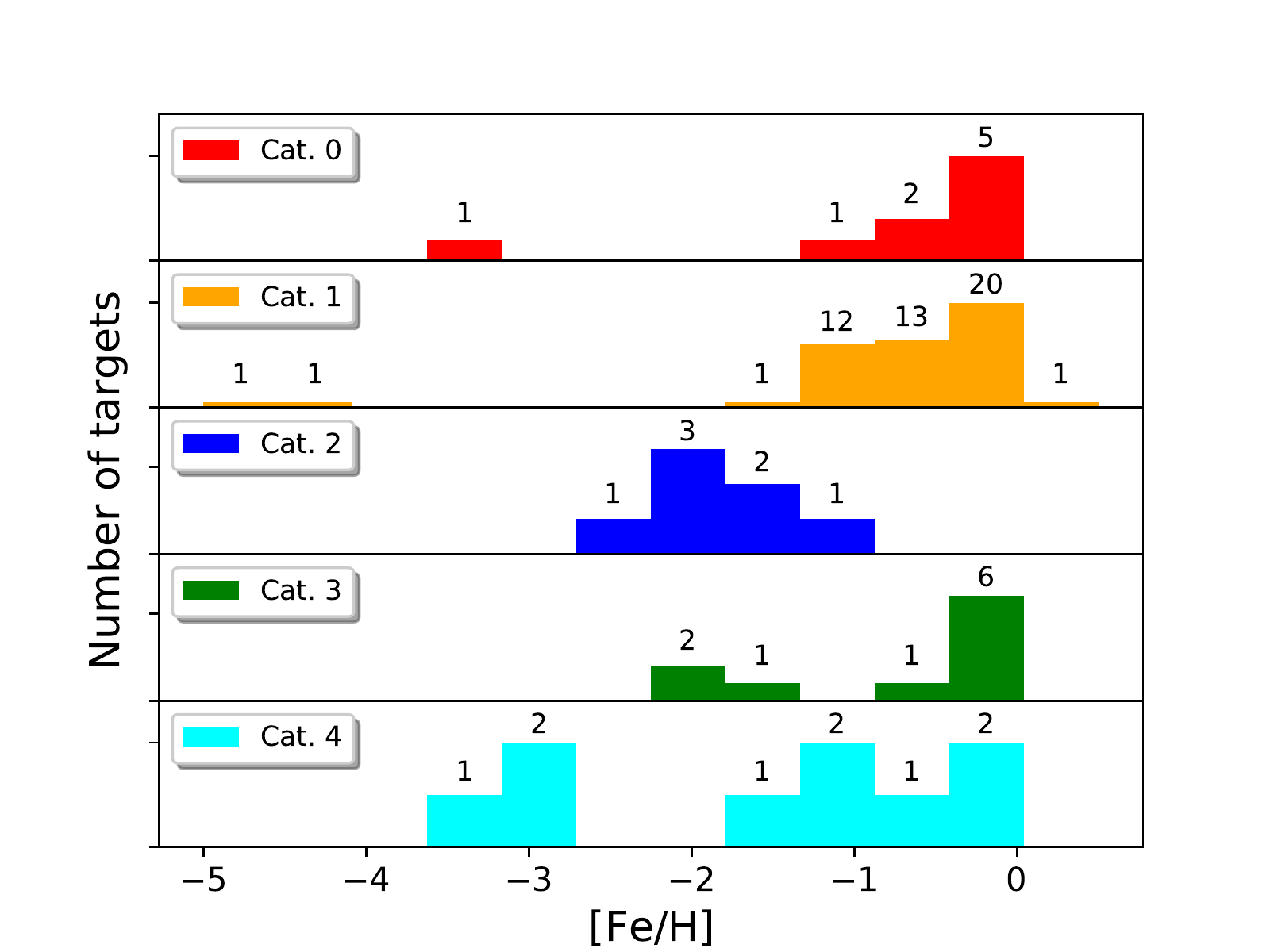}
    \includegraphics[scale=0.55]{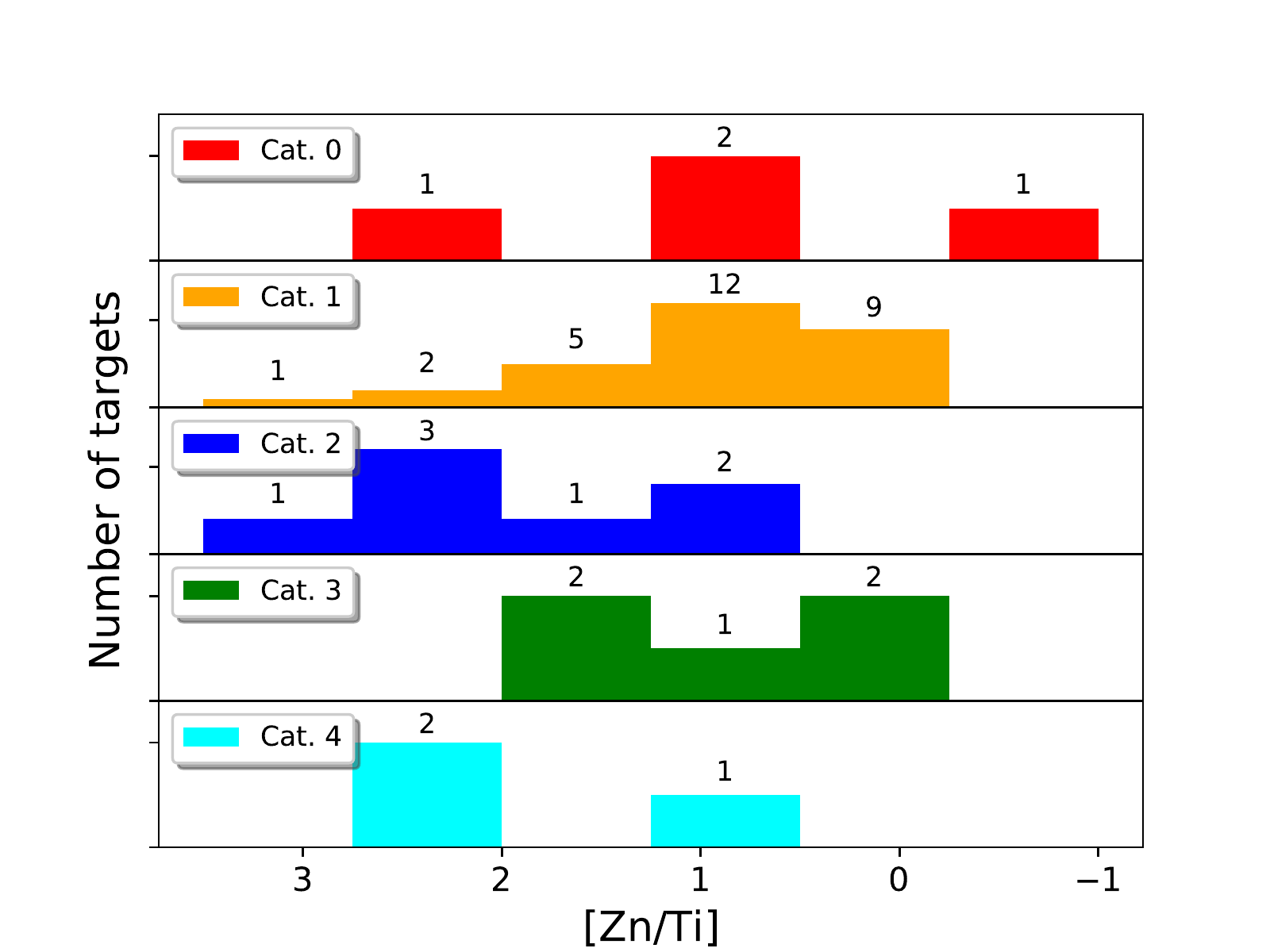}
    \includegraphics[scale=0.55]{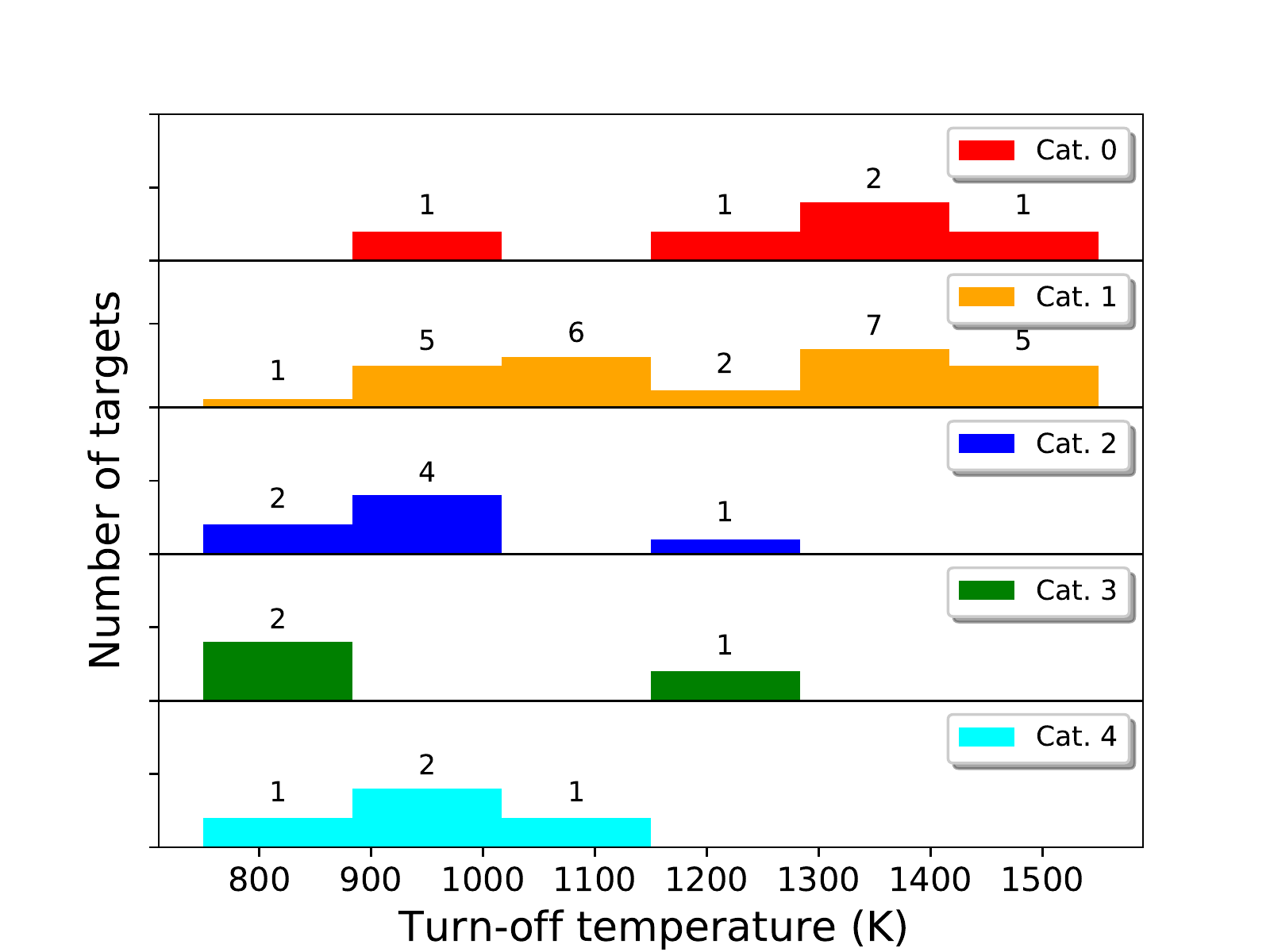}
    \caption{Histograms of [Fe/H] (top), [Zn/Ti] (middle), and turn-off temperature (bottom) for each category. }
    \label{fig:FeH}
\end{figure}

We investigated whether the chemical depletion observed on the surface of post-AGB binaries can be linked to the disk structure. 
We selected three depletion tracers: [Fe/H], which is observationally robust, [Zn/Ti], where both elements have similar nucleosynthetic origin and the turn-off temperature ($T_\mathrm{turn-off}$)  indicates the particular element condensation temperature for which the depletion pattern starts \citep{Oomen2019}.
The condensation temperature of a given element depends on various factors.
Here, they are defined following \citet{Lodders2003}.
The relative values of the condensation temperatures are more important than the absolute ones.

\begin{figure}[t]
    \centering
    \includegraphics[scale=0.6]{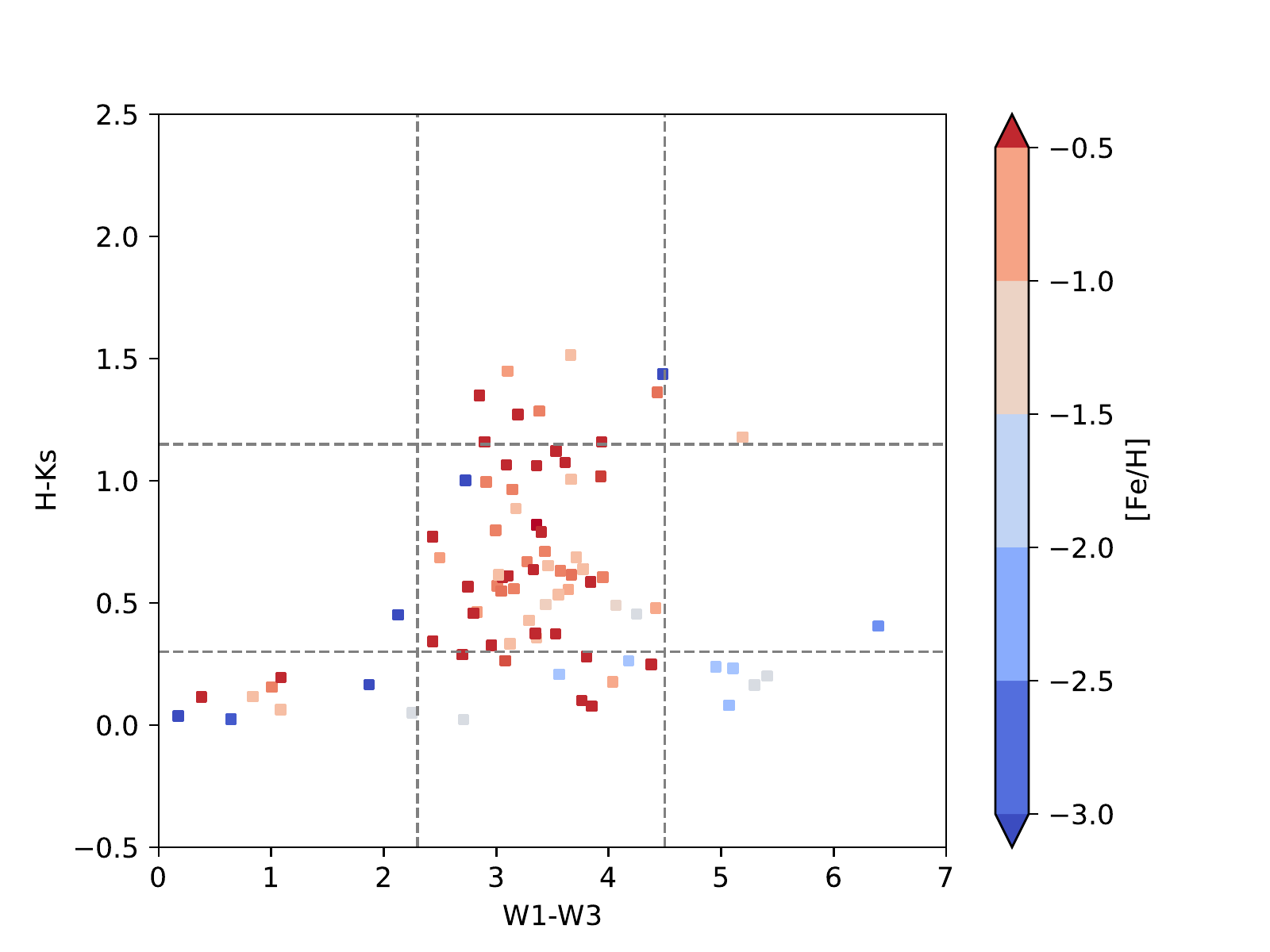}
    \caption{Infrared color-color diagram color-coded with [Fe/H] that traces the depletion of refractory materials on the surface of the post-AGB stars. }
    \label{fig:FeHCat}
\end{figure}

In Fig.\,\ref{fig:FeH} we show the histograms of [Fe/H] for each category.
We can see that all the objects in category 0 and 1 have values of [Fe/H] larger than $-$1.5 (with three exceptions that we discuss in Sect\,\ref{sec:extreme}).
For category 2 targets, we see a different distribution of [Fe/H], with values always lower or equal to $-$1.
For objects in category 3, there is a bi-modal distribution with values above and below [Fe/H]$\sim-$1.  
Objects in category 4 have values of [Fe/H] spread over a range from $-$4 to 0.
Consequently, there are clear differences in the distributions of [Fe/H] between the disk categories.
Figure\,\ref{fig:FeHCat} represents the distribution of the targets in the color-color diagram, color-coded with [Fe/H]. 
We can see that even without defining the categories, there are no objects with low values of [Fe/H] (-1>[Fe/H]>-3) in the most densely populated location of the color-color diagram with H-Ks$\lesssim$4 and W1-W3$\gtrsim$0.3.
We also note few objects with extremely low values of [Fe/H] ([Fe/H]<-3) are spread across the diagram, showing no evident link with their infrared colors. 

A similar category-sensitive distribution is seen for another depletion tracer: [Zn/Ti] (see Fig.\,\ref{fig:FeH} middle panel);
Zn has a low condensation temperature ($\sim$750\,K) relative to Ti \citep[$\sim$1600\,K;][]{Lodders2003} and both elements have a similar nucleosynthetic origin preventing any additional bias due to the stellar evolution history. Therefore,
[Zn/Ti] is a good tracer of depletion, even though the detection and characterization of the abundance of both elements is more difficult than for Fe.
This results in fewer targets having good estimation of [Zn/Ti].
Nevertheless, we see that categories 0, 1, and 3 have distributions that are centered around [Zn/Ti]$\sim$1, whereas category 2 targets have systematically higher values, which peak for [Zn/Ti]$>$2.
A similar trend might be seen in category 4 objects, despite the low number of targets having a measured [Zn/Ti].

In Fig.\,\ref{fig:FeH}, bottom panel, we can see that the distribution of T$_\mathrm{turn-off}$ is also sensitive to the disk categories.
While for categories 0 and 1, we have a wide distribution of turn-off temperatures, the distribution for categories 2 and 4 peaks at low values T$_\mathrm{turn-off}\sim$1000\,K.
Even if the number of targets that have adequately observed and well-constrained [Zn/Ti] and T$_\mathrm{turn-off}$ is lower that for [Fe/H], we observe a similar trend in all depletion tracers; namely, that category 2 and some targets in categories 3 and 4 are clearly more depleted than targets in categories 0 and 1.

\section{Discussion}
\label{sec:discussion}

In this discussion, we offer interpretations of the correlation analysis presented in Sect.\,\ref{sec:analysis}.
We discuss the localization of highly inclined disks in the color-color diagram (Sect.\,\ref{sec:inclination}), the origin of the link between depletion and disk structure (Sect.\,\ref{sec:SEDDepl}) and, finally, we propose an evolutionary scenario for the disk (Sect.\,\ref{sec:diskevol}).

\subsection{Category 0 targets have highly inclined disks}
\label{sec:inclination}

The common interpretation of the RVb phenomenon is that the disk is seen at high inclinations (close to edge-on), causing the line of sight to go through the disk atmosphere. The phenomenon reflects the variable extinction in the line-of-sight induced by the orbital motion: the object is faint and red when at inferior conjunction. The range of inclination values producing such a phenomenon depends on the disk scale height and flaring.
The disk inclination \modifLE{has}, therefore, a strong influence on the SED as the disk, as it is along the line of sight and can partially block the light of the central star(s), creating a SED dominated by the infrared disk emission.
We show in Sect.\,\ref{sec:RVb} that objects having a large amount of near-infrared excess -- mostly belonging to category 0 -- display the RVb phenomenon.
It thus confirms the common interpretation that this periodic variation is due to the change in total line-of-sight extinction during orbital revolution. 
The objects become faint and red when the line-of-sight experience more circumstellar extinction.
Spatially resolved near-infrared interferometric observations confirm this interpretation \citep{Kluska2019}.
Therefore, we use the RVb phenomenon as a proxy for highly inclined disks.

The Red Rectangle and QY\,Sge do not display the RVb phenomenon.  However, both are known to be edge-on \citep{Cohen2004,Rao2002}, confirming the high disk inclination nature of category 0 targets. We note, however, that in the Red Rectangle, the photometric change is due to a variable scattering efficiency \citep{Waelkens1996}, rather than periodical absorption of the light from the central stars.


The high disk inclination correlates with the fact that category 0 sources have an infrared excess luminosity higher than the apparent stellar luminosity (see Sect.\,\ref{sec:IRlum}) as it is expected when the disk absorbs stellar light.
It is thus very likely that the category 0 sources IRAS15556-5444 and IRAS18158-3445, for which we have no information on the inclination, are also highly inclined disks.
We can therefore conclude that Category 0 sources are composed of inclined disks.

Finally, the fact that there are fewer targets that display RVb phenomenon in category 2, 3, and 4 disks can be explained by the larger inner disk hole.
The RVb phenomenon to be observed requires a peculiar and higher inclination angle, which is less likely to be observed.

\subsection{Strong link between disk structure and depletion}
\label{sec:SEDDepl}

\begin{table*}
\caption{Transition disks targets and candidates.}              
\label{tab:TDs}      
\centering                                      
\begin{tabular}{l l c c c c c}          
\hline\hline                        
 IRAS & Alt. name & $H-K$s & W1-W3 & [Fe/H] & [Zn/Ti] & T$_\mathrm{turn-off}$  \\
 & & & & & & (K) \\
\hline                                   
\multicolumn{7}{c}{Category 2 targets}\\
\hline
06034+1354 & DY\,Ori & 0.23 & 5.1 & $-$2.0 & 2.1 & 1000 \\
06472-3713 & ST\,Pup & 0.17 & 5.3 & $-$1.5 & 2.1 & 800 \\
11472-0800 & AF\,Crt & 0.41 & 6.4 & $-$2.5 & 3.4 & 1000 \\
12067-4508 & RU\,Cen & 0.24 & 5.0 & $-$2.0 & 1.0 & 800 \\
17233-4330 & V1504\,Sco & 1.18 & 5.2 & $-$1.0 & 1.4 & 1000 \\
18281+2149 & AC\,Her & 0.20 & 5.4 & $-$1.5 & 1.0 & 1200 \\
18564-0814 & AD\,Aql & 0.08 & 5.1 & $-$2.1 & 2.5 & 1000 \\
\hline                                             
\multicolumn{7}{c}{Depleted category 3 targets}\\
\hline
06072+0953 & CT\,Ori & 0.26 & 4.2 & $-$2.0 & 1.9 & 1200 \\
16278-5526 & GZ\,Nor & 0.21 & 3.6 & $-$2.0 & 0.8 & 800 \\
19163+2745 & EP\,Lyr & 0.02 & 2.7 & $-$1.5 & 1.3 & 800 \\
\end{tabular}
\end{table*}

As the convective envelope of the post-AGB star is thin, the depletion pattern observed at the surface of post-AGB binaries is interpreted to be due to the accretion of matter from the circumbinary disk \citep[e.g.,][]{Oomen2019}. 
In many targets, we can see strong evidence that the composition of the accreted matter is depleted from refractory elements.
It is important to note that not all objects exhibited a solar composition when they were main-sequence objects, but the original metallicity is difficult to probe. 
The objects of our study are mainly Galactic thick disk objects \citep{VanWinckel2018} and they do not have Galactic halo kinematics.  
Consequently, the suspected spread in the original metallicity is unlikely to be large and we assume that the dominant cause of the low [Fe/H] values is depletion.
Moreover, as mentioned, the abundances of different elements are not linked to their nucleosynthetic history but are correlated with the condensation temperature \citep{Maas2003,Oomen2019}.
It is, therefore, very likely that the observed abundances are tracing the accreted material and not the nucleosynthetic history of the evolved star.

\subsubsection{Stars surrounded by transition disks show chemical depletion}

In Sect.\,\ref{sec:depletion}, we show that there is a strong link between depletion tracers and the absence of a strong near-infrared excess, indicating the presence of a large disk cavity or low dust density in the disk inner regions.
The properties of category 2 targets, surrounded by such transition disks, are listed in Table\,\ref{tab:TDs}. 
We also added category 3 objects that are on the mode of the [Fe/H] distribution with lower values.
All targets share similar values of depletion tracers: [Fe/H]$\sim$$-$2.0, [Zn/Ti]$\sim$2.0 and T$_\mathrm{turn-off}\sim$1000\,K.
It is likely that these stars are affected by the same mechanism, causing both low $H-Ks$ colors and the observed depletion pattern.
We postulate that the category 3 objects with higher depletion are similar to category 2 objects while those without strong depletion are more likely full disks without the over-resolved emission.
In that case, the disks around depleted category 3 targets are transition disks candidates.
The proportion of transition disks to the whole disk population is, thus, between 12\% (10/85) and 8\% (7/85), with and without including the category 3 transition disk candidates, respectively.
This is remarkably similar to the proportion of transition disks found around young stars \citep[e.g.,][]{vanderMarel2016}.

Our selection process did not influence this correlation. 
While we used all Galactic objects identified as such in the literature,  the sample was defined by their infrared properties and is therefore flux limited in the infrared.
Targets below the detection limit in the infrared are typically missed. Additionally, the selected objects must also be optically sufficiently bright, thus, we do miss the obscured or embedded post-AGB stars. There is, however, no evidence for the selection criteria to favor transition disks objects with high depletion.

The link between transition disks and depletion points toward a mechanism that depletes refractory material from volatiles within the disk beyond the dust sublimation radius. 
It is also indicated by the relatively lower values for the turn-off temperatures (T$_\mathrm{turn-off}\lesssim$1200\,K) for stars surrounded by transition disks.

\subsubsection{A population of ``extreme'' stars}
\label{sec:extreme}

\begin{figure}[t]
    \centering
    \includegraphics[scale=0.45]{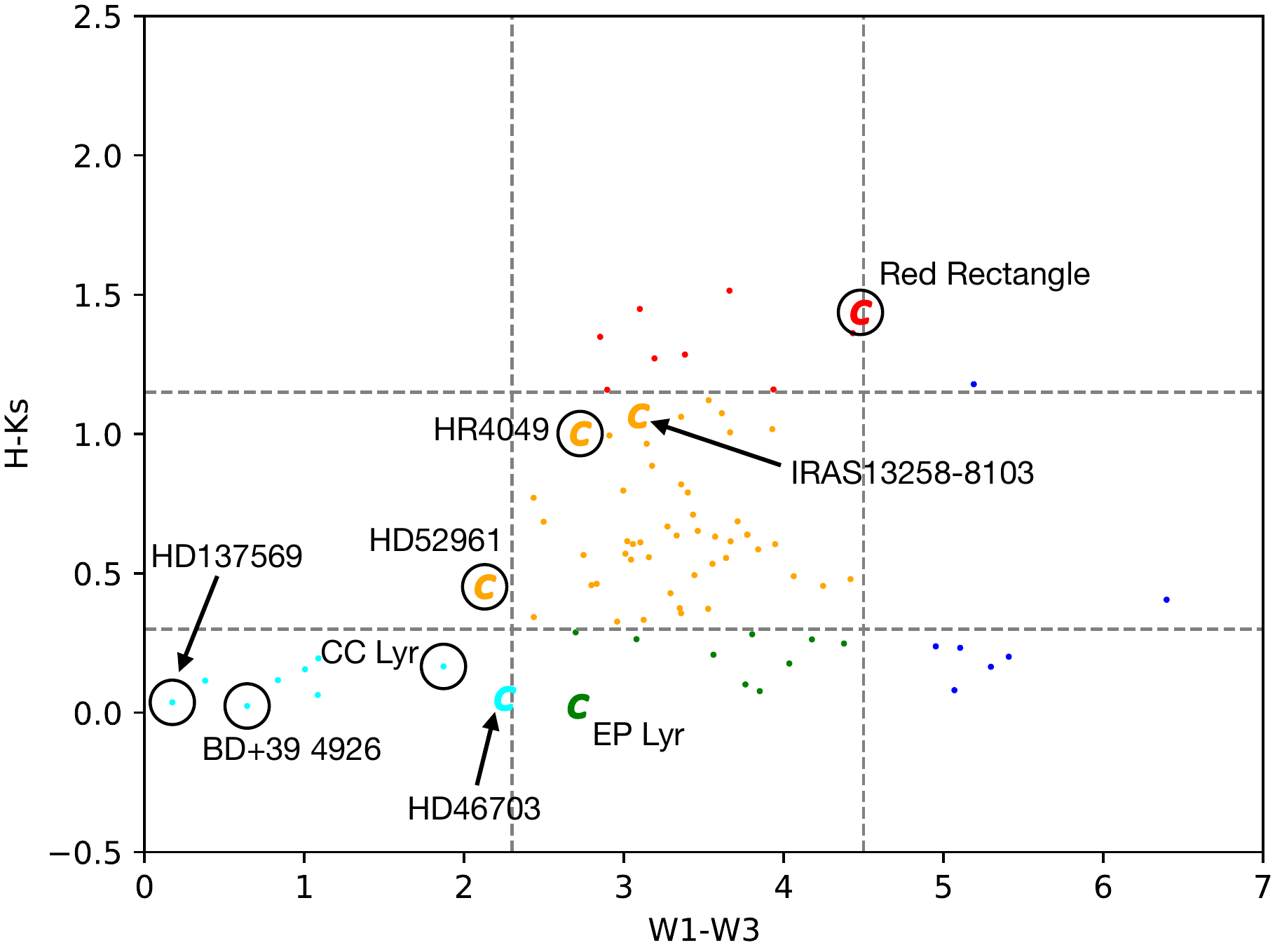}
    \caption{Color-color diagram including extreme stars (black empty circles) and targets with detected signatures of carbonaceous species (indicated with letter C). }
    \label{fig:extreme}
\end{figure}

\begin{table*}
\caption{Extreme targets and other targets with carbonaceous dust detected in their mid-infrared spectra \citep{Gielen2011b}.}              
\label{tab:extreme}      
\centering                                      
\begin{tabular}{c l l c c c c c c c }          
\hline\hline                        
Cat. & IRAS & Alt. name(s) & $H-K$s & W1-W3 & [Fe/H] & [Zn/Ti] & T$_\mathrm{turn-off}$ & IR lum. & Carbon \\ 
& & & & & & & (K) & & \\
\hline                                   
\multicolumn{9}{c}{``Extreme'' targets}\\
\hline
0 & 06176-1036 & Red Rectangle, HD\,44179 & 1.44 & 4.5 & $-$3.3 & \ldots & 1000\,K & >100\% & PAHs \\      
1 & 10158-2844 & HR\,4049 & 1.00 & 2.7 & $-$4.5 & \ldots & 800\,K & 26\% & PAHs  \\
1 & 07008+1050 & HD\,52961 & 0.45 & 2.1 & $-$4.8 & 3.0 & 1100\,K & 16\% & PAHs \& C$_{60}$     \\
4 & \ldots & CC Lyr & 0.17 & 1.9 & $-$3.5 & \ldots & 1000\,K & 3\% & \ldots  \\
4 & F15240+1452 & HD\,137569 & 0.04 & 0.17 & $-$3.0 & \ldots & \ldots & 0\% & \ldots \\
4 & \ldots & BD+39 4926 & 0.02 & 0.6 & $-$2.9 & 2.0 & 1000\,K & 0\% &  \ldots    \\
\hline  
\multicolumn{9}{c}{Other targets with detected signatures of carbonaceous dust}\\
\hline
1 & 13258-8103 & \ldots & 1.07 & 3.1 & 0 & \ldots & \ldots & >100\%  & PAHs   \\
3 & 19163+2745 & EP\,Lyr & 0.02  & 2.7 & $-$1.5 & 1.3 & 800\,K & 2\% & PAHs \\
4 & 06338+5333 & HD\,46703 & 0.05 & 2.3 & $-$1.5 & 0.9 & 800\,K & 2\% & PAHs \& C$_{60}$ \\
\end{tabular}
\tablefoot{ IR lum. is the ratio of the infrared luminosity to stellar luminosity.}
\end{table*}

We also note that there is a population of extremely depleted stars (e.g., [Fe/H]$\leq-$3.0) that belong to category 0, 1, and 4 (see Table\,\ref{tab:extreme}); thus, stars not belonging to categories 2 or 3.
However, they seem to have a low mid-infrared color ($W1-W3\lesssim$3, Fig\,\ref{fig:extreme}).
The only exception is the Red Rectangle, but its edge-on disk orientation changes the interpretation of the colors.
It is remarkable to note that three out of six of these targets have carbonaceous material detected in their surroundings in the form of PAHs and sometimes C$_{60}$ Fullerenes, whereas only six post-AGB targets are known to show such spectral features \citep{Gielen2011b}.
In Table\,\ref{tab:extreme}, we added the remaining three targets showing carbonaceous material for comparison.
In particular, two of them, EP\,Lyr and HD\,46703, are also relatively depleted with [Fe/H]=$-$1.5 and [Zn/Ti]$\sim$1.
All these targets have low turn-off temperatures (T$_\mathrm{turn-off} \lesssim$ 1000K) also suggesting a depletion mechanism acting within the disk.
The three extreme targets with nondetections of carbonaceous material are category 4 objects with very low infrared excess luminosity to stellar luminosity ratio (0\% and 2\% for BD+39~4926 and CC\,Lyr respectively).
The lack of infrared excess indicates that there is a low amount of dust left in the system and could hence explain the nondetection of carbonaceous material.
However, the detection of carbonaceous dust around the category 4 target HD\,46703 seems to contradict this.
Nevertheless, this target lies at the boundary between category 3 and 4 (Fig.\,\ref{fig:extreme}) and have a amount of infrared excess luminosity (2\%) comparable to CC\,Lyr.

There is no obvious link with transition disks, which are also depleted, even if EP\,Lyr is a category 3 transition disk candidate.
Nevertheless, IRAS13258-8103 also displays carbonaceous material in its mid-infrared spectra but is not depleted in iron ([Fe/H]=0.0), preventing us from drawing definitive conclusions from this sample of ``extreme'' targets.
Despite some degree of similarity between ``extreme'' targets and transition disks targets, it is not straightforward to assume that the same depletion mechanism is at play in both populations.
The origin of these ``extreme'' targets remains unknown.

\subsubsection{Giant planets: Most likely explanation for depletion and transition disks}

\label{sec:planet}

Hereafter, we discuss the depletion mechanism that could be responsible for targets identified as transition disks.
The depletion pattern observed in post-AGB disks is thought to be due to accretion of gas which was subject to dust formation \citep[e.g.,][]{Waters1992}.
This accreted matter was hypothesized to be depleted of dust by radiation pressure on the dust grains which separates the gas from the dust. 
This separation is followed by accretion of the gas component only \citep[e.g.,][]{Waters1992}. 
The impact of this gas accretion was modeled by \cite{Oomen2019, Oomen2020} who introduced the concept of turn-off temperature because of the different abundance profiles being observed. 
Assuming that this process of gas-dust separation takes place at the inner rim, it would imply that all targets accrete material with a similar composition and the observed depletion is due to the quantity of accreted material from the circumbinary disk.

However, here we show that the depletion pattern depends on the disk category as well, which illustrates that the gas-dust separation and accretion process is linked to the structure of the disk. 
Thus, the depletion pattern cannot be directly used to quantify the accretion history of a given target. 
Moreover, while the high radiation pressure present in these high luminosity targets acts on dust grains on the disk surface, previous studies have shown that the disk is optically thick in the radial directions shielding most of the dust grains from stellar radiation. Finally, such a depletion mechanism may be at play at the disk inner dust rims, however, it does not explain the presence of transition disks and why \modifLE{such a mechanism} would be more effective in such a disk configuration.

The strong link between depletion and infrared excess colors and, thus, disk morphology points toward a mechanism that separates the gas from dust within the disk while shaping the disk structure.
The disks are expected to be weakly or not magnetized because of the lack of high energy photons; nor are they driven by magneto-rotational instability. 
A presence of a dead-zone \citep[i.e., a nonionized zone of the disk suppressing accretion via the magneto-rotational instability;][]{Balbus1991,Lesur2014} is, therefore, unlikely to produce the transition disks, especially as it does not rule out the presence of dust close to the dust sublimation radius \citep[e.g.,][]{Pinilla2016DZ}.

A natural explanation would be the presence of a third component in the system that carves the disk hole in dust by acting as a filter, trapping the dust in the outer disk but still letting the depleted gas to be accreted onto the central stars. 
It was proposed by \citet{Kama2015} that giant planets (with masses of 1-10\,M$_\mathrm{jup}$) can be responsible for the observed depletion pattern in Herbig stars.
This hypothesis was further backed up by evolutionary disk models to explain the depletion of Solar twins by the presence of a giant planet in the protoplanetary disk \citep{Booth2020}.
It was shown that planets can trap the dust outside their orbit and can deplete the inner disks of dust  \citep{Pinilla2012,Pinilla2016,deJuanOvelar2013,Owen2014,vanderMarel2015,vanderMarel2016b} more efficiently than photo-evaporation, for example \citep{Gorti2015}.
Further studies have found correlations between the near-infrared continuum emission, depletion, and gas emission as traced by CO lines in the mid-infrared that are well explained by the presence of a giant planet producing the large inner hole \citep{Garufi2017,Banzatti2018,Bosman2019}.
Consequently, this planetary scenario would naturally explain the link between depletion and transition disks seen in post-AGB binary systems.
While we favor such a scenario to explain depletion in objects with transition disks, we do lack high spatial resolution studies of gas tracers and the dust sub-structures in the inner regions of the circumbinary disks to firmly draw a conclusion about the planetary origin of the observed target characteristics.

The post-AGB disks seem to be massive and dense enough to possibly form planets by themselves \citep{Sahai2011,Bujarrabal2013,Hillen2014,Kluska2018}.
While the post-AGB disk lifetime is expected to be short (10$^4$-10$^5$\,yrs), it was recently shown that planet formation can indeed occur on very short timescales as well \citep[$\sim10^5$yrs,][]{Sheehan2018,SeguraCox2020,Tychoniec2020,Lichtenberg2021}.
However, it seems more plausible that if the transition disks were produced by a planet, it would be a first-generation planet that survived the binary interaction phase.
This scenario seems to be backed up by the proportion of transition disks in our catalog that are close to the occurrence of giant planets \citep{Fulton2021}, assuming that the population of planets around binary systems is similar to the one around single stars.
Indeed, it seems that circumbinary planet are as frequent as planets around single stars, despite the relatively low number of detections \citep{Winn2015, Martin2018}.
If such a planet is present from the start in the disk, it will efficiently deplete the inner disk from refractory material and produce the depletion pattern observed on the surface of the star \citep{Booth2020}.


\subsection{Scenario for post-AGB disk evolution}
\label{sec:diskevol}

\begin{figure*}[t]
    \centering
    \includegraphics[scale=0.65]{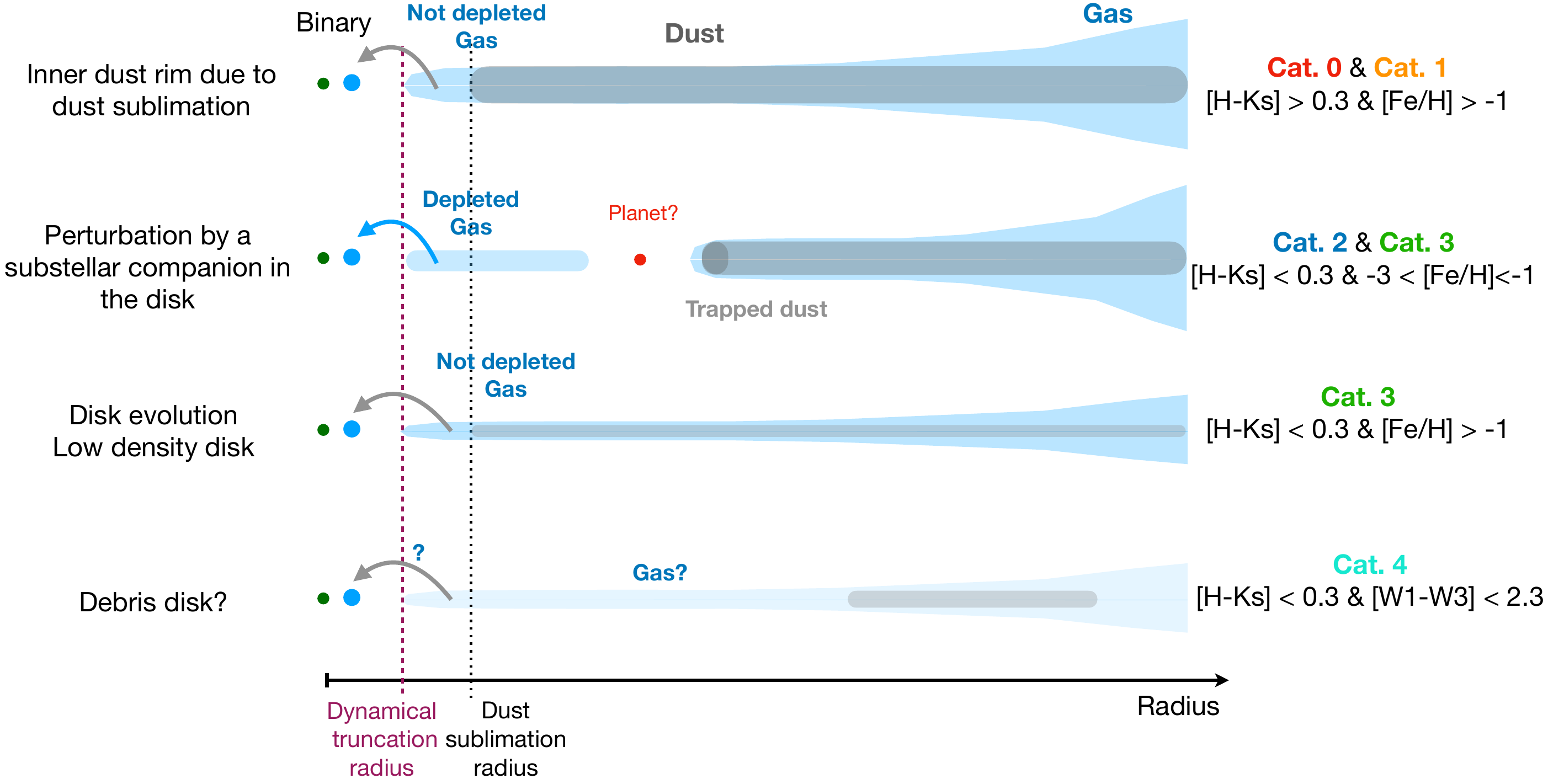}
    \caption{Scenario for the evolution of disks around post-AGB binaries. Schematic representation that is not to scale.}
    \label{fig:evol}
\end{figure*}

It is complex exercise to establish a scenario for disk evolution without a reliable proxy for the age of the systems.
We also do not have reliable distances to the targets that would help us better constrain the star and disk masses. The binary nature of our targets is likely biasing the Hipparcos and Gaia parallaxes.

Nevertheless, we can establish a tentative evolutionary scenario based on the results of this work.
We illustrated the different evolutionary stages on Fig.\,\ref{fig:evol}. 
After an episode of binary interaction, the circumbinary disk is formed without being depleted making category 0 and 1 disks seen at different inclinations.
Then, the disk will viscously evolve and will become less and less dense because of viscous spreading and accretion onto the stars becoming a nondepleted category 3 target with larger dust sizes.
This is illustrated by the change in the location of full disk models in the color-color diagram with respect to the disk mass and dust grain sizes (Fig.\,\ref{fig:CCeffect}).
Finally, the disk would become depleted of most of its gas (accreted onto the system or lost through disk winds) and with residual dust similar to debris disks (category 4).

Alternatively, if there is a first-generation giant planet in the disk, it will efficiently deplete the inner disk and the stellar surface of refractory elements \citep[as seen in Sect.\,\ref{sec:planet};][]{Booth2020} producing transition disks of category 2 and 3. 
This step would be avoided by disks without such a giant planet.
These disks would then evolve, producing depleted category 4 objects.

As mentioned, we do not expect the evolution of the disks to be ruled by magneto-rotational instability as there is low evidence for disk ionization from cosmic rays or high energy photons from the star \citep[even though the recent first detection of X-rays in a post-AGB binary, U\,Monocerotis, may suggest the contrary;][]{Vega2021}.
We also do not expect the disk to be photo-evaporated until the post-AGB star itself contracts and reaches a high enough temperature to produce high-energy photons, which is not (yet) the case in the targets of our sample.
The accretion of matter from the circumbinary disk onto the post-AGB stars can delay the star contraction and, therefore, prolong the disk lifetime by a factor of 2-5 \citep{Oomen2019}.
We do not have a direct tracer for accretion, but given that circumcompanion jets are commonly observed \citep{Bollen2020}, accretion appears to be ongoing in almost all systems.
Our scenario is not complete as we do not include the ``extreme'' objects.
The origin of these targets is unknown and requires further investigation.

Finally, if the planetary explanation is correct for transition disks, it would mean that there is a pressure maximum outside the planet's orbit, trapping the dust and creating a favorable environment for second-generation planet formation \citep[as in e.g.,][but already including a planet in the disk skipping the first step creating an initial radial over-density]{Chaterjee2014}.
The possibility of second-generation planet formation is further supported by the detection of two giant planets around the white-dwarf binary system NN\,Ser.
These planets are candidates for having been formed in such second-generation disks \citep{Beuermann2010,Mustill2013,Parsons2014,Schleicher2014,Volschow2014,Hardy2016}.
If the planetary scenario is confirmed, these disks would become a promising site for studying second-generation planet formation and, hence, planet formation scenarios in an unprecedented parameter space.

%


\section{Conclusions}
\label{sec:conclusion}

We list the main findings of this study below:
  \begin{enumerate}
      \item We present a catalog of disks around Galactic post-AGB binaries that we classified in five categories depending on their infrared colors.
      \item 8-12\% of the targets belong to a population of transition disks, namely, disks with large inner holes.
      \item We found a strong connection between the depletion of chemical elements observed at the surface of the star (-3$\mathbf{\lesssim}$[Fe/H]$\lesssim$-1) and transition disks.
      \item We conclude that there is a mechanism separating the dust and gas within the circumbinary disk  that is contrary to what has been postulated until now.
   \end{enumerate}
 
 These observational results lead us to draw the following interpretations:
   \begin{enumerate}
      \item We advocate for an alternative depletion mechanism induced by a first-generation giant planet that produces a disk cavity in dust and a gap in gas. If true, it would provide the first evidence of the presence of a planet in these disks.
      \item We propose a disk evolutionary scenario based on our findings with two evolutionary branches depending on the presence of a first- or second-generation giant planet carving a disk hole.
      \item There is a population of ``extreme'' targets that are strongly depleted and that do not seem to be directly linked to the disk type. There is a tentative link between these ``extreme'' targets and the detection of carbonaceous material, that is, PAHs or Fullerenes, in their direct circumbinary environment.
      \item If the planetary scenario is confirmed, these disks would become a unique laboratory for studying potential second-generation planet formation and testing  planet formation scenarios in an an unprecedented parameter space.
   \end{enumerate}
   
This work allows us to open up new directions of study.
First, we need to investigate the transition disks identified in this work to probe the gas in the inner disk, especially in the dust inner hole.
This is possible with current observatories and instruments such as ALMA in the sub-mm and CRIRES+ and MATISSE in the mid-infrared.
Second, the study of the gas should be complemented with the study of dust substructures at the disk inner rim.
It would then be possible to investigate whether a planet is indeed able to create such gas and dust structures and be at the origin of transition disks.
The advent of the third data release of Gaia with binary solutions will allow us to obtain reliable distances to our targets and locate them in the Hertzsprung-Russel diagram. We will be able to estimate their stellar masses and trace the ages of the systems. Third, we have no direct tracer for accretion rates in these targets so far.
Finding such a tracer would help to link the depletion pattern to the accretion history of the system and also to the chemical composition of the depleted inner disk parts.
Finally, tracing the continuum with ALMA and MATISSE would allow us to look for the presence of substructures in the disk that would favor second-generation planet formation around these intriguing disks surrounding evolved binaries.

\longtab[1]{
\begin{landscape}
\begin{longtable}{l|l|l|l|c|c|c|c|c|c|c|c|c|c}

\caption{\label{Table:list} Full list of Galactic sources. We list the IRAS name, alternative identifications, coordinates and the reference to the paper which was used to assemble this list. A complete table with all the objects characteristics used in the study will be available online.}\\
\hline\hline
IRAS & Name & RA & DEC & T$_\mathrm{eff}$ & P$_\mathrm{orb}$ & $e$ & H-K$_\mathrm{s}$ & W1-W3 & [Fe/H] & [Zn/Ti] & T$_\mathrm{Turn-off}$ & IR Lum. & Ref. \\
 & & & & (K) & (days) & & & & & & (K) & (\%) & \\
\hline
\endfirsthead
\caption{continued.}\\
\hline\hline
IRAS & Name & RA & DEC & T$_\mathrm{eff}$ & P$_\mathrm{orb}$ & $e$ & H-K$_\mathrm{s}$ & W1-W3 & [Fe/H] & [Zn/Ti] & T$_\mathrm{Turn-off}$ & IR Lum. & Ref. \\
 & & & & (K) & (days) & & & & & & (K) & (\%) & \\
\hline
\endhead
\hline
\endfoot
\hline
\multicolumn{10}{l}{Category 0}\\
\hline
06108+2743 & SU\,Gem  & 06 14 00.0 & +27 42 12 & 5750 & 690 & \ldots & 1.45$\pm$0.04 & 3.10$\pm$0.08 & -0.7 & 0.5 & 1400 & 310 & a \\
06176-1036 & HD\,44179, & 06 19 58.2 & -10 38 15 & 7500 & 317.6 & 0.27 & 1.44$\pm$0.25 & 4.48$\pm$0.28 & -3.3 & \ldots & 1000 & 1034 & a \\
 & Red Rectangle &&&&&&&&&&&& \\
08011-3627 &  AR\,Pup  & 08 03 01.1 & -36 35 47 & 6000 & 1250 & \ldots & 1.51$\pm$0.05 & 3.66$\pm$0.40 & -1.0 & \ldots & 1200 & 639 & a \\
15556-5444 &  \ldots  & 15 59 32.1 & -54 53 18 & 6000 & \ldots & \ldots & 1.16$\pm$0.04 & 3.94$\pm$0.08 & 0.0 & \ldots & \ldots & 215 & a \\
18158-3445 &  \ldots  & 18 19 13.6 & -34 44 32 & 6500 & \ldots & \ldots & 1.27$\pm$0.04 & 3.19$\pm$0.04 & 0.0 & \ldots & \ldots & 126 & a \\
19125+0343 & BD+03\,3950 & 19 15 01.2 & +03 48 43 & 7750 & 519.7 & 0.24 & 1.29$\pm$0.04 & 3.38$\pm$0.14 & -0.5 & 2.3 & 1400 & 119 & a \\
19472+4254 & DF\,Cyg & 19 48 53.9 & +43 02 15 & 4800 & 784 & \ldots & 1.16$\pm$0.02 & 2.90$\pm$0.04 &0.0& -0.7 & \ldots & 76 & b \\
19548+1951 & RS\,Sge & 19 57 06.4 & +19 59 44 & 6000 & \ldots & \ldots & 1.35$\pm$0.04 & 2.85$\pm$0.05 &0.0& \ldots & \ldots & 79 & b \\
20056+1834 & QY\,Sge  & 20 07 54.8 & +18 42 57 & 5850 & \ldots & \ldots & 1.36$\pm$0.04 & 4.43$\pm$0.07 & -0.4 & 1.2 & 1500 & 829 & a \\
\hline
\multicolumn{10}{l}{Category 1}\\
\hline 
01427+4633 & BD+46.442 & 01 45 47.0 & +46 49 01 & 6250 & 140.82 &0.0& 0.46$\pm$0.07 & 2.83$\pm$0.05 & -0.7 & -0.2 & \ldots & 27 & h \\
04166+5719 & TW Cam  & 04 20 48.1 & +57 26 26 & 4800 & 662.2 & 0.25 & 0.57$\pm$0.05 & 3.01$\pm$0.08 & -0.5 & 0.3 & 1100 & 49 & a \\
04440+2605 &  RV Tau  & 04 47 06.8 & +26 10 44 & 4500 & 1198 & \ldots & 0.61$\pm$0.03 & 3.67$\pm$0.09 & -0.4 & 0.5 & 1100 & 48 & a \\
06160-1701 &  UY\,CMa  & 06 18 16.4 & -17 02 35 & 5500 & \ldots & \ldots & 0.59$\pm$0.06 & 3.84$\pm$0.04 & 0.0 & 1.8 & 1000 & 71 & a \\
\ldots & EZ\,Gem & 06 46 04.5 & +13 05 02 & 6555 & \ldots & \ldots & 0.77 & 2.43$\pm$0.03 & 0.0 & \ldots & \ldots & 11 & b \\
06489-0118 & SZ Mon & 06 51 27.8 & -01 22 16 & 4700 & \ldots & \ldots & 0.55$\pm$0.07 & 3.05$\pm$0.04 & -0.4 & 0.8 & 1500 & 37 & b \\
07008+1050 & HD\,52961 & 07 03 39.6 & +10 46 13 & 6000 & 1288.6 & 0.23 & 0.45$\pm$0.04 & 2.13$\pm$0.08 & -4.8 & 3.0 & 1100 & 15 & a \\
07140-2321 & SAO\,173329 & 07 16 08.3 & -23 27 02 & 7000 & 115.9 &0.0& 0.56$\pm$0.04 & 3.64$\pm$0.04 & -0.8 & -0.1 & \ldots & 47 & a \\
08005-2356 & V510 Pup & 08 00 32.6 & -23 56 15 & 7000 & 2654 & 0.36 & 1.01$\pm$0.04 & 3.66$\pm$0.09 & -1.0 & \ldots & \ldots & 67 & m \\
08544-4431 & V390 Vel & 08 56 14.2 & -44 43 11 & 7250 & 501.1 & 0.2 & 0.97$\pm$0.25 & 3.14$\pm$0.40 & -0.5 & 0.9 & 1200 & 60 & a \\
09060-2807 &  V*BZPyx & 09 08 10.1 & -28 19 10 & 6500 & 372 & \ldots & 0.99$\pm$0.05 & 2.91$\pm$0.06 & -0.5 & 0.2 & \ldots & 347 & a \\
09144-4933 &  \ldots  & 09 16 09.1 & -49 46 06 & 5750 & 1762 & 0.3 & 0.71$\pm$0.04 & 3.43$\pm$0.08 & -0.5 & \ldots & 1400 & 67 & a \\
09256-6324 &  IW\,Car  & 09 26 53.4 & -63 37 48 & 6700 & 1449 & \ldots & 0.69$\pm$0.26 & 3.71$\pm$0.20 & -1.0 & 2.1 & 1100 & 69 & a \\
09538-7622 &  GP\,Cha & 09 53 58.5 & -76 36 53 & 5500 & \ldots & \ldots & 0.80$\pm$0.05 & 3.00$\pm$0.05 & -0.5 & 0.0 & \ldots & 110 & a \\
10158-2844 & HR\,4049 & 10 18 07.6 & -28 59 31 & 7500 & 430.6 & 0.3 & 1.00$\pm$0.30 & 2.73$\pm$0.12 & -4.5 & \ldots & 800 & 18 & a \\
11000-6153 & HD\,95767 & 11 02 04.3 & -62 09 43 & 7600 & 1989 & 0.25 & 0.82$\pm$0.05 & 3.36$\pm$0.10 & 0.1 & \ldots & \ldots & 56 & a \\
11118-5726 &  GK\,Car  & 11 14 01.3 & -57 43 09 & 5500 & \ldots & \ldots & 0.49$\pm$0.05 & 4.06$\pm$0.03 & -1.3 & 1.2 & 1000 & 64 & a \\
11235-5921 & CN\,Cen & 11 25 47.9 & -59 37 56 & 5000 & \ldots & \ldots & 0.43$\pm$0.03 & 3.29$\pm$0.03 & -1.0 & \ldots & \ldots & 45 & b \\
11385-5517 & HD\,101584 & 11 40 58.8 & -55 34 26 & 7000 & \ldots & \ldots & 1.12$\pm$0.26 & 3.53$\pm$0.19 & 0.0 & \ldots & \ldots & 174 & j \\
12185-4856 &  SX\,Cen  & 12 21 12.6 & -49 12 41 & 6000 & 564.3 &0.0& 0.65$\pm$0.04 & 3.46$\pm$0.06 & -1.0 & 1.5 & 1100 & 45 & a \\
12222-4652 & HD\,108015 & 12 24 53.5 & -47 09 08 & 7000 & 906.3 &0.0& 1.02$\pm$0.05 & 3.93$\pm$0.09 & -0.1 & 0.1 & \ldots & 109 & a \\
13085-6747 & PX Mus & 13 11 57.5 & -68 03 35 & 5000 & 770 & \ldots & 0.46$\pm$0.03 & 2.80$\pm$0.03 & 0.0 & \ldots & \ldots & 11 & b \\
13258-8103 &  \ldots  & 13 31 07.1 & -81 18 30 & 6000 & \ldots & \ldots & 1.06$\pm$0.03 & 3.09$\pm$0.04 & 0.0 & \ldots & \ldots & 139 & a \\
14524-6838 & EN TrA, HD131356 & 14 57 00.7 & -68 50 23 & 6000 & 1488 & 0.32 & 0.63$\pm$0.04 & 3.57$\pm$0.08 & -0.5 & 0.6 & 1300 & 67 & a \\
15469-5311 &  \ldots  & 15 50 44.0 & -53 20 44 & 7500 & 390.2 & 0.08 & 1.06$\pm$0.05 & 3.36$\pm$0.09 & 0.0 & 1.8 & 1300 & 68 & a \\
16230-3410 &  \ldots  & 16 26 20.3 & -34 17 12 & 6250 & 649.8 &0.0& 0.60$\pm$0.05 & 3.95$\pm$0.04 & -0.5 & 1.0 & 1500 & 42 & a \\
17038-4815 &  \ldots  & 17 07 36.3 & -48 19 08 & 4750 & 1394 & 0.63 & 0.46$\pm$0.04 & 4.25$\pm$0.06 & -1.5 & 0.7 & 1400 & 85 & a \\
17111-1819 & HD155720 & 17 14 03.1 & -18 22 40 & 4250 & 712 & \ldots & 0.36$\pm$0.04 & 3.36$\pm$0.05 & -1 & \ldots & \ldots & 19 &n\\
17243-4348 &  LR\,Sco  & 17 27 56.1 & -43 50 48 & 6250 & 475 & \ldots & 0.79$\pm$0.04 & 3.40$\pm$0.07 & 0.0 & 0.8 & 1400 & 91 & a \\
17250-5951 & UY\,Ara & 17 29 29.0 & -59 54 02 & 5500 & \ldots & \ldots & 0.53$\pm$0.03 & 3.55$\pm$0.03 & -1.0 & 1.4 & 1000 & 63 & b \\
17279-1119 & HD\,158616, V340\,Sgr & 17 30 46.9 & -11 22 08 & 7400 & 363.3 &0.0& 0.69$\pm$0.05 & 2.50$\pm$0.06 & -0.7 & 0.0 & \ldots & 13 & b \\
17534+2603 & 89 Her & 17 55 25.2 & +26 02 60 & 6500 & 289.1 & 0.29 & 0.61$\pm$0.28 & 3.06$\pm$- & 0.0 & 0.6 & 1500 & 42 & a \\
17530-3348 &  AI\,Sco  & 17 56 18.5 & -33 48 47 & 5000 & 977 & \ldots & 0.61$\pm$0.04 & 3.10$\pm$0.08 & 0.0 & 0.3 & 1100 & 63 & a \\
18123+0511 &  \ldots  & 18 14 49.4 & +05 12 55 & 5000 & \ldots & \ldots & 0.64$\pm$0.04 & 3.77$\pm$0.07 & 0.0 & \ldots & \ldots & 128 & a \\
18548-0552 & BZ Sct & 18 57 33.4 & -05 47 56 & 6250 & \ldots & \ldots & 0.48$\pm$0.03 & 4.42$\pm$0.03 & -0.8 & 1.3 & 1000 & 42 & k \\
19135+3937 & \ldots & 19 15 12.5 & +39 42 51 & 6000 & 126.97 & 0.13 & 0.56$\pm$0.03 & 3.16$\pm$0.04 & -0.5 & 0.5 & 1500 & 20 & l \\
19157-0247 & V1801 Aql & 19 18 22.5 & -02 42 09 & 7750 & 119.8 & 0.34 & 1.07$\pm$0.04 & 3.61$\pm$0.07 & 0.0 & \ldots & 1500 & 73 & a \\
19199+3950 & HP Lyr & 19 21 39.1 & +39 56 08 & 6300 & 1818 & 0.2 & 0.64$\pm$0.03 & 3.77$\pm$0.05 & -1.0 & 2.6 & 1300 & 29 & i \\
19370+0829 & EZ\,Aql & 19 39 29.7 & +08 36 28 & 5500 & \ldots & \ldots & 0.62$\pm$0.03 & 3.02$\pm$0.03 & -1.0 & \ldots & \ldots & 36 & b \\
19410+3733 & HD\,186438 & 19 42 52.9 & +37 40 42 & 6500 & \ldots & \ldots & 0.33$\pm$0.02 & 3.12$\pm$0.06 & -1.0 & \ldots & \ldots & 27 & b \\
19587+3928 & GK\,Cyg & 20 00 34.5 & +39 36 36 & 5250 & \ldots & \ldots & 0.33$\pm$0.03 & 2.96$\pm$0.03 & 0.0 & \ldots & \ldots & 46 & b \\
19580-3038 & V1711\,Sgr & 20 01 08.0 & -30 30 39 & 5000 & \ldots & \ldots & 0.49$\pm$0.06 & 3.44$\pm$0.03 & -1.2 & \ldots & \ldots & 38 & b \\
20094+3721 & \ldots & 20 11 16.8 & +37 30 52 & 5750 & \ldots & \ldots & 0.37$\pm$0.03 & 3.53$\pm$0.04 & 0.0 & \ldots & \ldots & 18 &n\\
20117+1634 &  R\,Sge & 20 14 03.8 & +16 43 35 & 5750 & \ldots & \ldots & 0.67$\pm$0.03 & 3.27$\pm$0.10 & -0.5 & 1.1 & 1200 & 97 & a \\
20131+2554 & EF\,Vul & 20 15 18.6 & +26 03 38 & 4250 & \ldots & \ldots & 0.38$\pm$0.06 & 3.35$\pm$0.03 & 0.0 & \ldots & \ldots & 35 &n\\
20343+2625 & V\,Vul  & 20 36 31.8 & +26 36 17 & 5250 & \ldots & \ldots & 0.57$\pm$0.06 & 2.75$\pm$0.08 &0.0& -0.1 & \ldots & 36 & a \\
20401+2550 & HD\,340949 & 20 42 20.7 & +26 01 22 & 5000 & \ldots & \ldots & 0.34$\pm$0.03 & 2.44$\pm$0.11 & 0.0 & \ldots & \ldots & 15 & n \\
22223+5556 & BT\,Lac & 22 24 13.8 & +56 11 33 & 5250 & 650 & \ldots & 0.64$\pm$0.04 & 3.33$\pm$0.03 & 0.0 & 0.5 & 1400 & 36 & b \\
22327-1731 & HD\,213985 & 22 35 27.5 & +39 56 08 & 8250 & 259.6 & 0.21 & 0.89$\pm$0.05 & 3.18$\pm$0.07 & -1.0 & \ldots & 1000 & 26 & a \\
\hline
\multicolumn{10}{l}{Category 2}\\
\hline 
06034+1354 &  DY\,Ori  & 06 06 14.9 & +13 54 19 & 6000 & 1248 & 0.22 & 0.23$\pm$0.02 & 5.10$\pm$0.05 & -2.0 & 2.1 & 1000 & 53 & a \\
06472-3713 &  ST\,Pup  & 06 48 56.4 & -37 16 33 & 5750 & 417.4 &0.0& 0.16$\pm$0.06 & 5.30$\pm$0.03 & -1.5 & 2.1 & 800 & 75 & a \\
11472-0800 &  AF\,Crt & 11 49 48.5 & -08 17 21 & 5750 & \ldots & \ldots & 0.41$\pm$0.03 & 6.40$\pm$0.02 & -2.5 & 3.4 & 1000 & 59 & a \\
12067-4508 &  RU\,Cen  & 12 09 23.7 & -45 25 35 & 6000 & 1489 & 0.62 & 0.24$\pm$0.05 & 4.95$\pm$0.05 & -2.0 & 1.0 & 800 & 39 & a\\
17233-4330 & V1504\,Sco & 17 26 57.5 & -43 33 13 & 6250 & 735 & \ldots & 1.18$\pm$0.04 & 5.19$\pm$0.03 & -1.0 & 1.4 & 1000 & 509 & a \\
18281+2149 &  AC\,Her  & 18 30 16.2 & +21 52 00 & 5500 & 1188.9 &0.0& 0.20$\pm$0.02 & 5.41$\pm$0.08 & -1.5 & 1.0 & 1200 & 37 & a \\
18564-0814 &  AD\,Aql  & 18 59 08.1 & -08 10 14 & 6300 & \ldots & \ldots & 0.08$\pm$0.03 & 5.07$\pm$0.03 & -2.1 & 2.5 & 1000 & 35 & a \\
	\hline
\multicolumn{10}{l}{Category 3}\\
\hline 
05208-2035 &  \ldots  & 05 22 59.4 & -20 32 53 & 4000 & 234.38 &0.0& 0.25$\pm$0.05 & 4.38$\pm$0.05 & 0.0 & \ldots & \ldots & 44 & a \\
06072+0953 &  CT\,Ori  & 06 09 58.0 & +09 52 32 & 5500 & \ldots & \ldots & 0.26$\pm$0.05 & 4.18$\pm$0.05 & -2.0 & 1.9 & 1200 & 56 & a \\
06165+3158 & \ldots & 06 19 50.6 & +31 56 58 & 6555 & 262.6 &0.0& 0.08$\pm$0.03 & 3.85$\pm$0.04 & 0.0 & 0.0 & \ldots & 12 & b \\
06452-3456 & \ldots & 06 47 00.5 & -34 59 53 & 6000 & 215.4 &0.0& 0.29$\pm$0.04 & 2.70$\pm$0.06 & 0.0 & \ldots & \ldots & 11 & i \\
07284-0940 &  U\,Mon  & 07 30 47.5 & -09 46 37 & 5000 & 2550 & 0.25 & 0.18$\pm$0.32 & 4.04$\pm$0.32 & -0.8 & 0.0 & \ldots & 57 & a \\
07331+0021 & AI CMi & 07 35 41.2 & +00 14 58 & 6000 & \ldots & \ldots & 0.28$\pm$0.03 & 3.80$\pm$0.09 &0.0& \ldots & \ldots & 18 & e \\
16278-5526 & GZ Nor & 16 31 53.2 & -55 33 08 & 4750 & \ldots & \ldots & 0.21$\pm$0.03 & 3.56$\pm$0.03 & -2.0 & 0.8 & 800 & 20 & c \\
19163+2745 & EP Lyr  & 19 18 19.5 & +27 51 03 & 7000 & 1151 & 0.39 & 0.02$\pm$0.04 & 2.71$\pm$0.03 & -1.5 & 1.3 & 800 & 3 & a \\
21216+1803 & AU Peg & 21 24 00.2 & +18 16 44 & 5750 & 53 & \ldots & 0.26$\pm$0.03 & 3.08$\pm$0.05 & -0.2 & \ldots & \ldots & 15 & n \\
22251+5406 & \ldots & 22 27 05.1 & +54 21 29 & 6125 & \ldots & \ldots & 0.10$\pm$0.05 & 3.76$\pm$0.04 & 0.0 & \ldots & \ldots & 13 & n \\
\hline
\multicolumn{10}{l}{Category 4}\\
\hline 
06054+2237 & SS\,Gem & 06 08 35.1 & +22 37 02 & 5400 & \ldots & \ldots & 0.12$\pm$0.04 & 0.83$\pm$0.05 & -1.0 & 2.0 & 1100 & 2 & b \\
06338+5333 & HD\,46703 & 06 37 52.4 & +53 31 02 & 6250 & 597.4 & 0.3 & 0.05$\pm$0.02 & 2.25$\pm$0.03 & -1.5 & 0.9 & 800 & 2 & a \\
13110-5425 & HD\,114855 & 13 14 08.3 & -54 41 35 & 6000 & \ldots & \ldots & 0.11$\pm$0.05 & 0.38$\pm$0.07 & 0.0 & \ldots & \ldots & 2 & g \\
F15240+1452 & HD\,137569 & 15 26 20.8 & +14 41 36 & 12000 & 530 & \ldots & 0.04$\pm$0.02 & 0.17$\pm$0.03 & -3.0 & \ldots & \ldots & 0 & d \\
16399-6247 & CPD-62\,5428 & 16 44 35.4 & -62 53 26 & 7500 & \ldots & \ldots & 0.06$\pm$0.03 & 1.08$\pm$0.03 & -1.0 & \ldots & \ldots & 6 & g \\
\ldots& CC\,Lyr & 18 33 57.4 & +31 38 24 & 6250 & \ldots & \ldots & 0.17$\pm$0.03 & 1.87$\pm$0.03 & -3.5 & \ldots & 1000 & 3 & f \\
18448-0545 & R\,Sct & 18 47 29.0 & -05 42 19 & 4500 & \ldots & \ldots & 0.16$\pm$0.37 & 1.00$\pm$0.31 & -0.5 & \ldots & \ldots & 6 & b \\
20160+2734 & AU\,Vul & 20 18 05.9 & +27 44 04 & 6000 & \ldots & \ldots & 0.20$\pm$0.05 & 1.09$\pm$0.06 &0.0& \ldots & \ldots & 5 & n \\
\ldots & BD+394926 & 22 46 11.2 & +40 06 26 & 7500 & 871.7 & \ldots & 0.02$\pm$0.02 & 0.64$\pm$0.03 & -2.9 & 2.0 & 1000 & 1 & a \\
\hline
\multicolumn{10}{l}{Uncategorized}\\
\hline 
10456-5712 & HD\,93662 & 10 47 38.4 & -57 28 03 & 4250 & 572 & \ldots & 0.35$\pm$0.30 & \ldots &  0.0 & \ldots & \ldots & 48 & a \\
\hline \\
\multicolumn{14}{l}{\tablebib{(a): \citet{DeRuyter2006}; (b) \citet{Gezer2015}; (c) \citet{Gezer2019}; (d) \citet{Giridhar2005a}; (e) \citet{Klochkova1996}; (f) \citet{Maas2007}; (g) \citet{Giridhar2010}; (h) \citet{Gorlova2012}; (i) \citet{Oomen2018}; (j) \citet{Olofsson2015}; (k) \citet{Giridhar2005b}; (l) \citet{Gorlova2015}; \textbf{ (m) \citet{Manick2021};}  (n) This work}}
\end{longtable}

\end{landscape}
}

\begin{acknowledgements}
We want to thank the referee for his/her valuable comments that helped us to improve the manuscript.
J.K. acknowledges support from FWO under the senior postdoctoral fellowship (1281121N). H.V.W. acknowledges support from KU Leuven under contract C1/4/17/082 and from the FWO under contract G097619N.
D.K. acknowledges the support of the Australian Research Council (ARC) Discovery Early Career Research Award (DECRA) grant (95213534). 
J.K. dedicates this work to his grandmother that passed away because of COVID during the writing of this paper.
\end{acknowledgements}

\bibliography{references} 
\bibliographystyle{aa} 

\begin{appendix}

\section{Spectral energy distributions of the target catalogs}

\begin{figure*}
   \centering
   \includegraphics[width= 6cm]{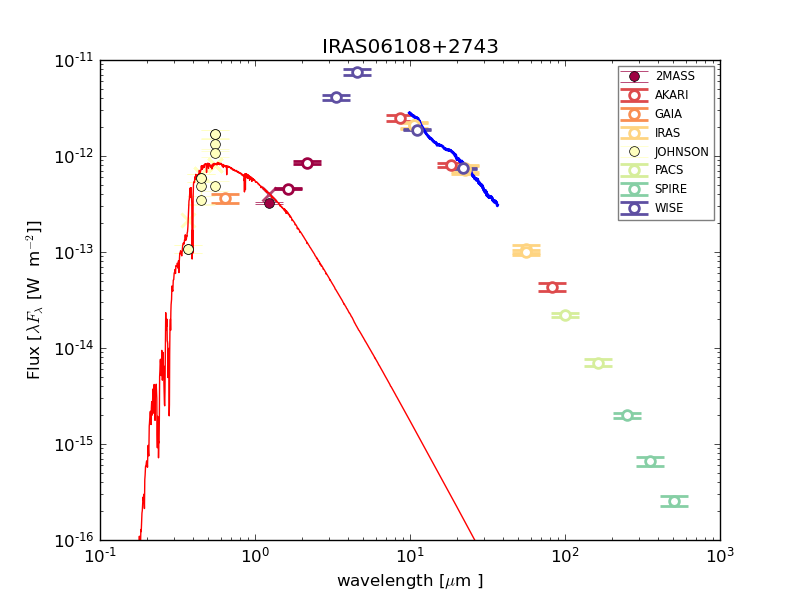}
   \includegraphics[width= 6cm]{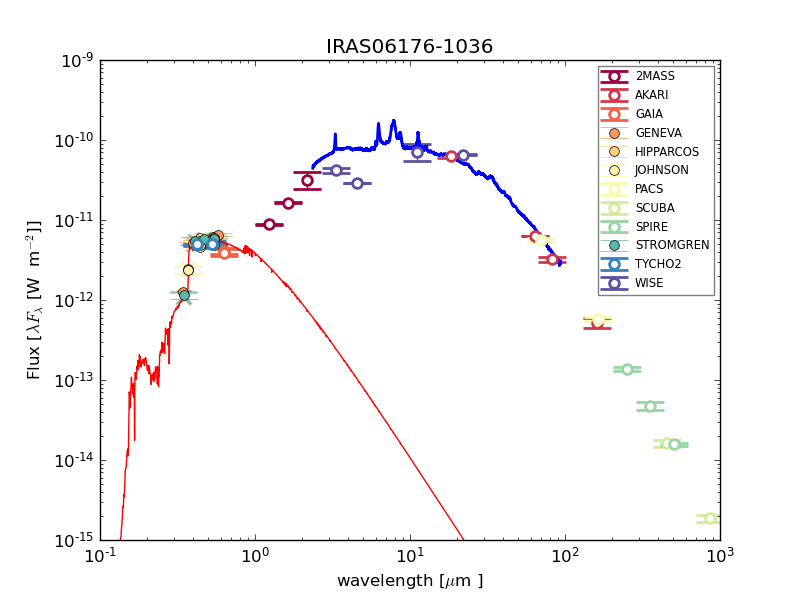}
   \includegraphics[width= 6cm]{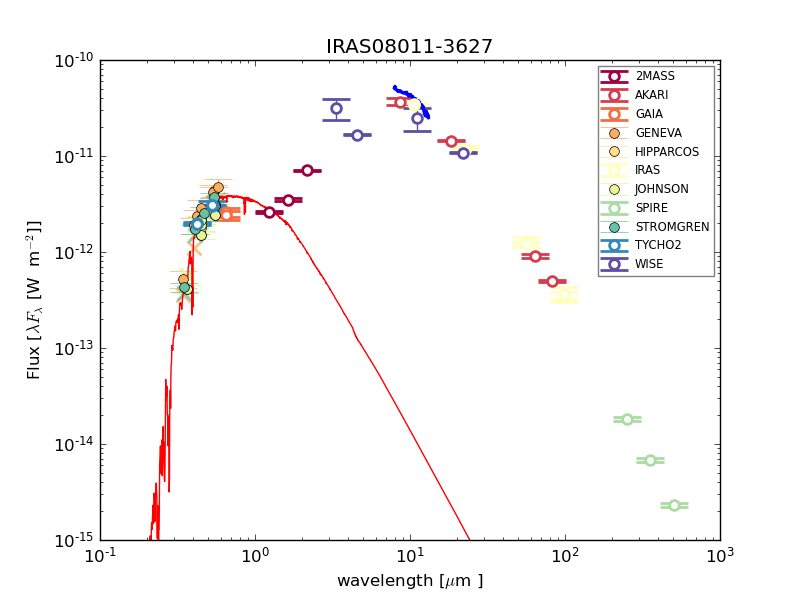}
   \includegraphics[width= 6cm]{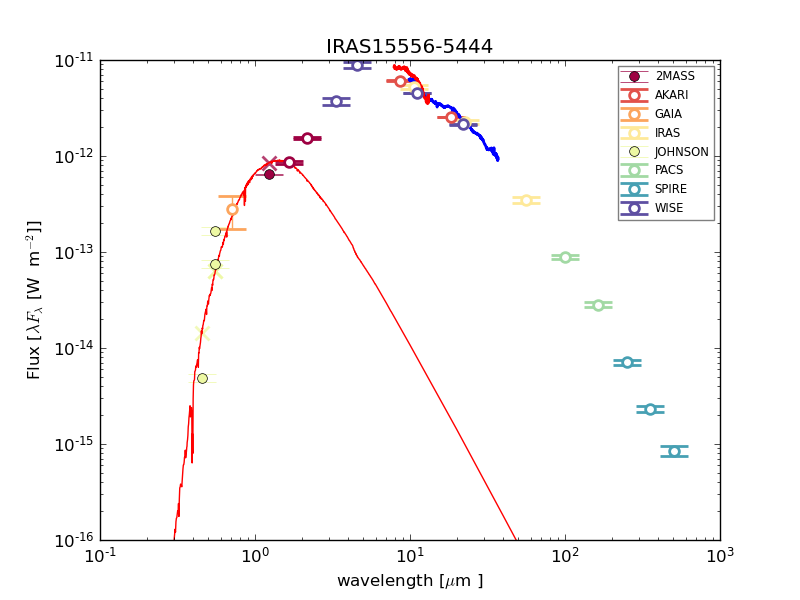}
    \includegraphics[width= 6cm]{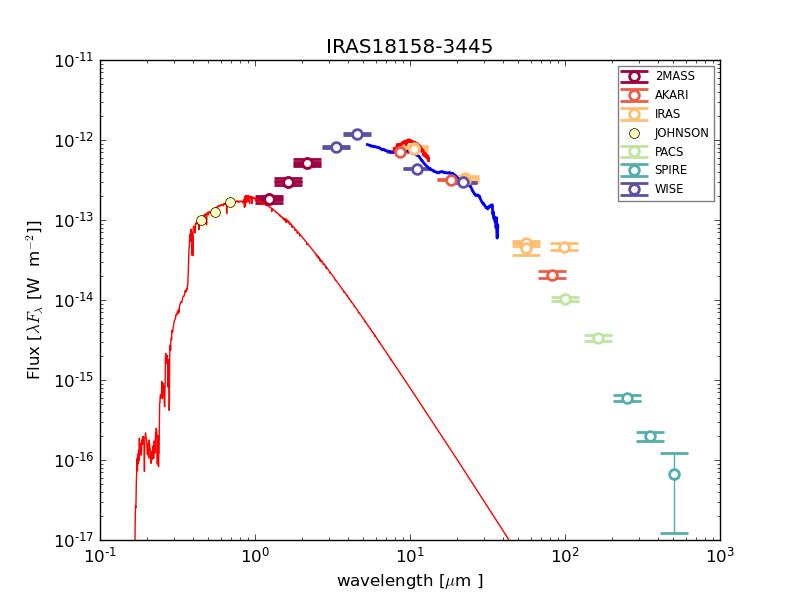}
   \includegraphics[width= 6cm]{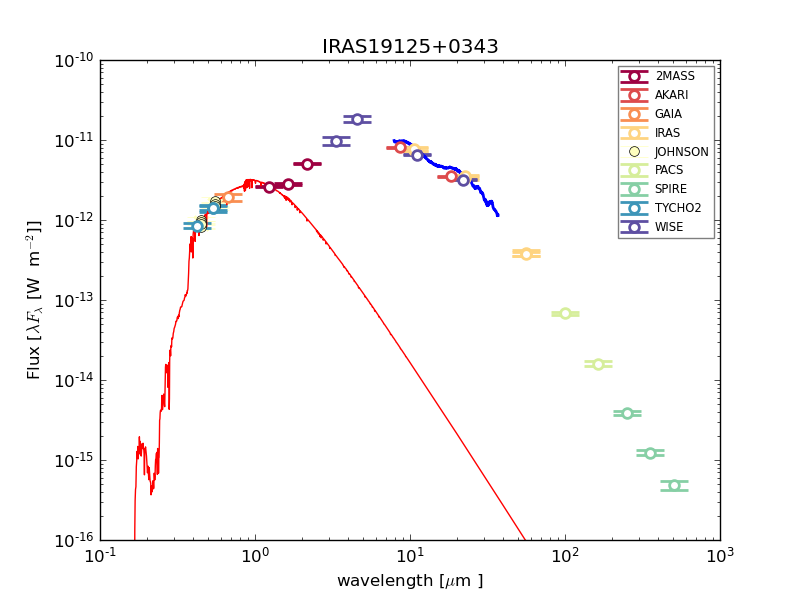}
   \includegraphics[width= 6cm]{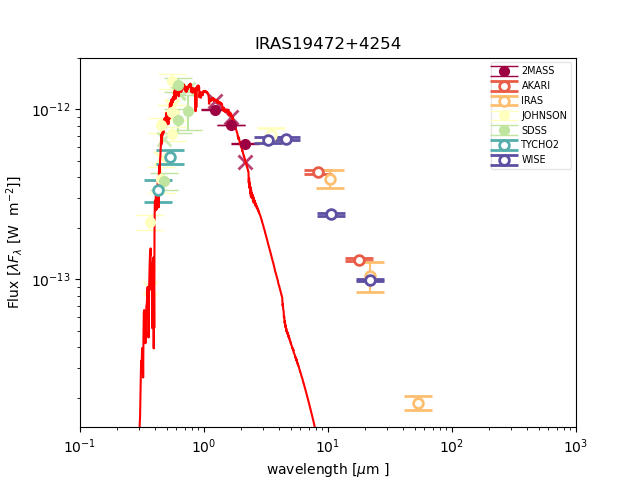}
   \includegraphics[width= 6cm]{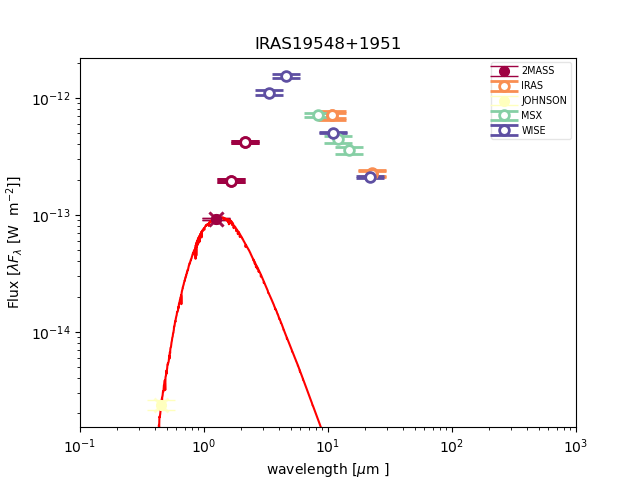}
   \includegraphics[width= 6cm]{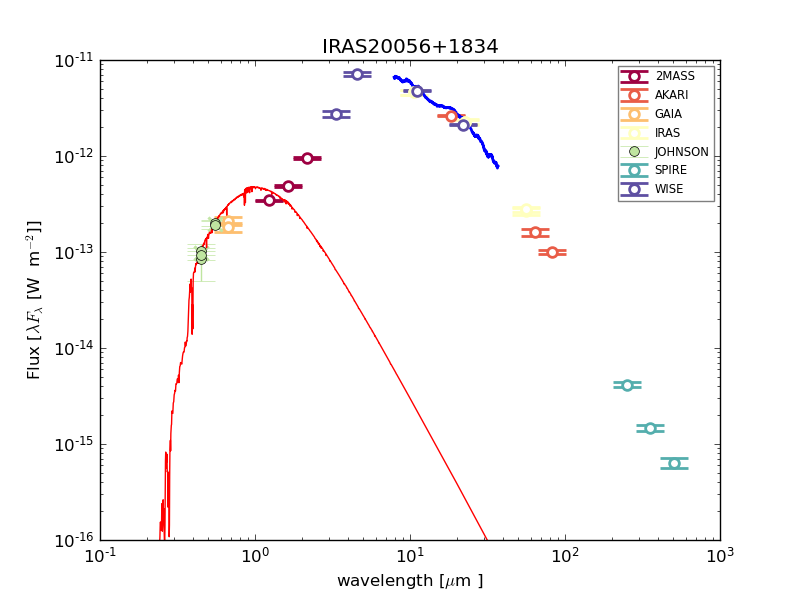}
   \caption{Spectral energy distributions (SED) of Category 0 targets. Red: Best fit photospheric models. Blue: SPITZER spectrum when available.}
   \label{fig:Cat0}
    \end{figure*}

\begin{figure*}
   \centering
    \includegraphics[width= 6cm]{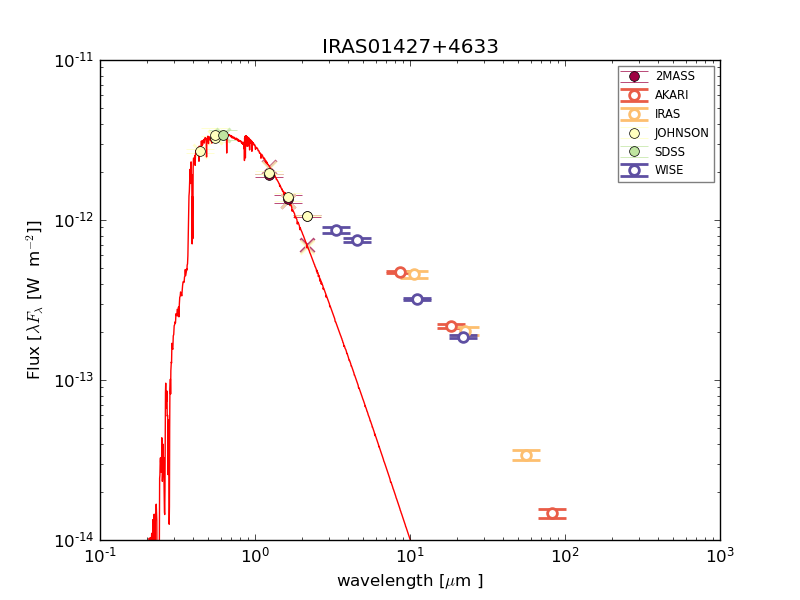}
   \includegraphics[width= 6cm]{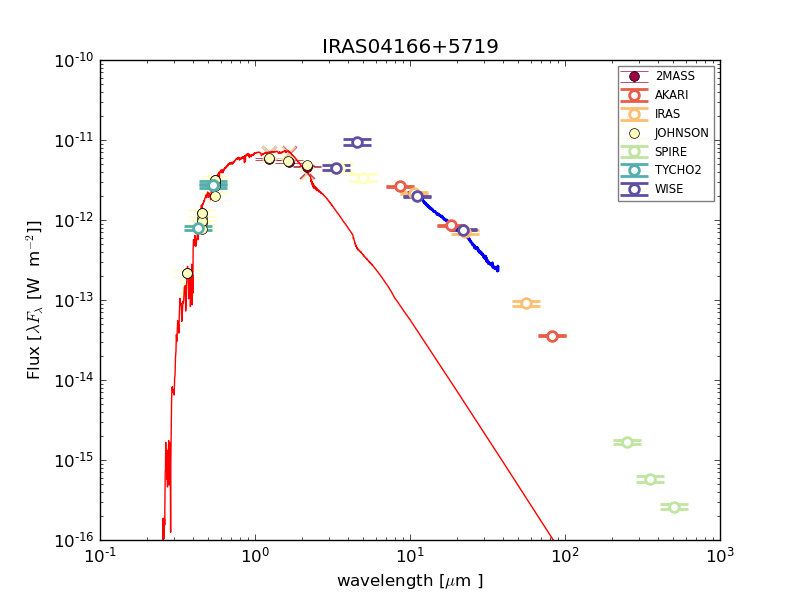}
   \includegraphics[width= 6cm]{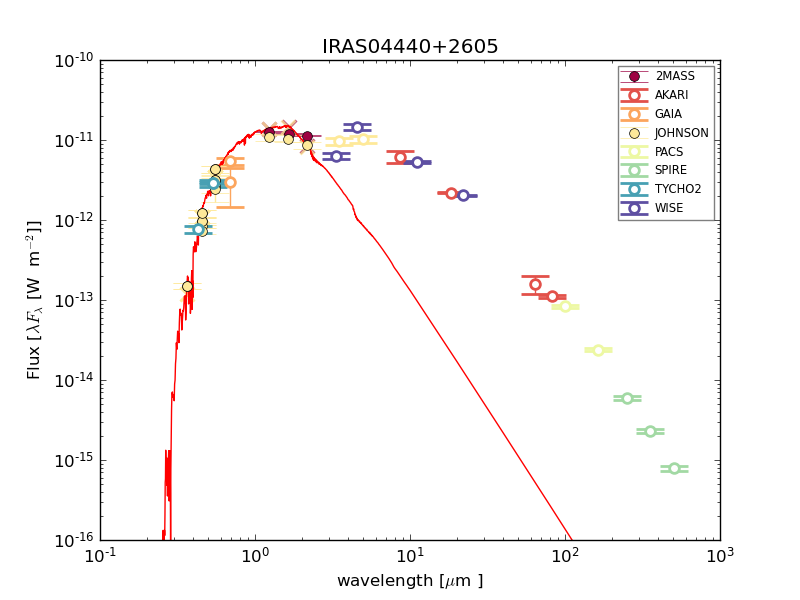}
   \includegraphics[width= 6cm]{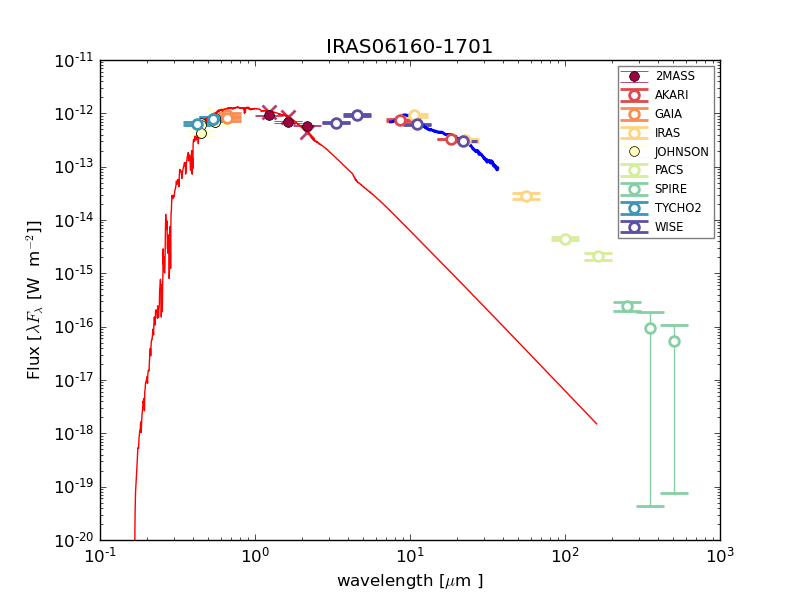}
    \includegraphics[width= 6cm]{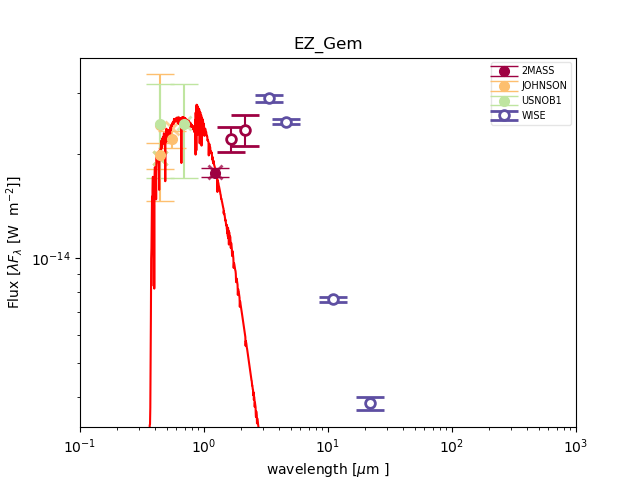}
   \includegraphics[width= 6cm]{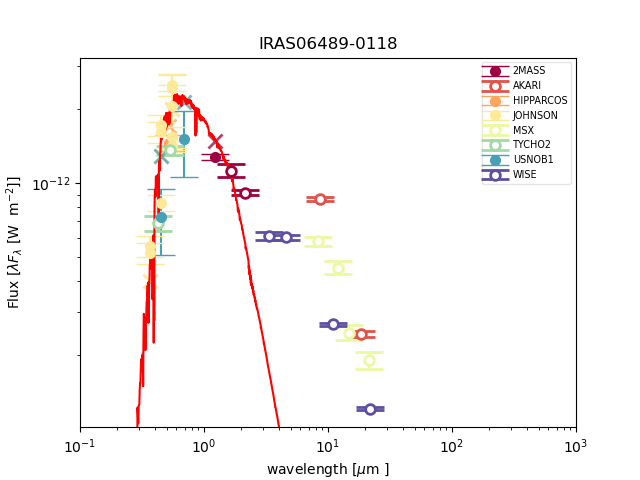}
   \includegraphics[width= 6cm]{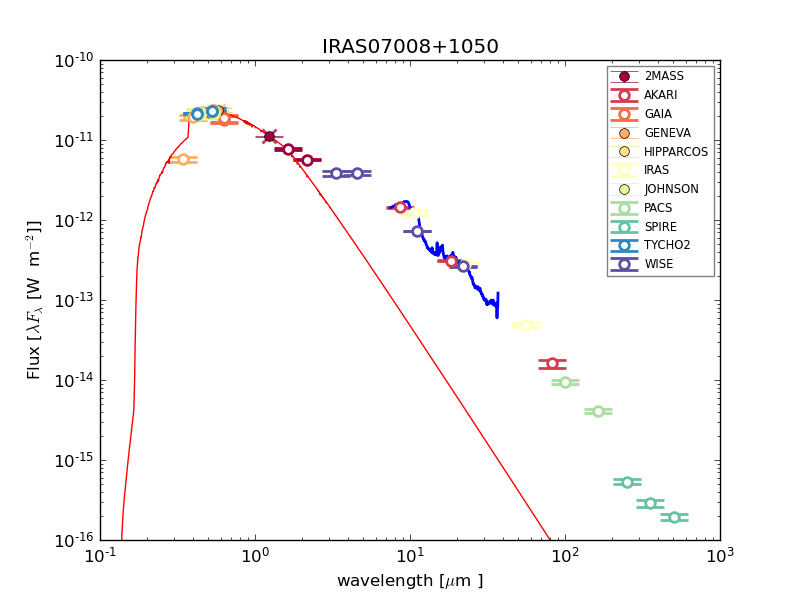}
      \includegraphics[width= 6cm]{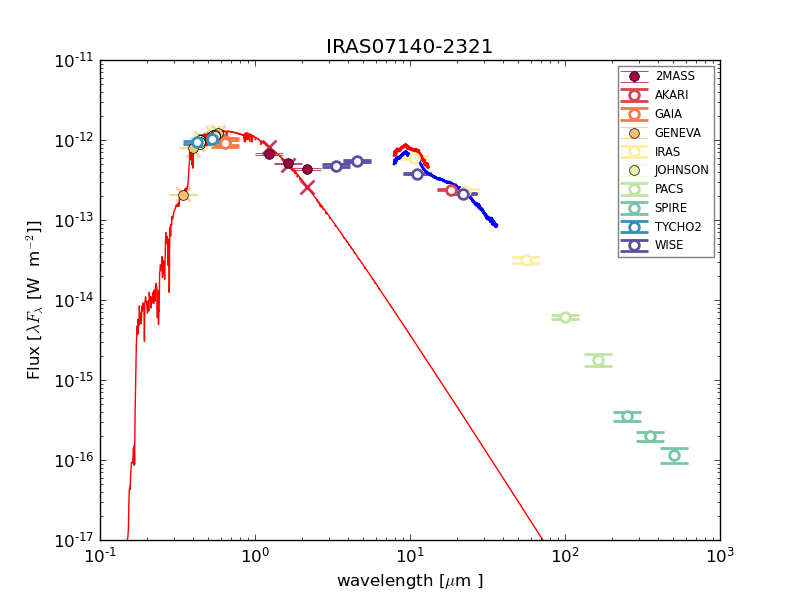}
   \includegraphics[width= 6cm]{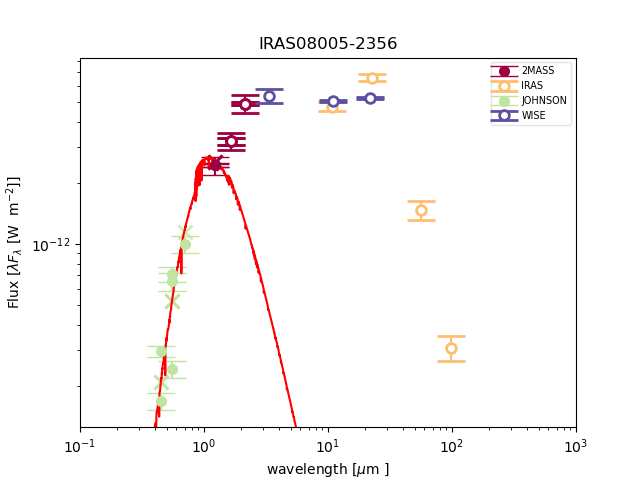}
\includegraphics[width= 6cm]{Fig/SED/IRAS08544-4431SED.png}
   \includegraphics[width= 6cm]{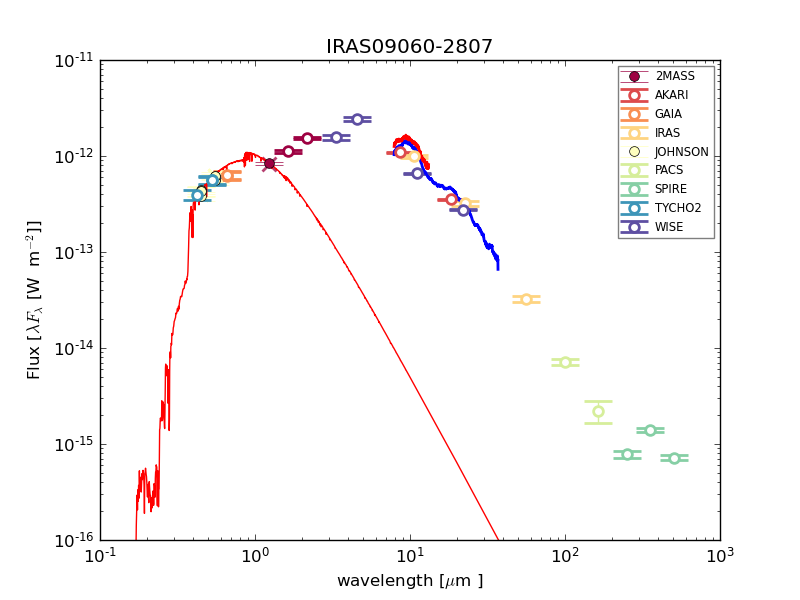}
   \includegraphics[width= 6cm]{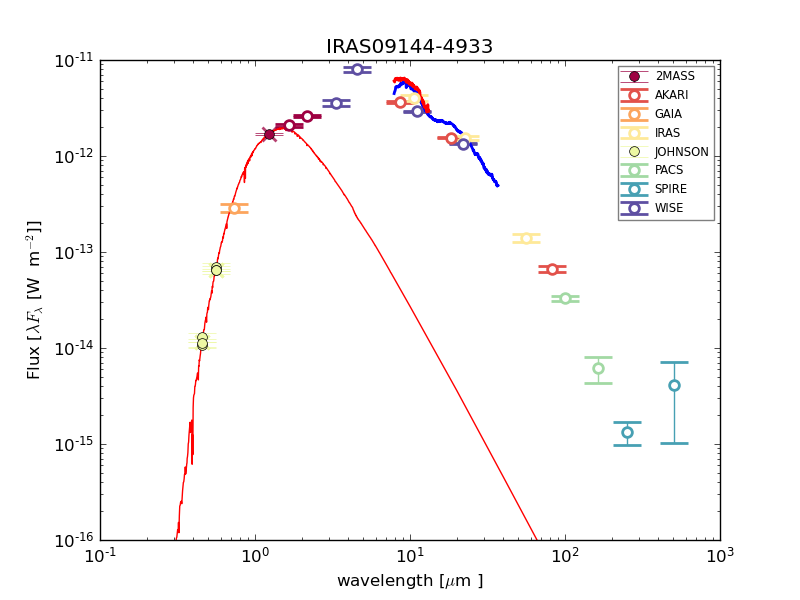}
   \includegraphics[width= 6cm]{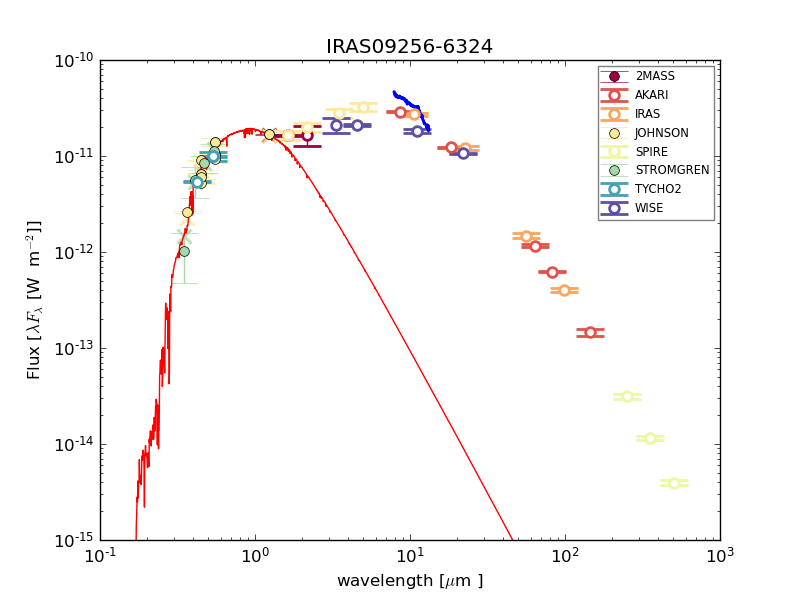}
    \includegraphics[width= 6cm]{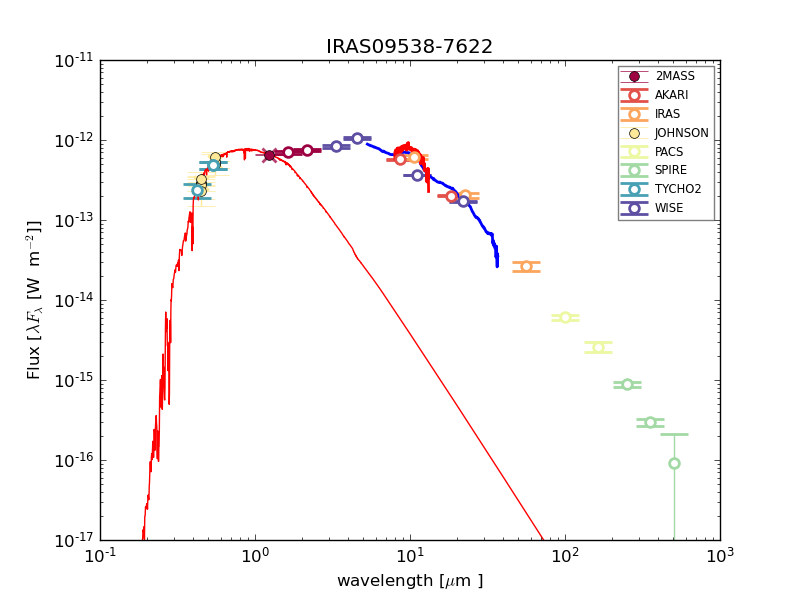}
   \includegraphics[width= 6cm]{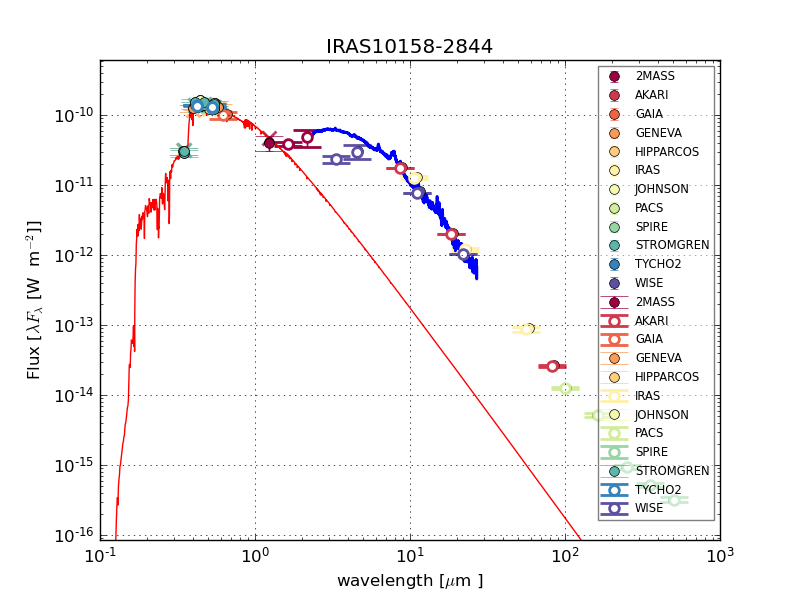}
  
    \caption{\modifLE{Spectral energy distributions (SED) of Category 1 targets. Red: Best fit photospheric models. Blue: SPITZER spectrum when available.} }
    \label{fig:Cat1}
  \end{figure*}

\begin{figure*}
   \centering
    \includegraphics[width= 6cm]{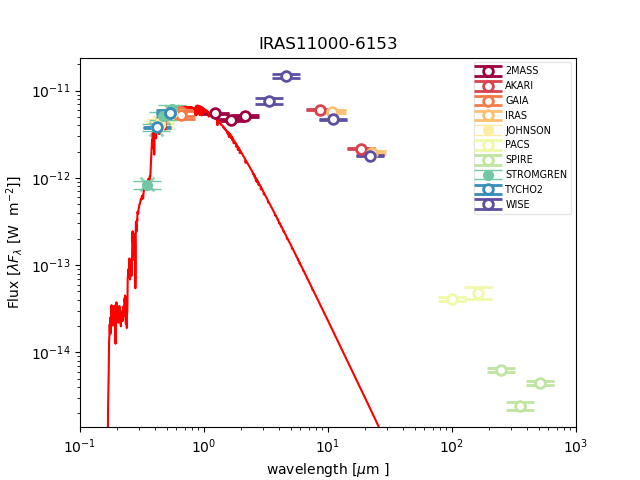}
     \includegraphics[width= 6cm]{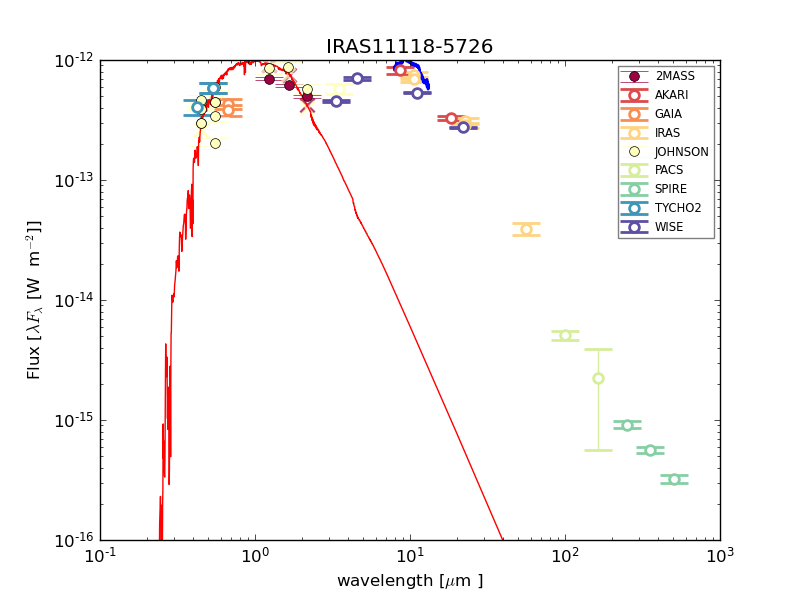}
      \includegraphics[width= 6cm]{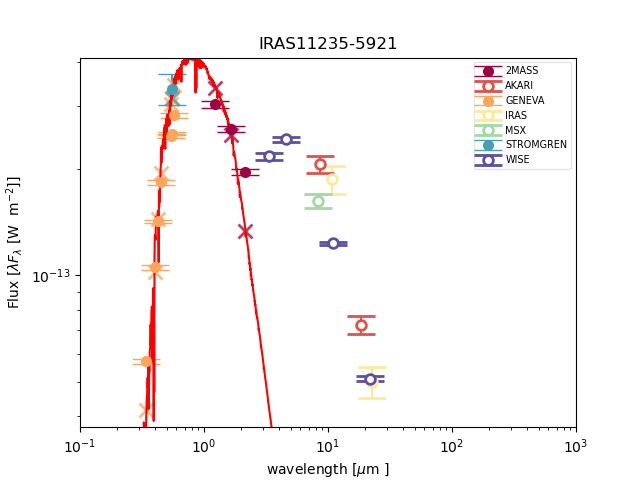}
    \includegraphics[width= 6cm]{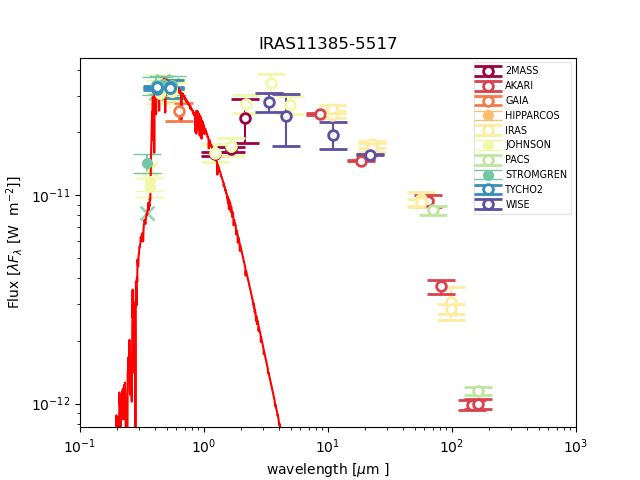}
   \includegraphics[width= 6cm]{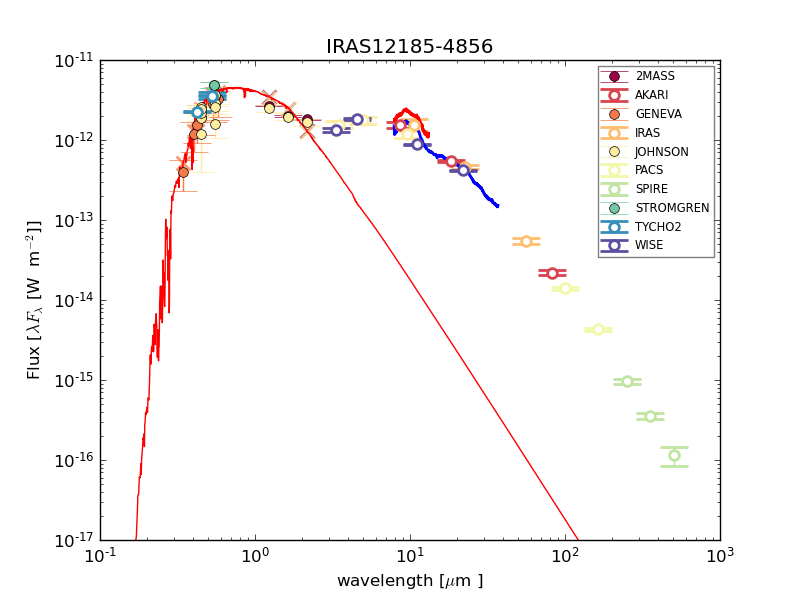}
   \includegraphics[width= 6cm]{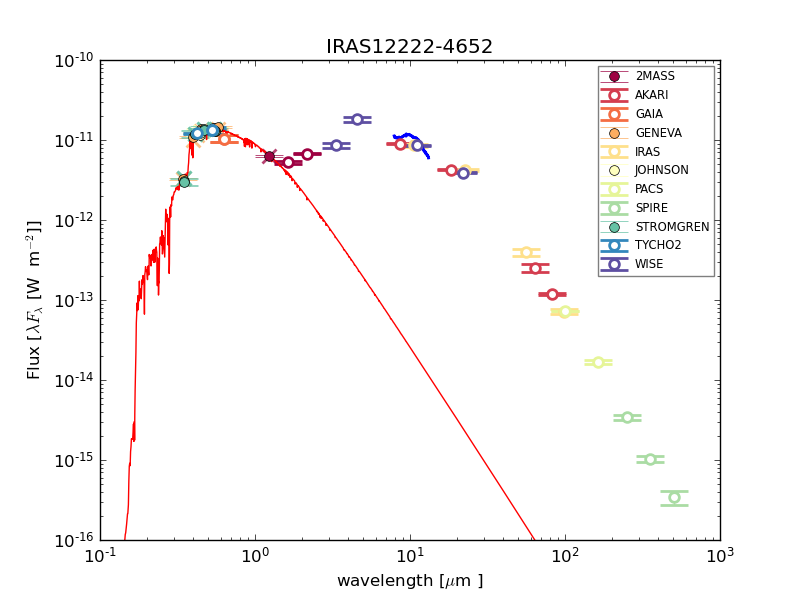}
   \includegraphics[width= 6cm]{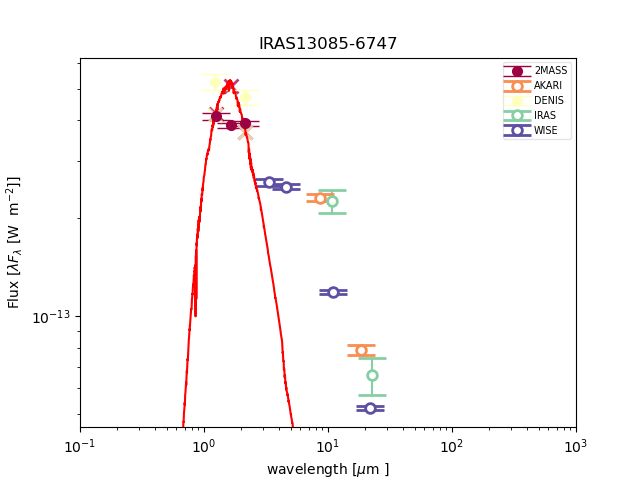}
    \includegraphics[width= 6cm]{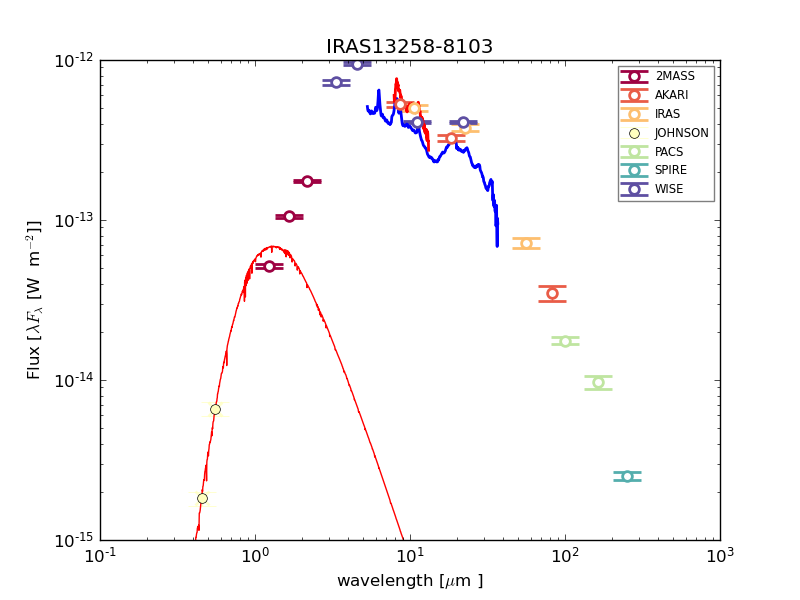}
   \includegraphics[width= 6cm]{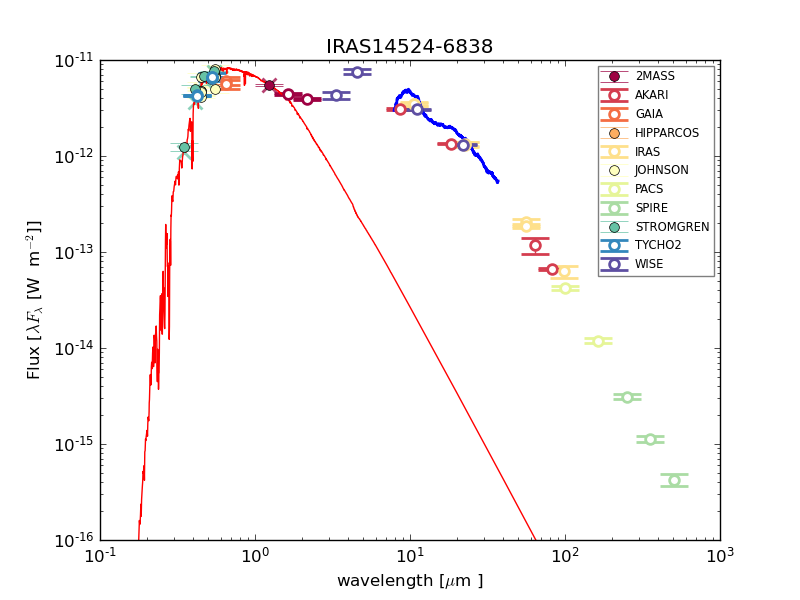}
   \includegraphics[width= 6cm]{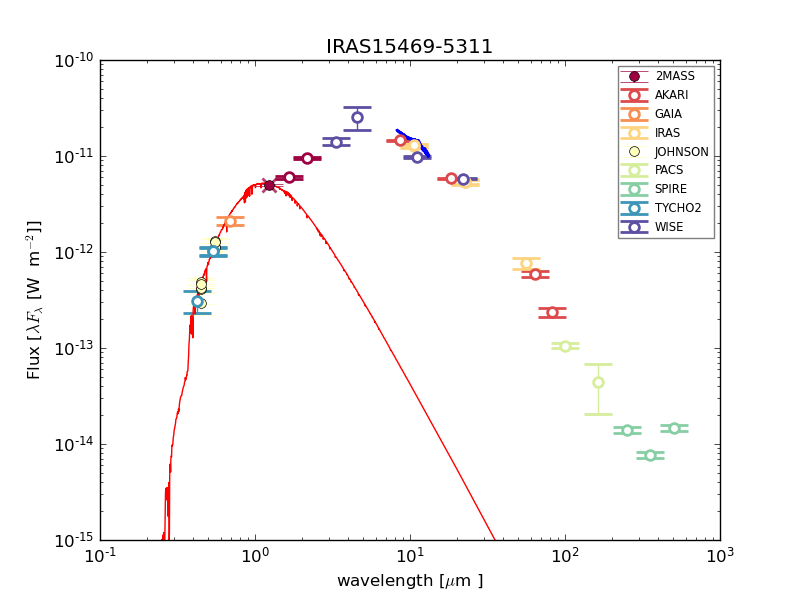}
   \includegraphics[width= 6cm]{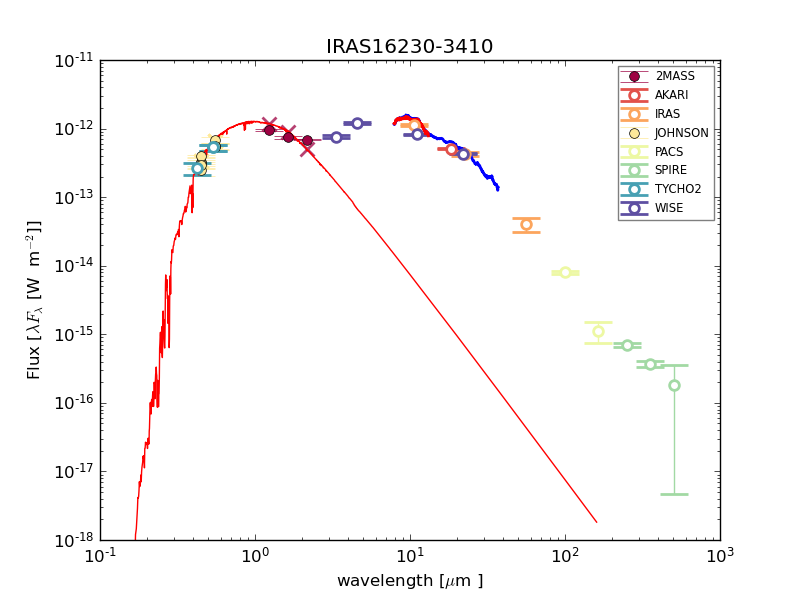}
   \includegraphics[width= 6cm]{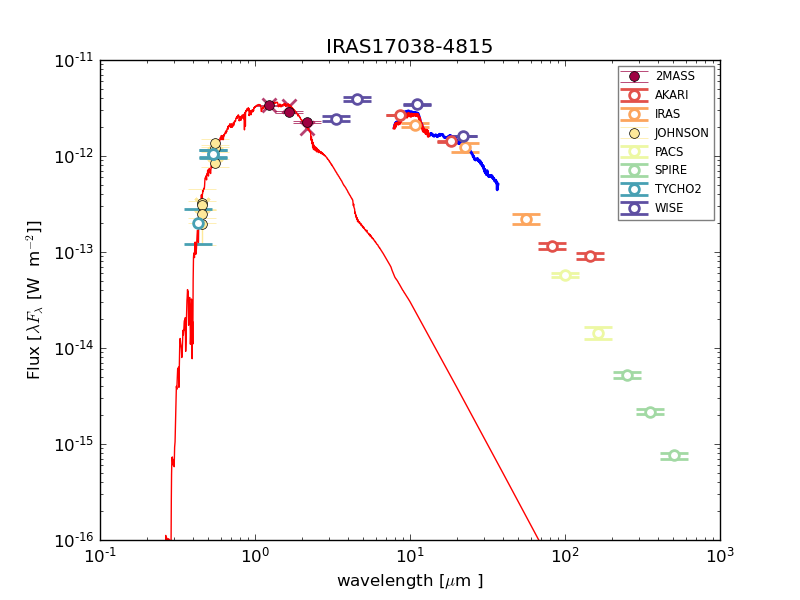}
   \includegraphics[width= 6cm]{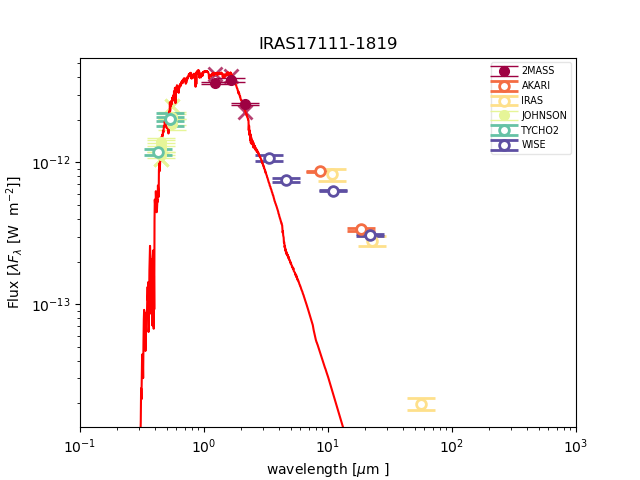}
   \includegraphics[width= 6cm]{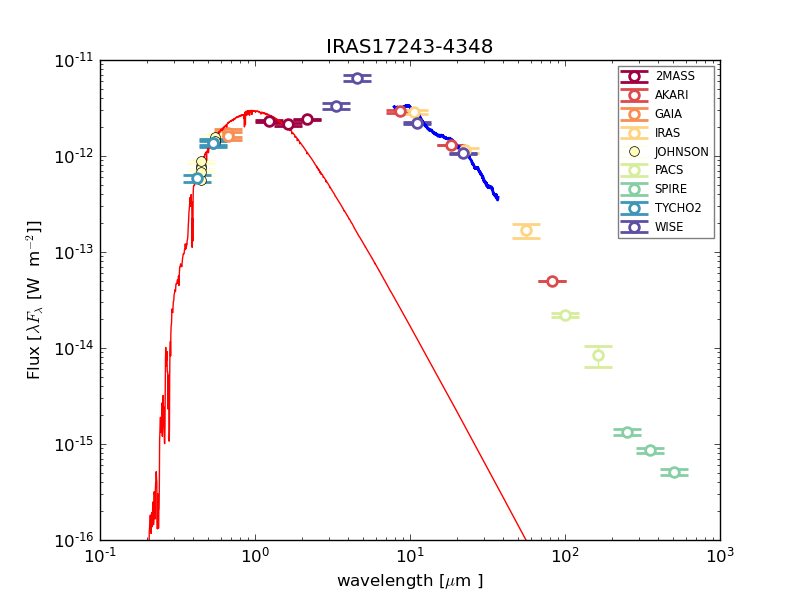}
   \includegraphics[width= 6cm]{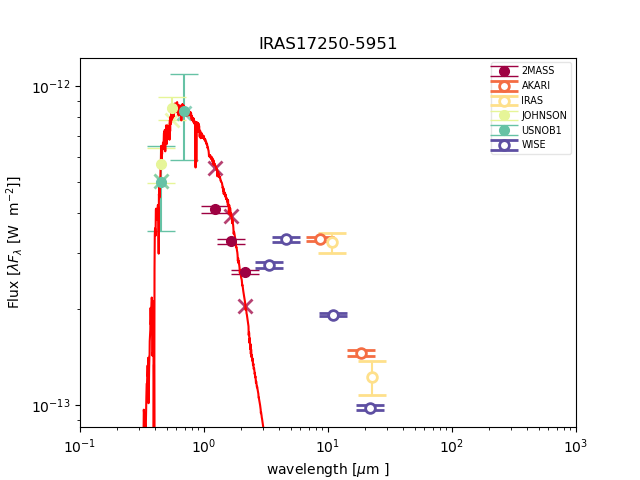}
     \caption{Fig.\,\ref{fig:Cat1} continued.}
    \label{fig:Cat12}
  \end{figure*}

\begin{figure*}
   \centering
      \includegraphics[width= 6cm]{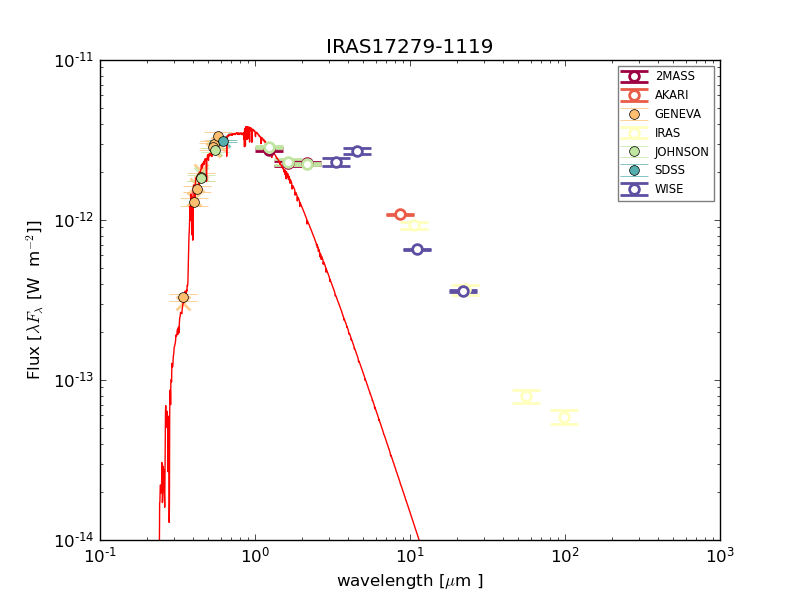}
      \includegraphics[width= 6cm]{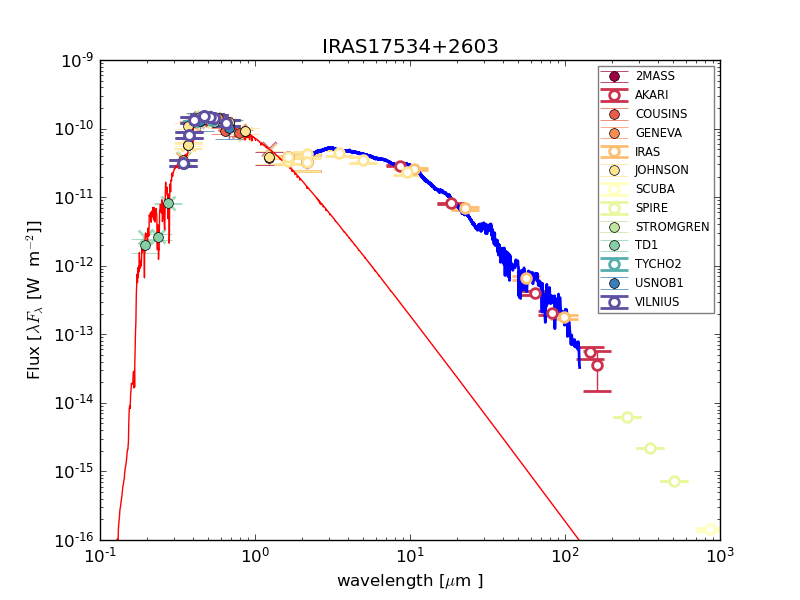}
      \includegraphics[width= 6cm]{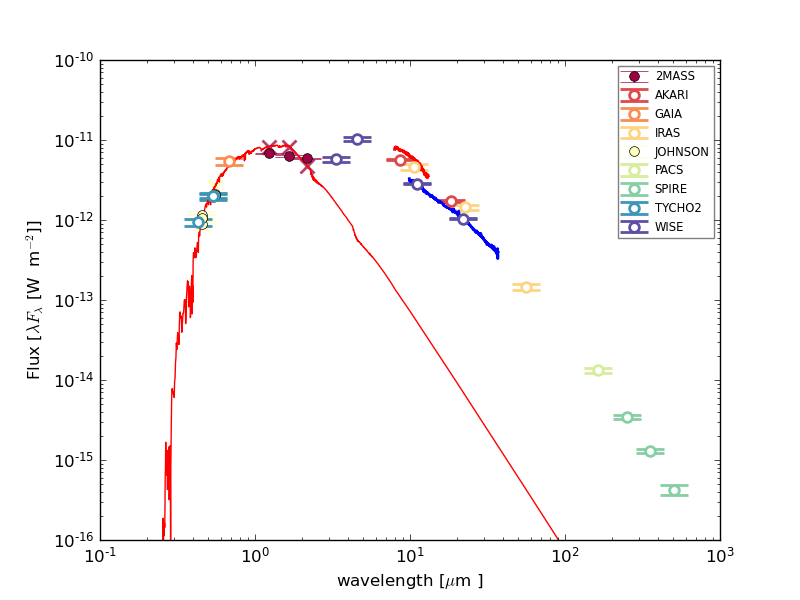}
   \includegraphics[width= 6cm]{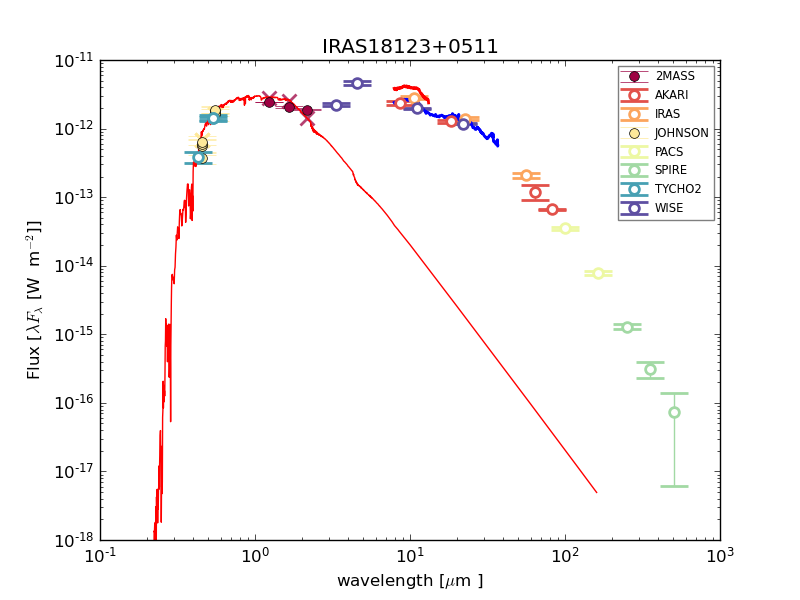}
    \includegraphics[width= 6cm]{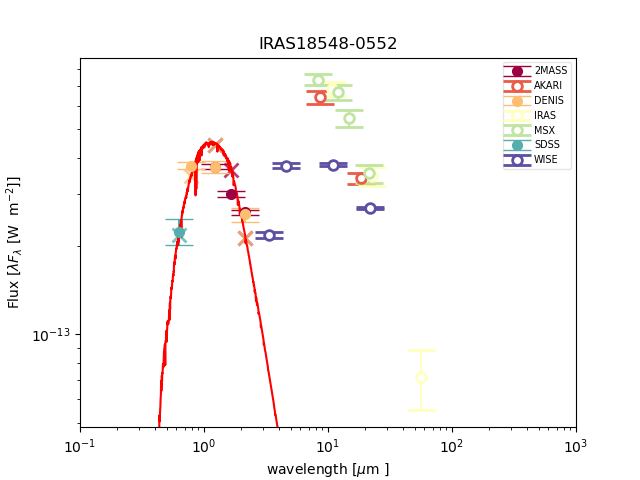}
   \includegraphics[width= 6cm]{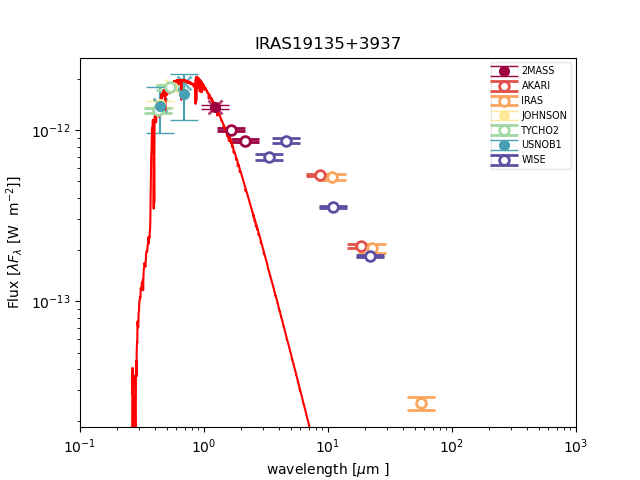}
   \includegraphics[width= 6cm]{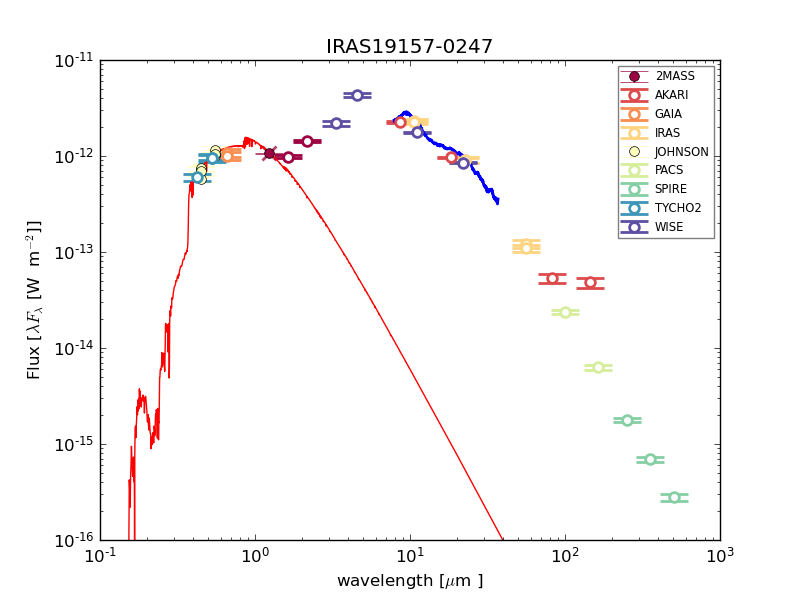}
   \includegraphics[width= 6cm]{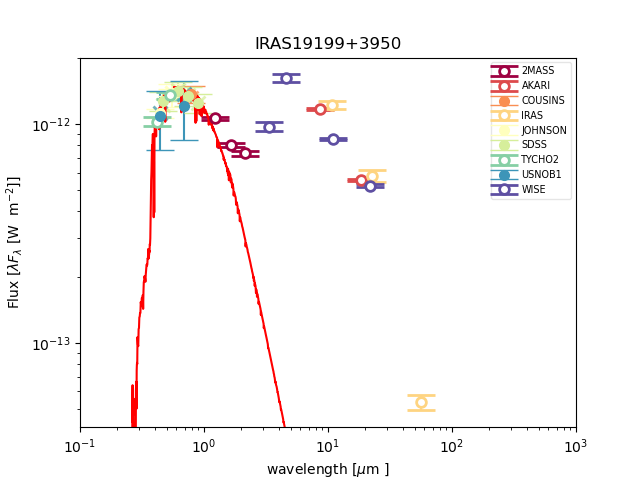}
    \includegraphics[width= 6cm]{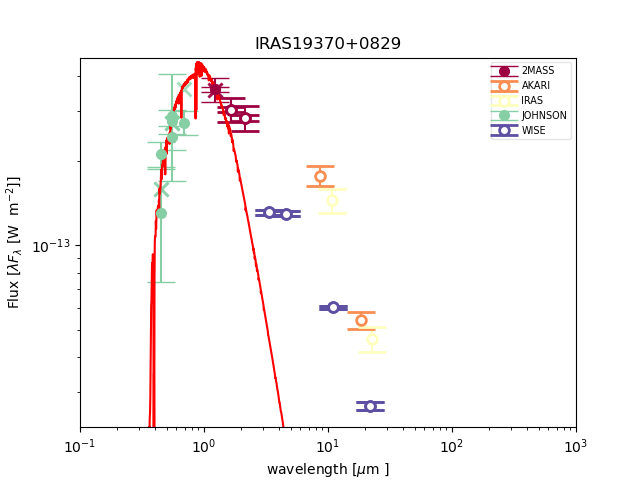}
   \includegraphics[width= 6cm]{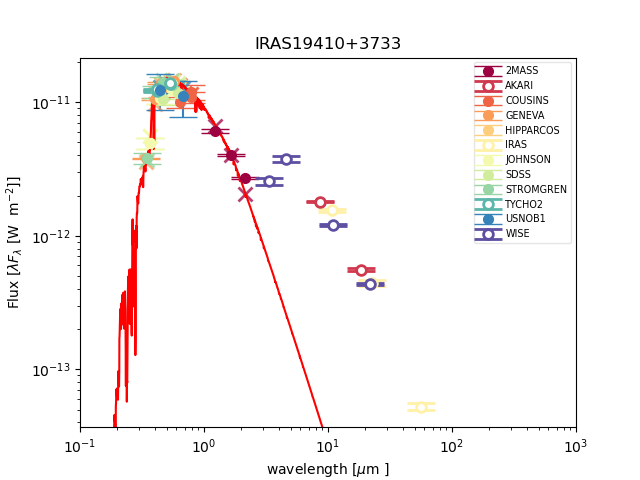}
   \includegraphics[width= 6cm]{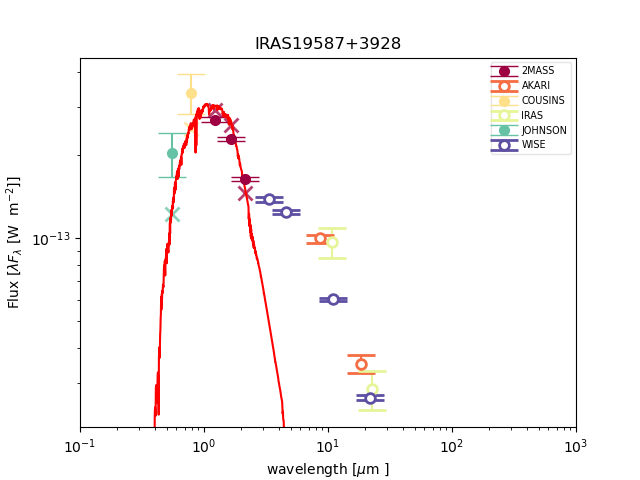}
   \includegraphics[width= 6cm]{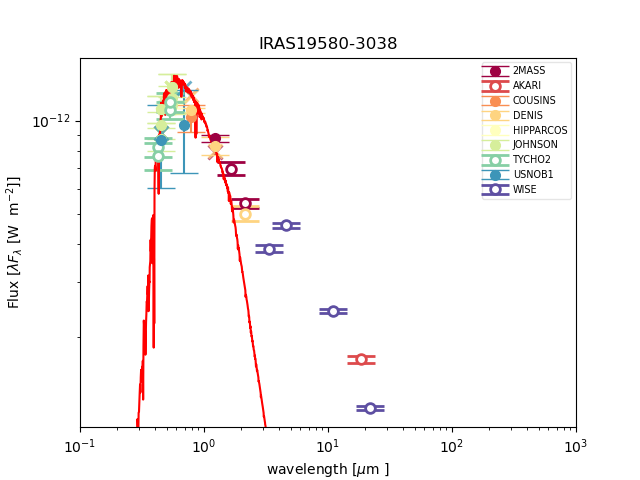}
\includegraphics[width= 6cm]{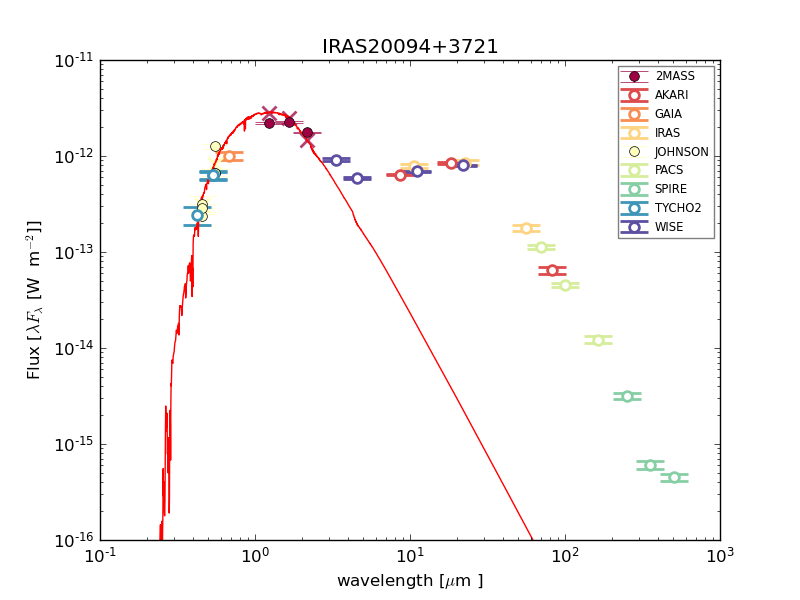}
   \includegraphics[width= 6cm]{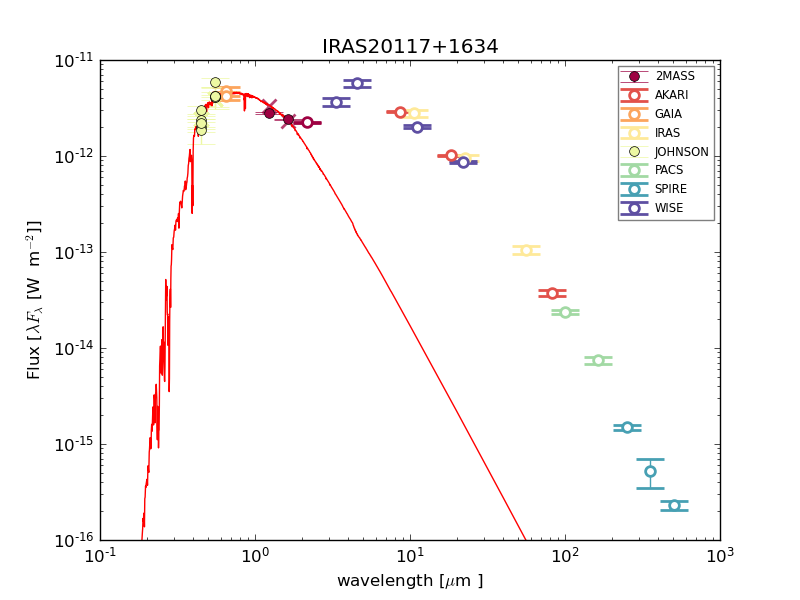}
   \includegraphics[width= 6cm]{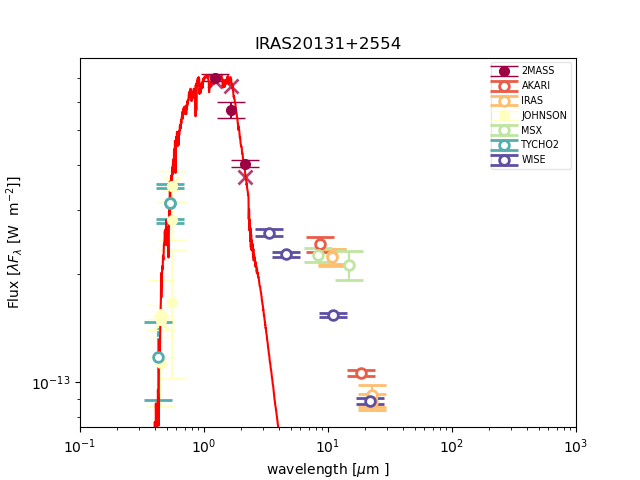}
        \caption{Fig.\,\ref{fig:Cat1} continued.}
    \label{fig:Cat13}
  \end{figure*}

\begin{figure*}
   \centering
      \includegraphics[width= 6cm]{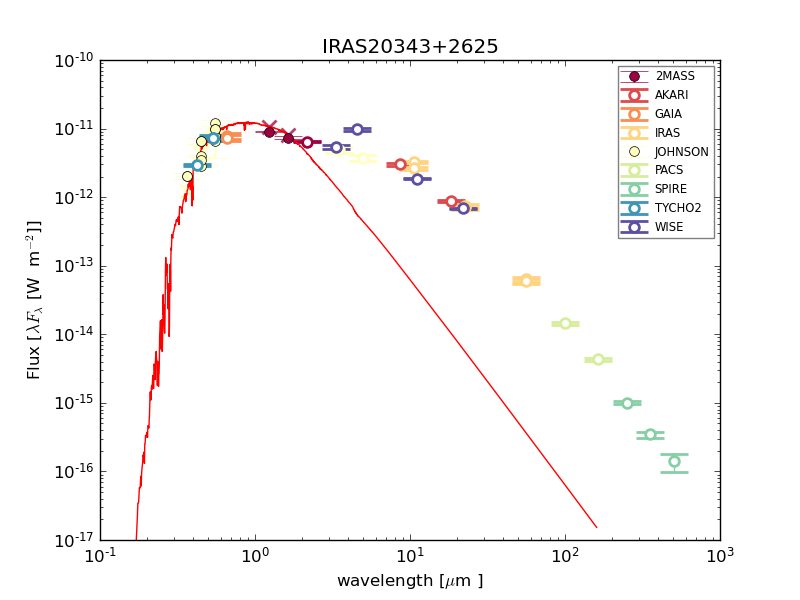}
       \includegraphics[width= 6cm]{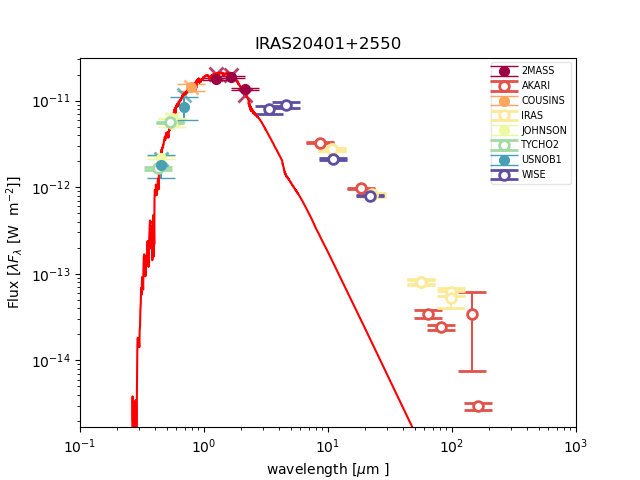}
   \includegraphics[width= 6cm]{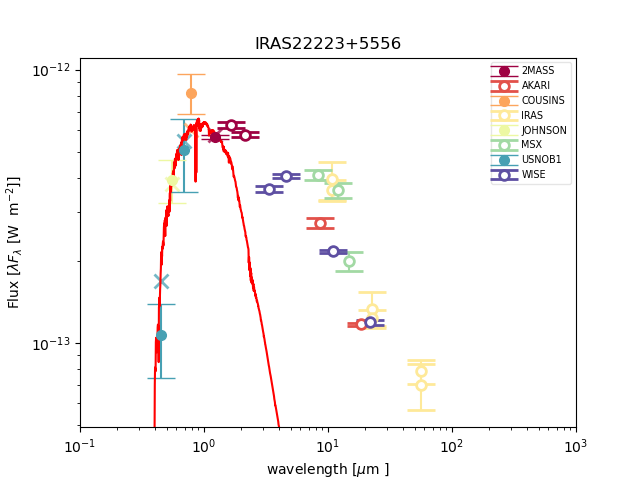}
   \includegraphics[width= 6cm]{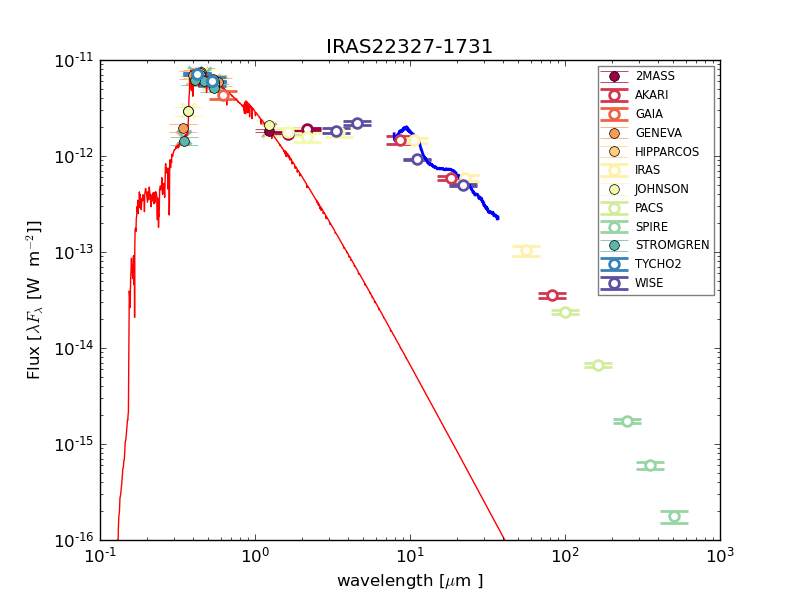}
    \caption{Fig.\,\ref{fig:Cat1} continued. }
    \label{fig:Cat14}
  \end{figure*}

\begin{figure*}
   \centering
    \includegraphics[width= 6cm]{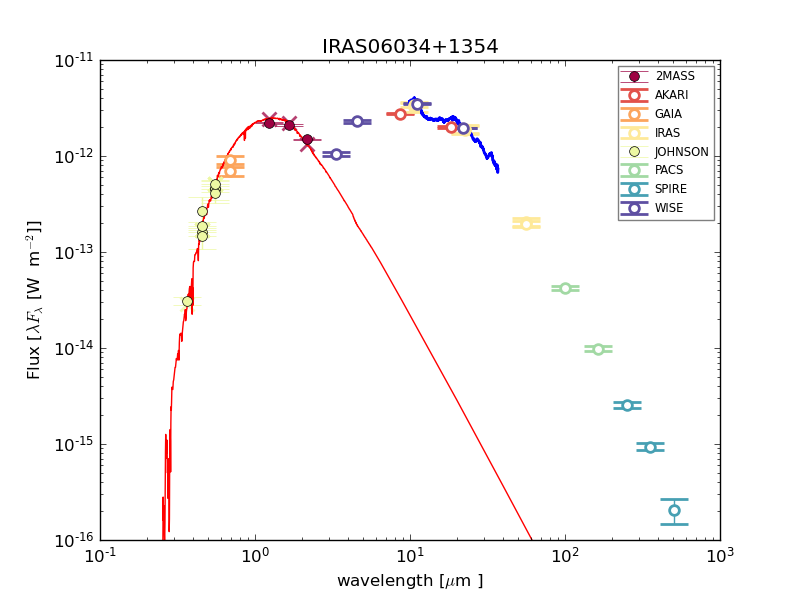}
   \includegraphics[width= 6cm]{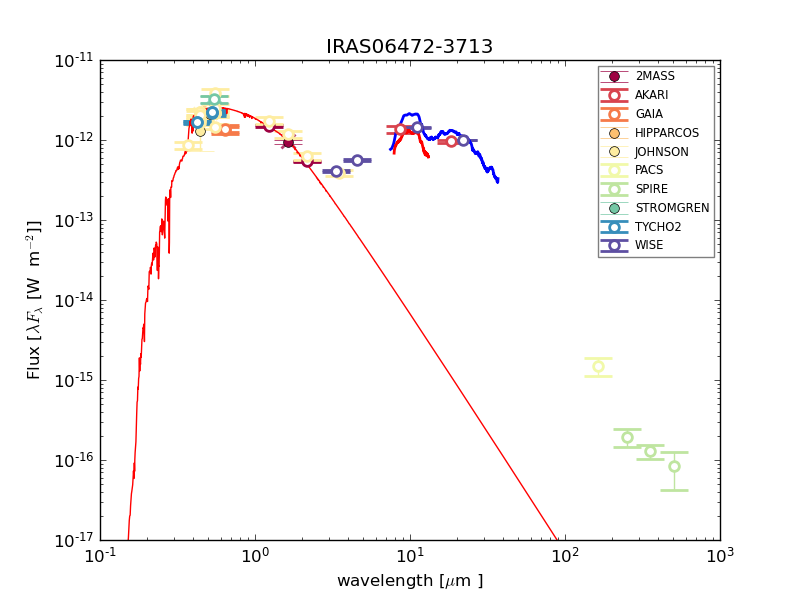}
   \includegraphics[width= 6cm]{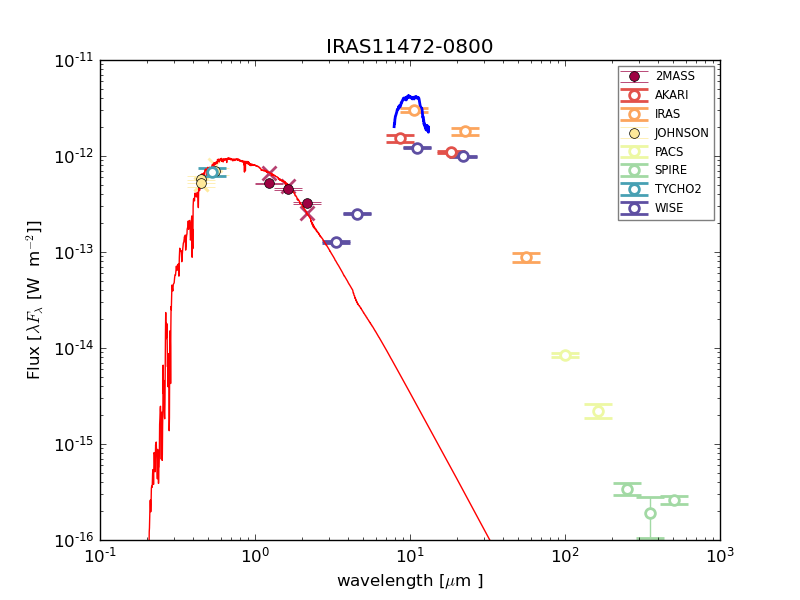}
   \includegraphics[width= 6cm]{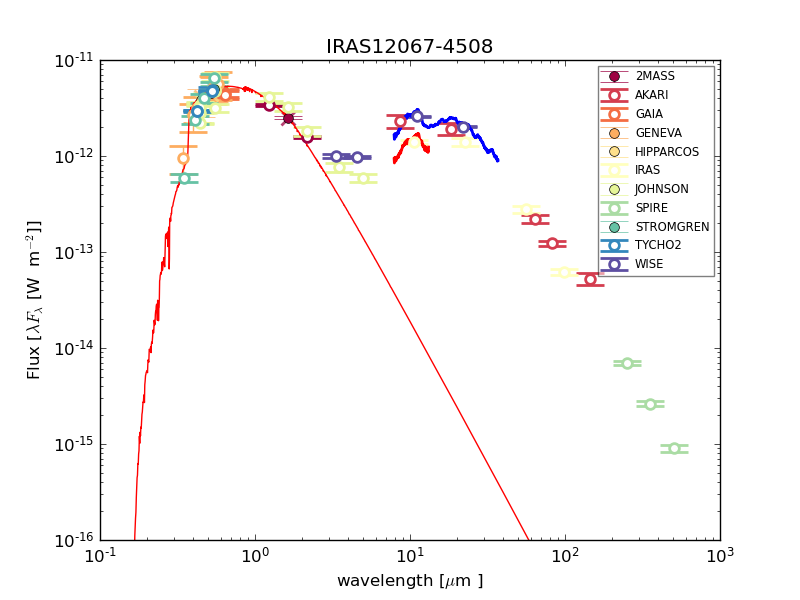}
    \includegraphics[width= 6cm]{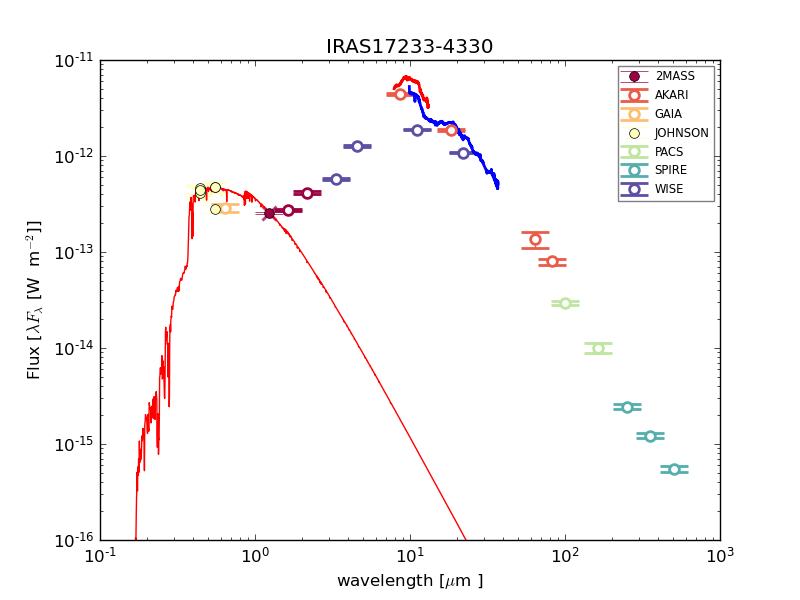}
   \includegraphics[width= 6cm]{Fig/SED/IRAS18281+2149SED.png}
   \includegraphics[width= 6cm]{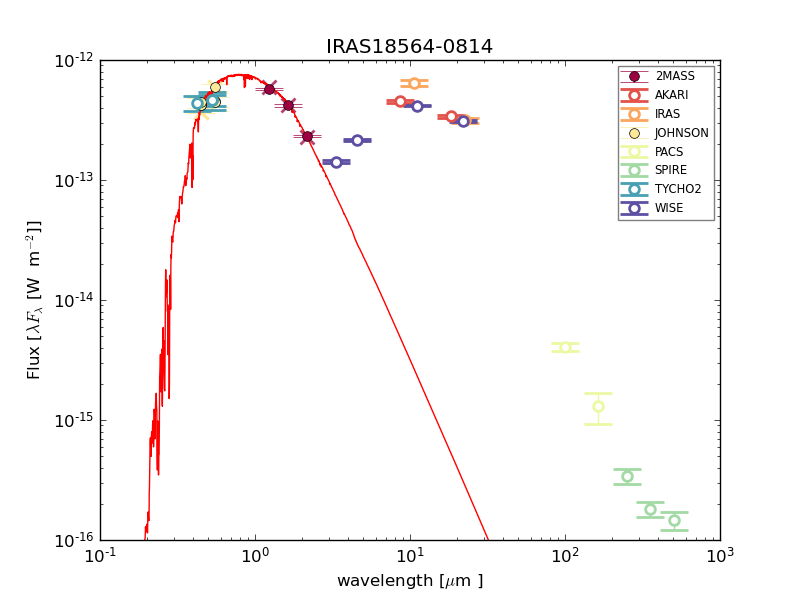}
    \caption{\modifLE{Spectral energy distributions (SED) of Category 2 targets. Red: Best fit photospheric models. Blue: SPITZER spectrum when available.}}
     \label{fig:Cat2}
  \end{figure*}

\begin{figure*}
   \centering
    \includegraphics[width= 6cm]{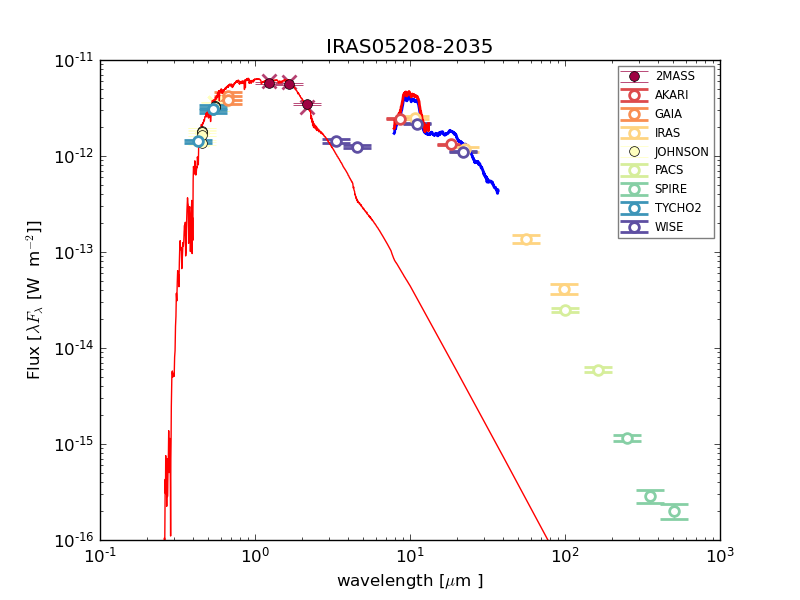}
   \includegraphics[width= 6cm]{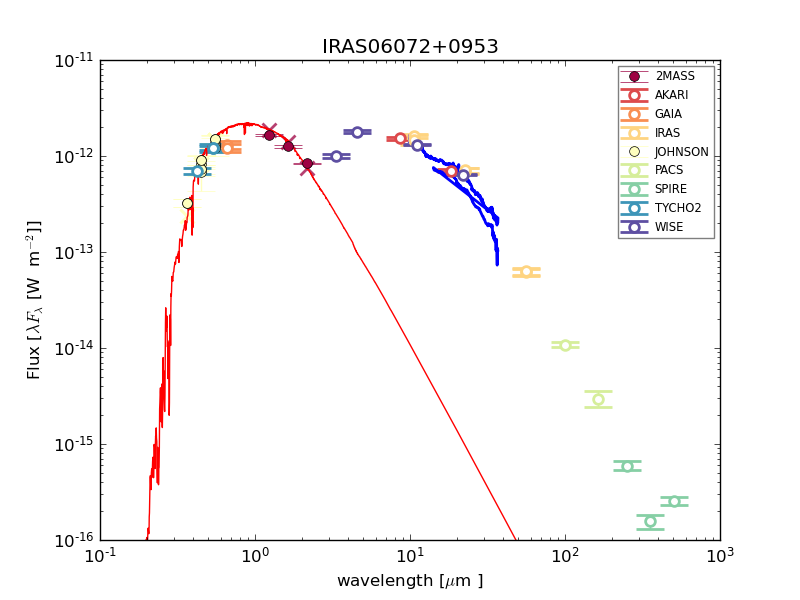}
   \includegraphics[width= 6cm]{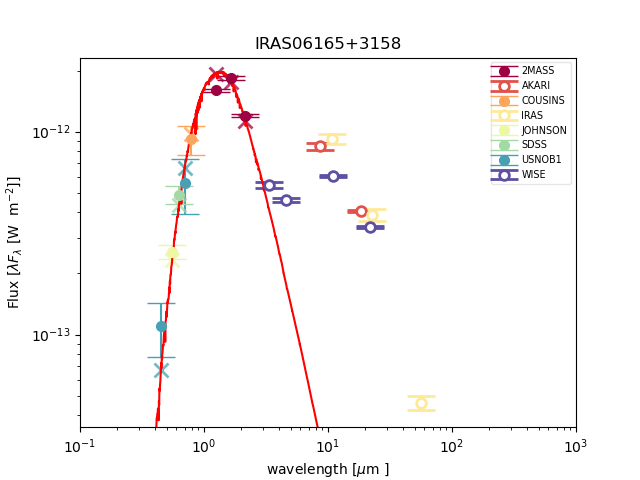}
   \includegraphics[width= 6cm]{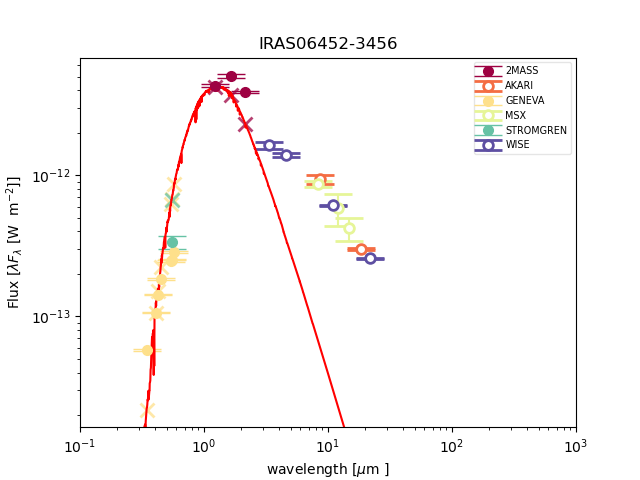}
    \includegraphics[width= 6cm]{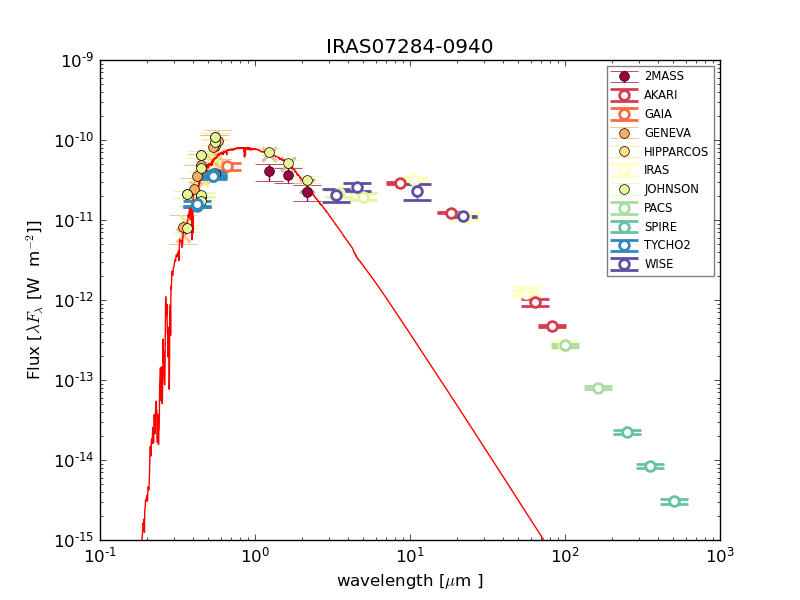}
   \includegraphics[width= 6cm]{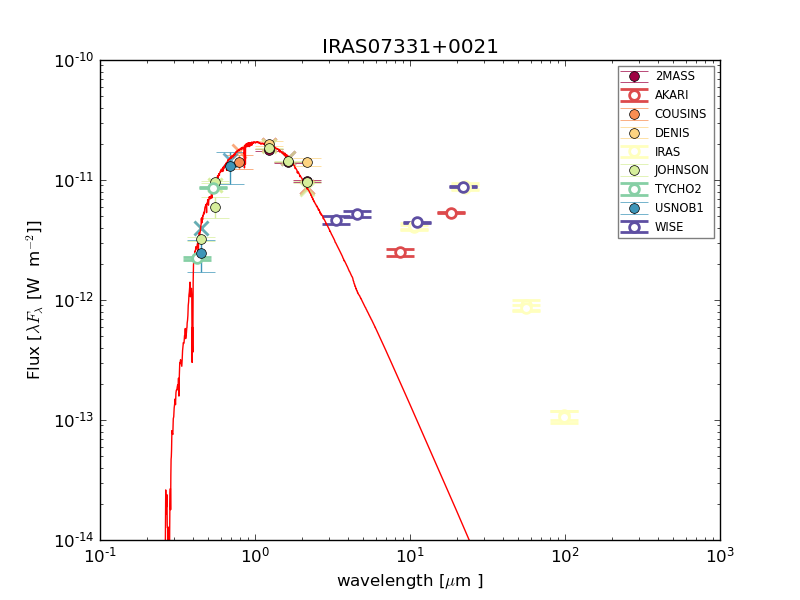}
   \includegraphics[width= 6cm]{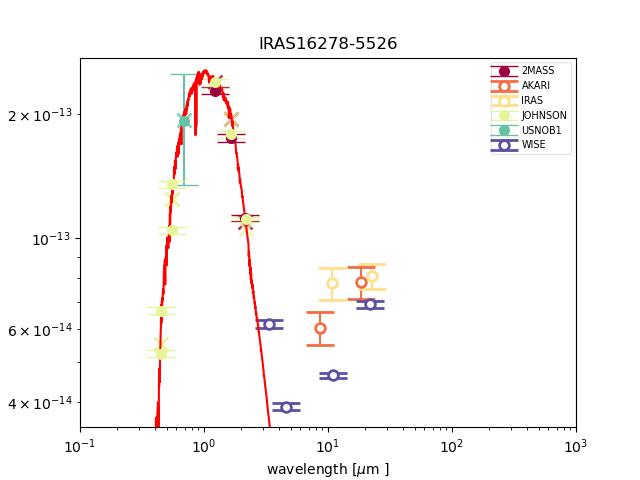}
   \includegraphics[width= 6cm]{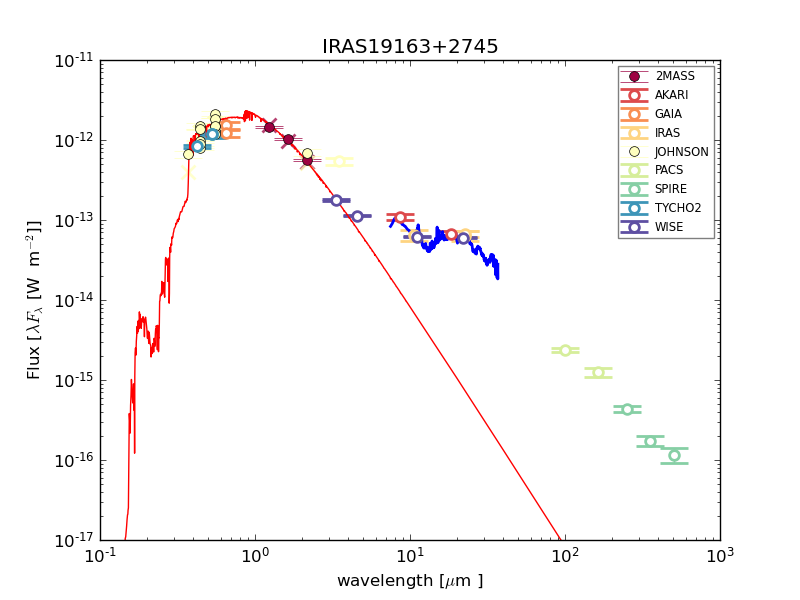}
\includegraphics[width= 6cm]{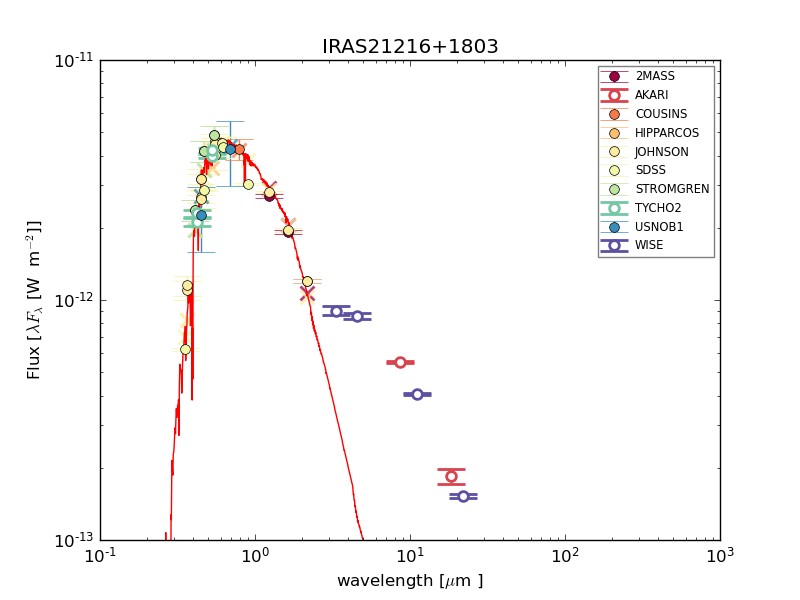}
   \includegraphics[width= 6cm]{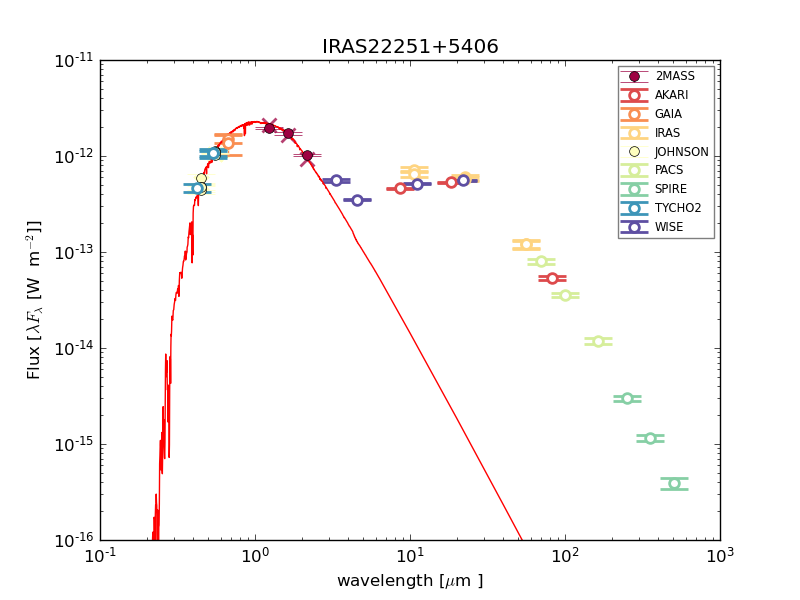}
 \caption{\modifLE{Spectral energy distributions (SED) of Category 3 targets. Red: Best fit photospheric models. Blue: SPITZER spectrum when available.}}
  \label{fig:Cat3}
  \end{figure*}

\begin{figure*}
   \centering
    \includegraphics[width= 6cm]{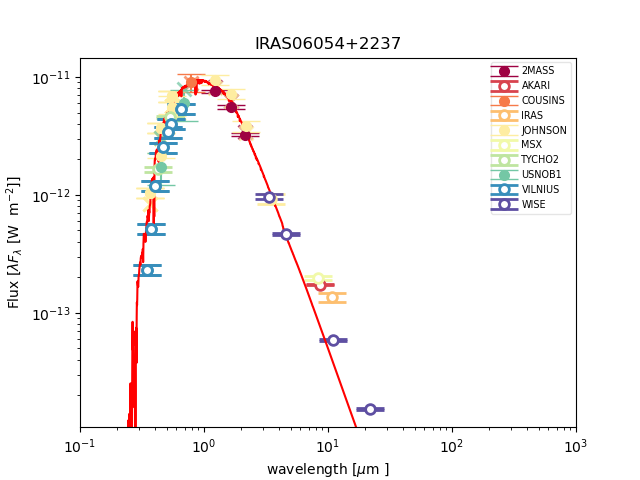}
   \includegraphics[width= 6cm]{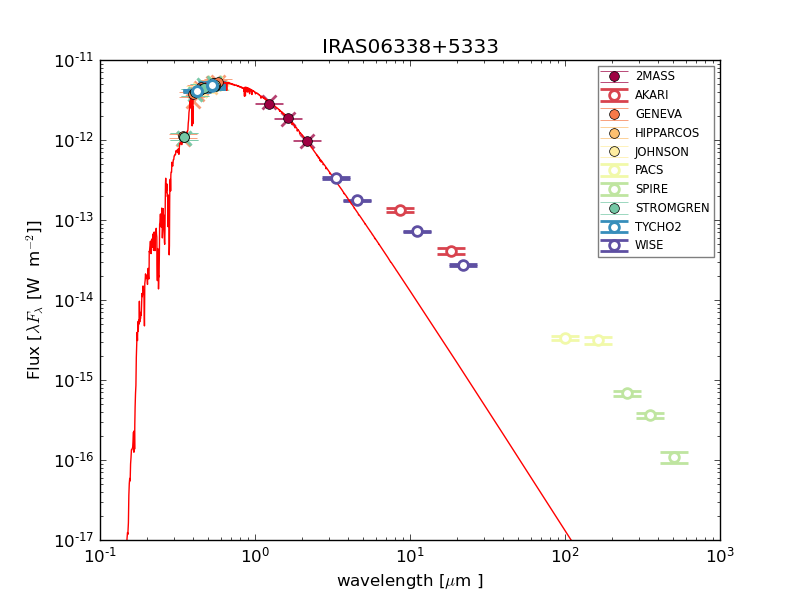}
   \includegraphics[width= 6cm]{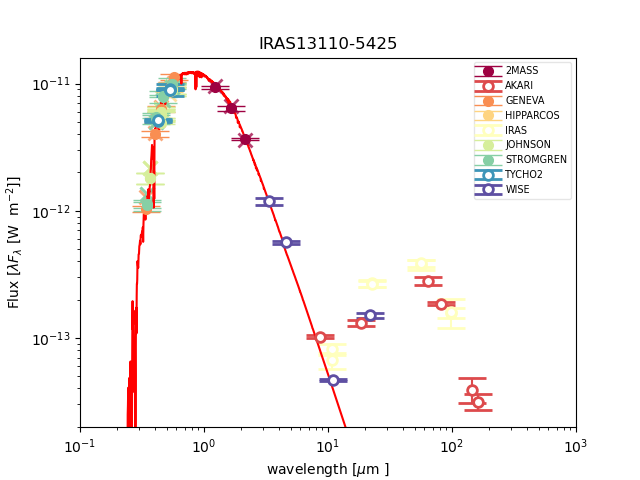}
   \includegraphics[width= 6cm]{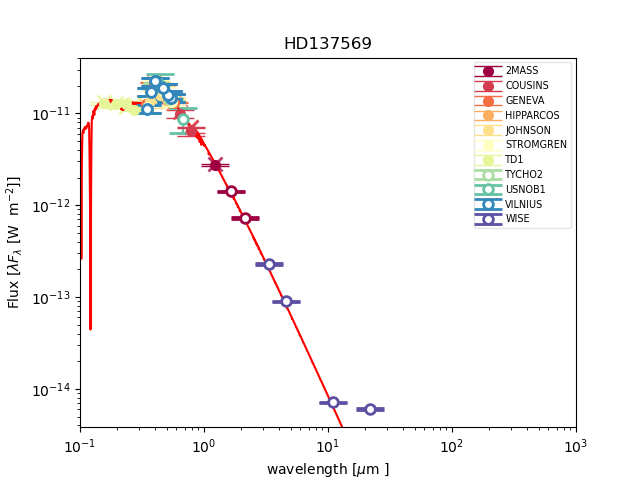}
    \includegraphics[width= 6cm]{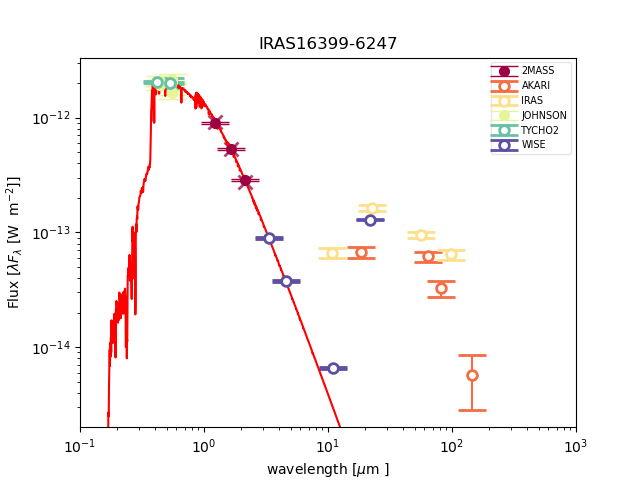}  
   \includegraphics[width= 6cm]{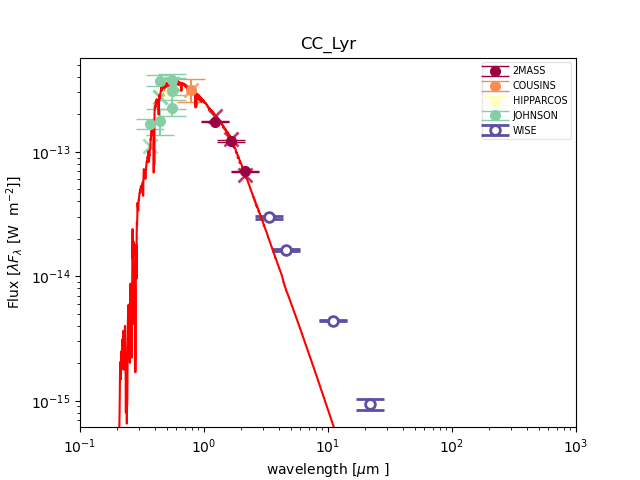}
    \includegraphics[width= 6cm]{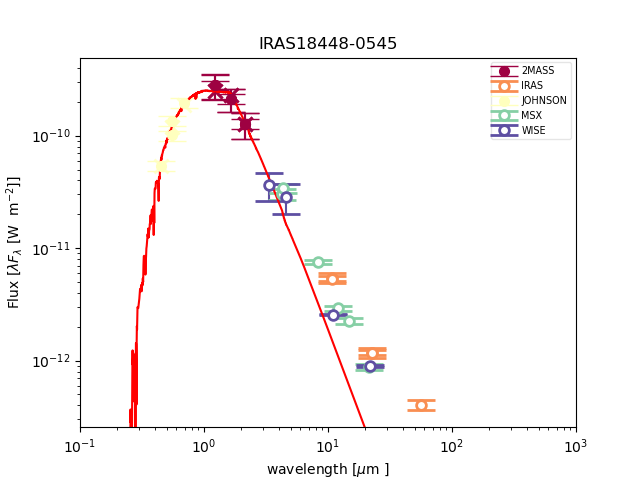}
   \includegraphics[width= 6cm]{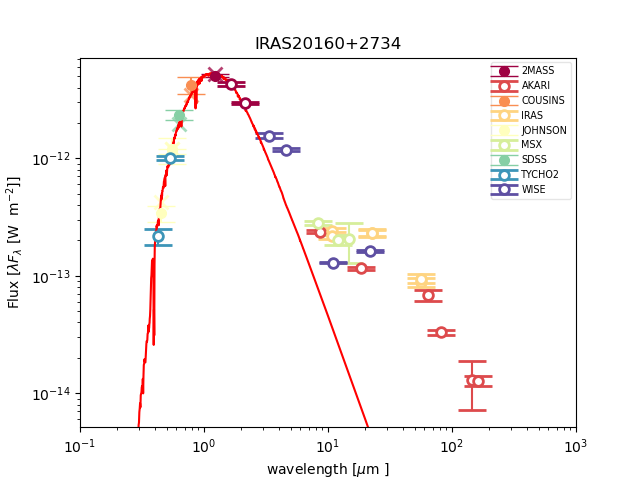}
   \includegraphics[width= 6cm]{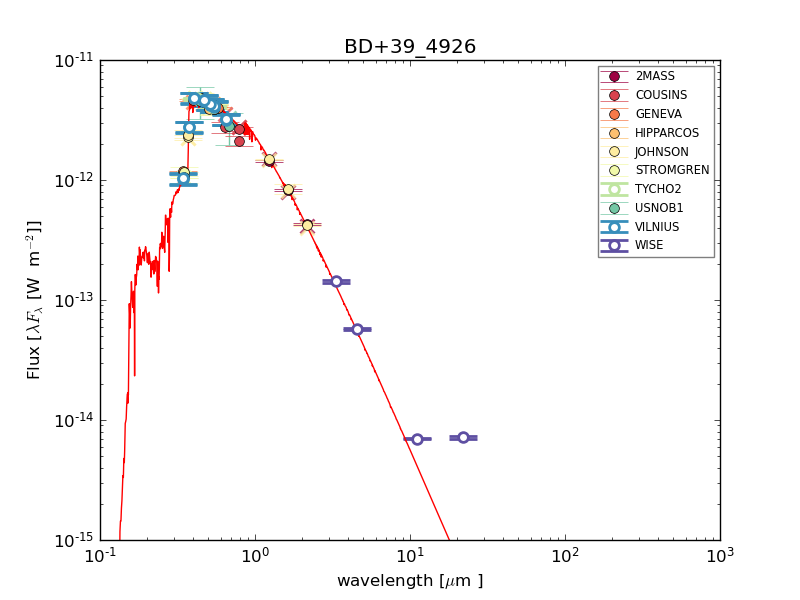}
   \caption{\modifLE{Spectral energy distributions (SED) of Category 4 targets. Red: Best fit photospheric models. Blue: SPITZER spectrum when available.}}
    \label{fig:Cat4}
  \end{figure*}

\begin{figure*}
   \centering
    \includegraphics[width= 6cm]{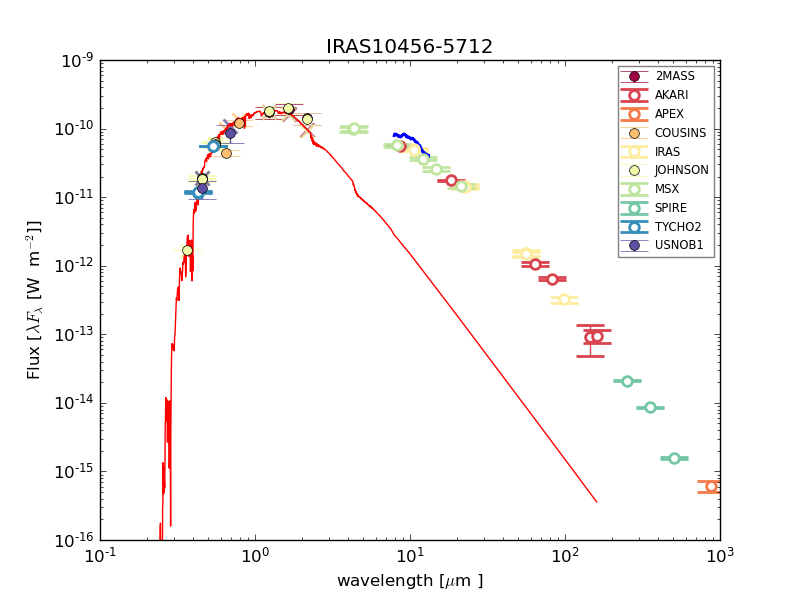}
   \caption{\modifLE{Spectral energy distribution (SED) for the unclassified target. Red: Best fit photospheric models. }}
    \label{fig:Uncategorized}
  \end{figure*}

Figs.\,\ref{fig:Cat0}, \ref{fig:Cat1}, \ref{fig:Cat12}, \ref{fig:Cat13}, \ref{fig:Cat14}, \ref{fig:Cat2}, \ref{fig:Cat3}, \ref{fig:Cat4}, and \ref{fig:Uncategorized} display the targets spectral energy distributions per category.


\section{Change of the location of models in the color-color diagram with respect to model parameters}

\begin{figure*} 
   \centering
   \includegraphics[width=6cm]{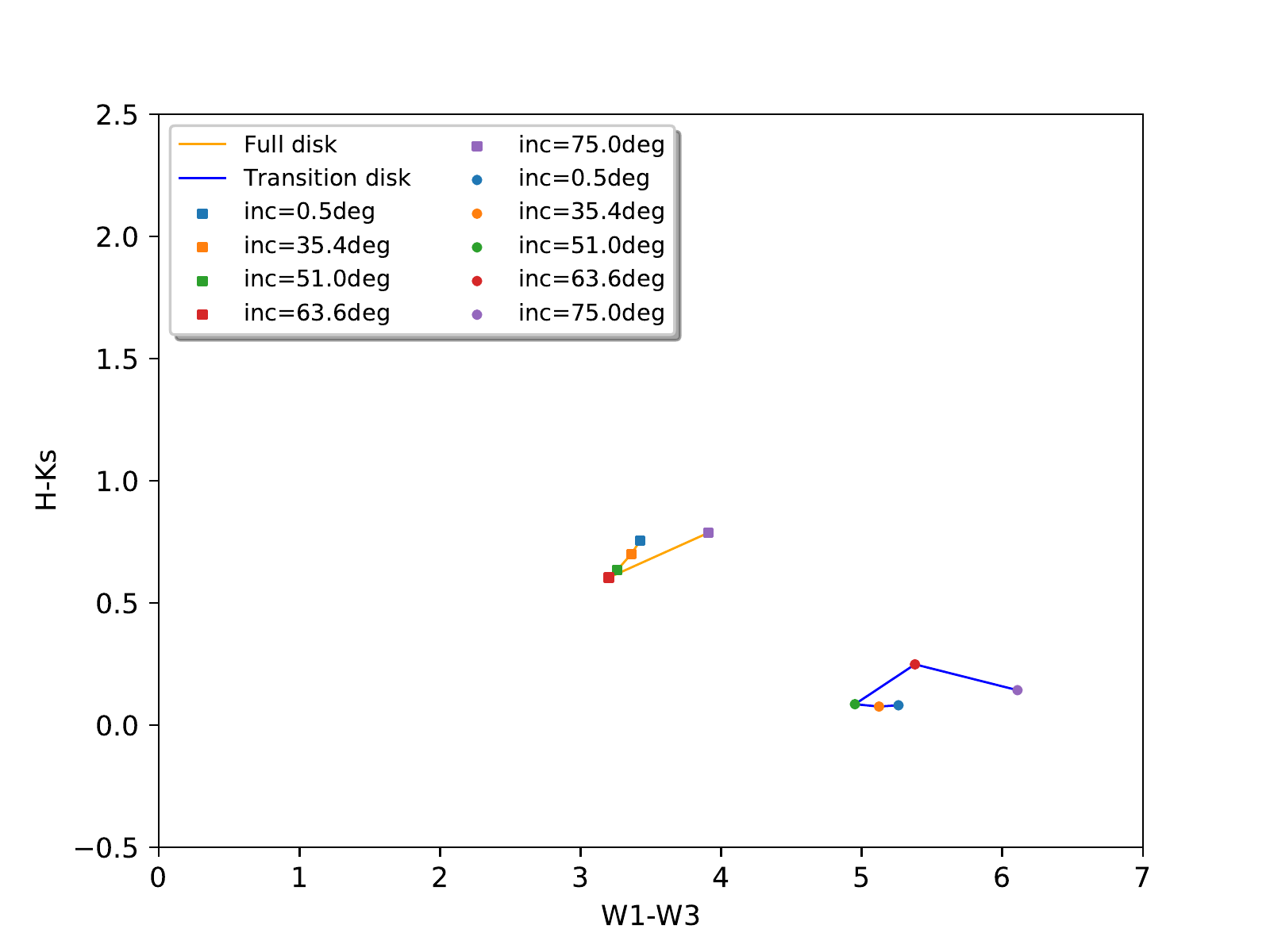}
   \includegraphics[width=6cm]{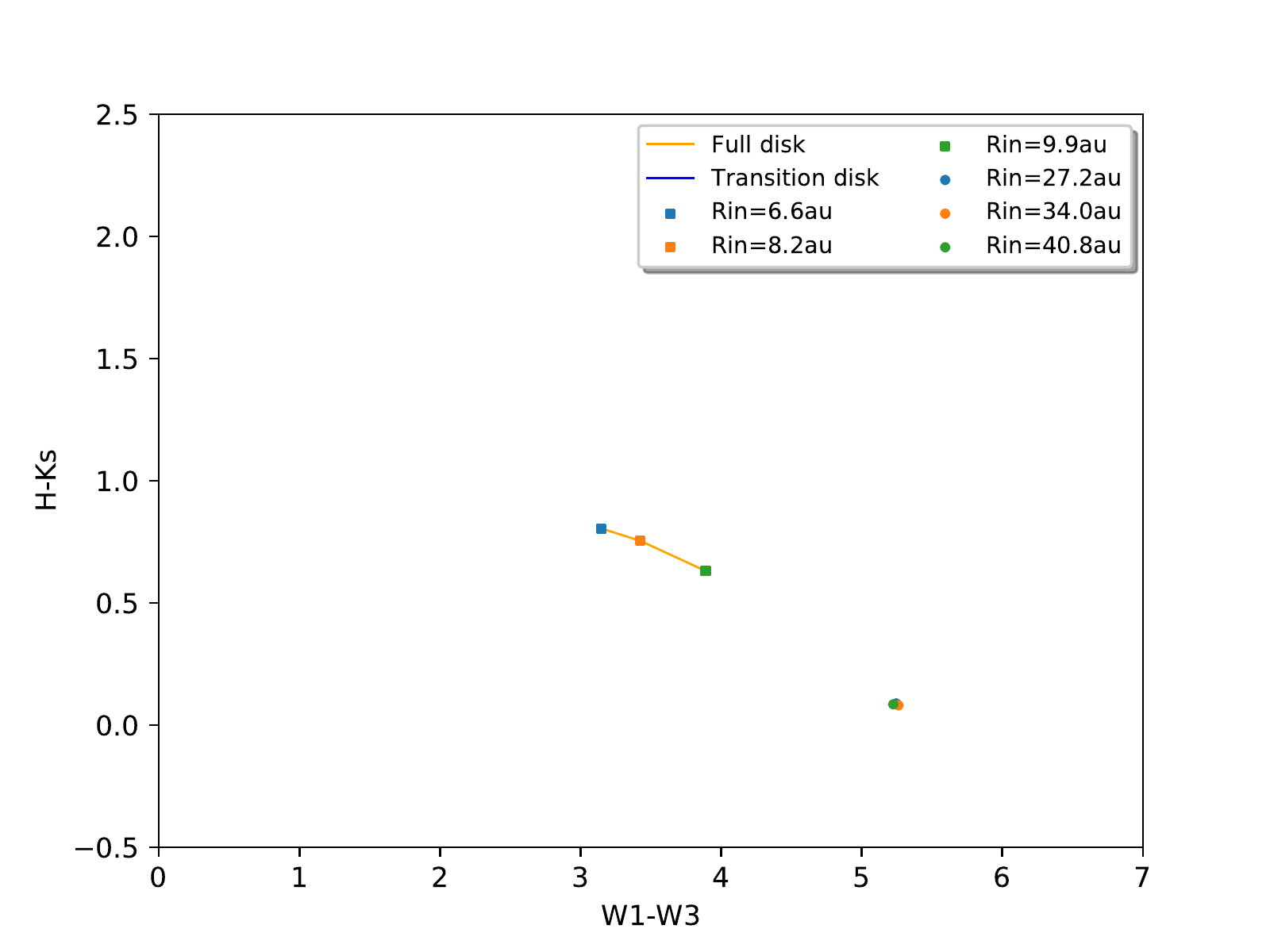}
   \includegraphics[width=6cm]{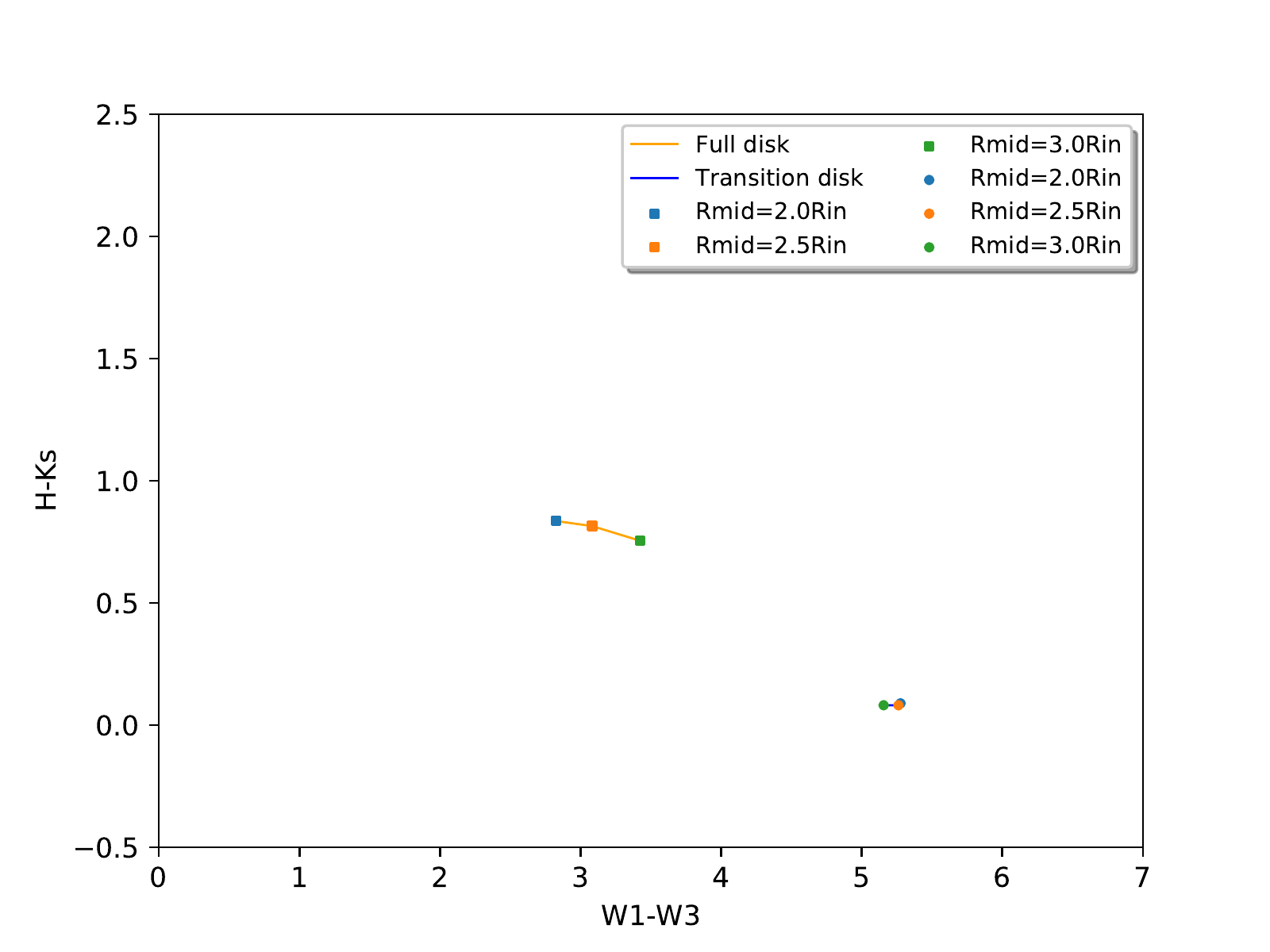}
   \includegraphics[width=6cm]{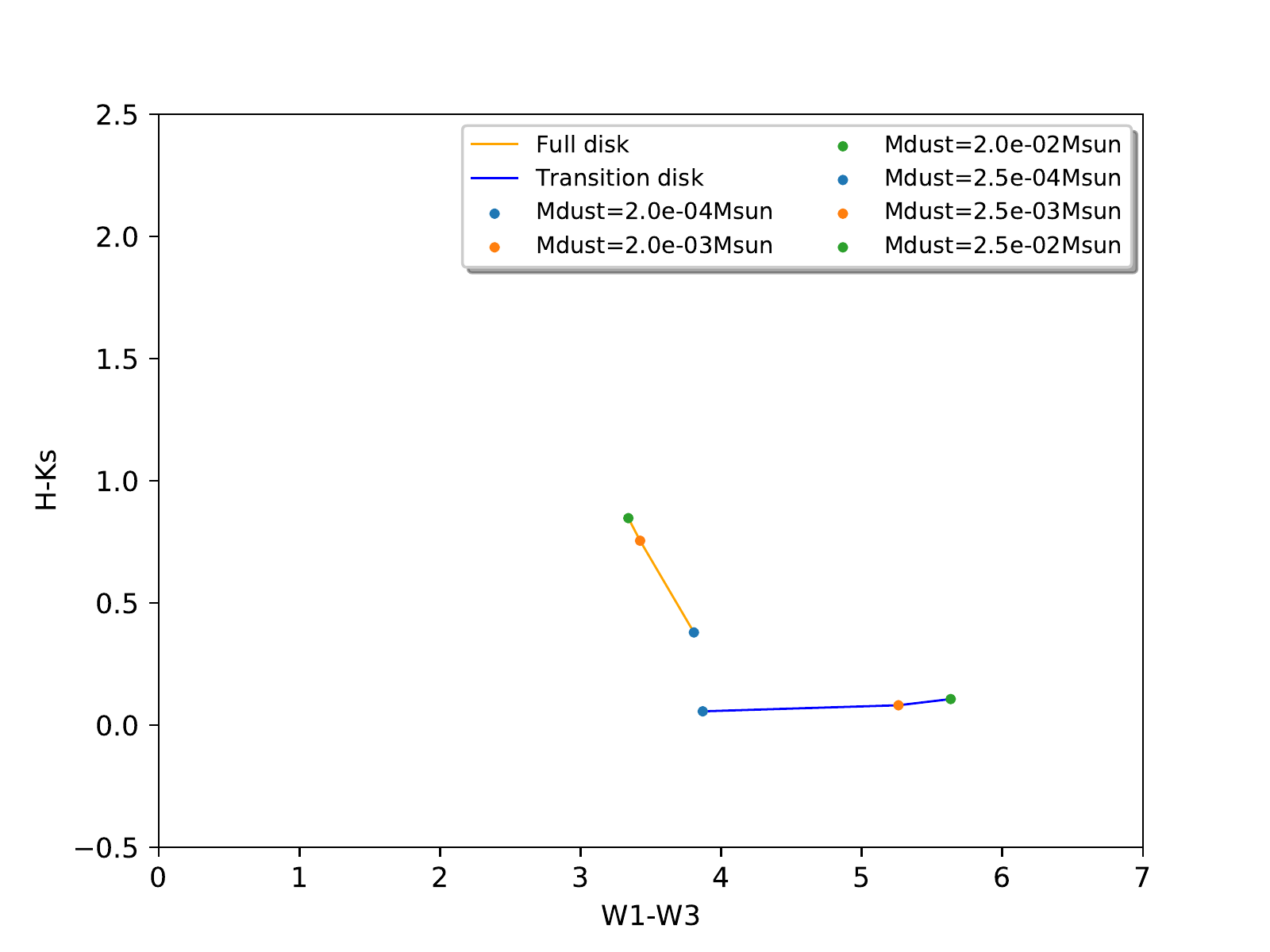}
   \includegraphics[width=6cm]{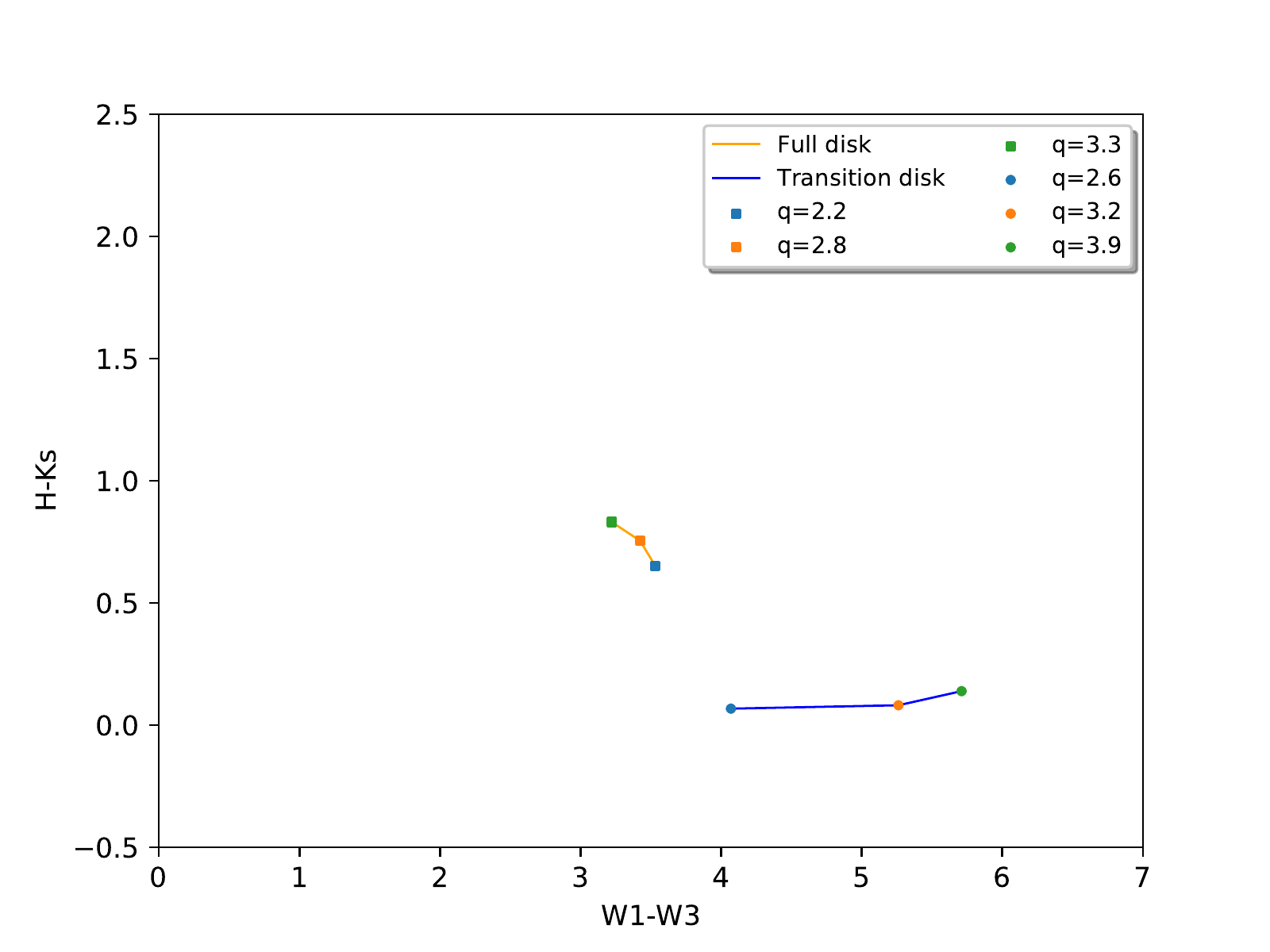}
   \includegraphics[width=6cm]{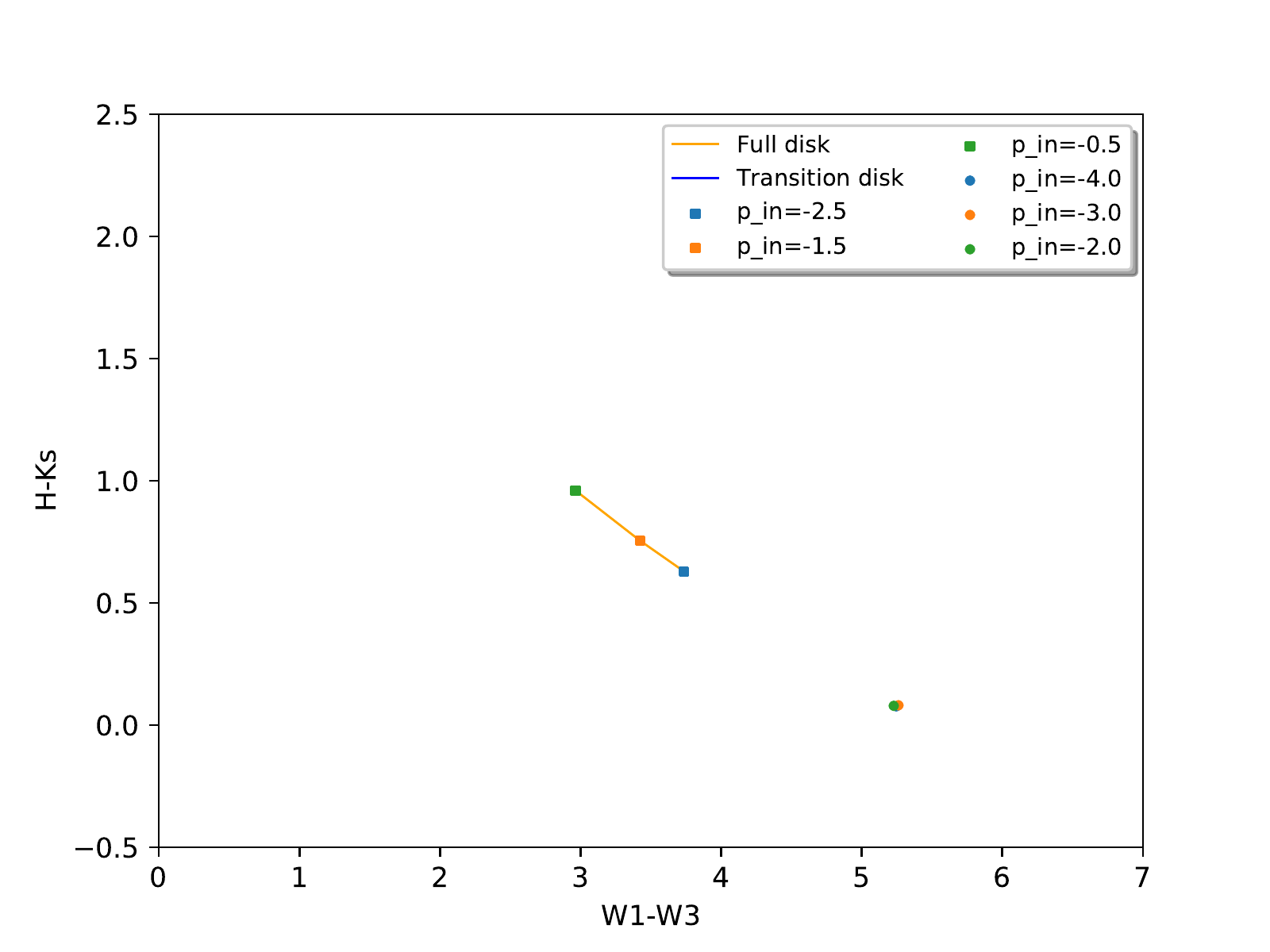}
   \includegraphics[width=6cm]{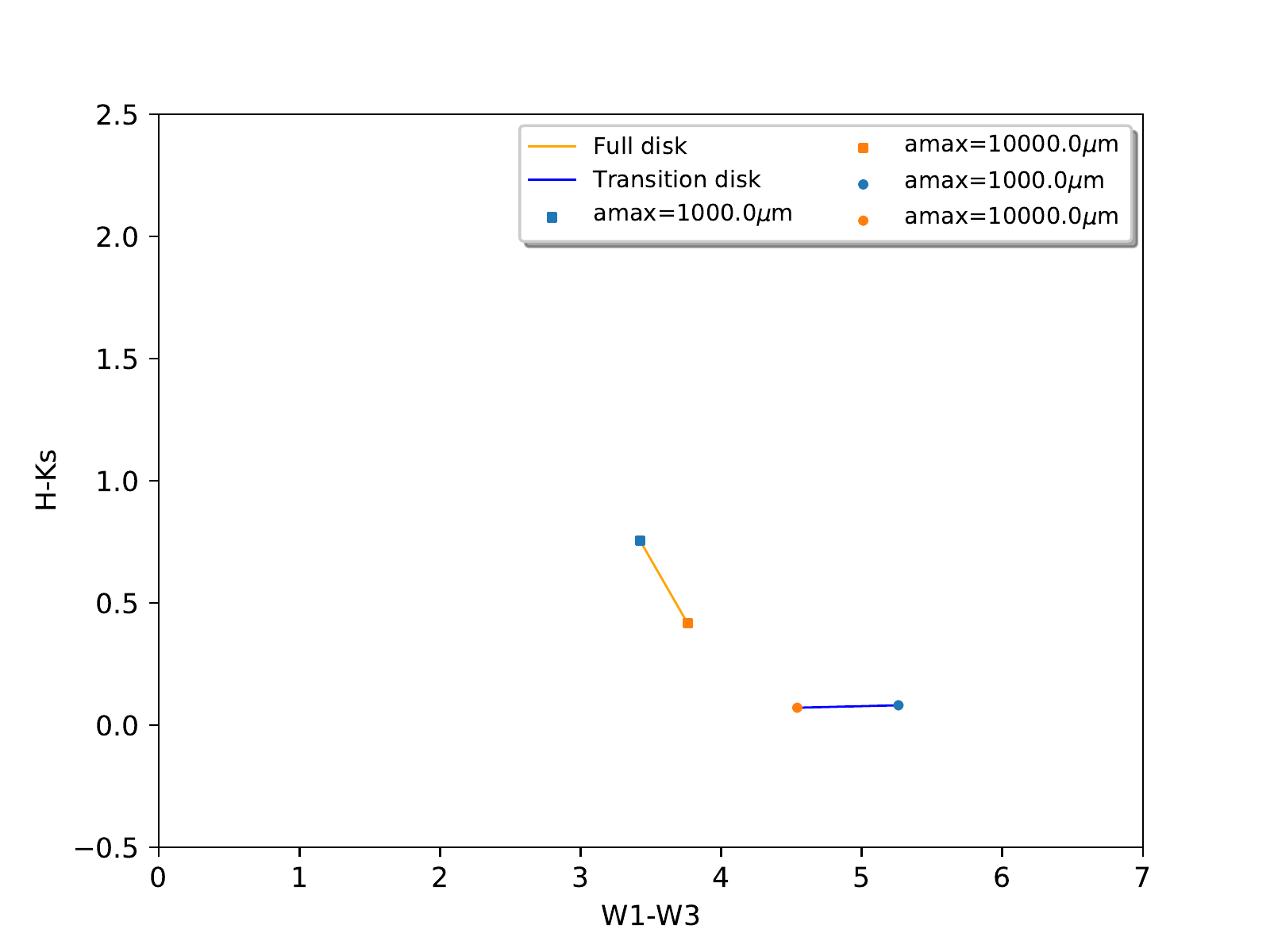}
   \includegraphics[width=6cm]{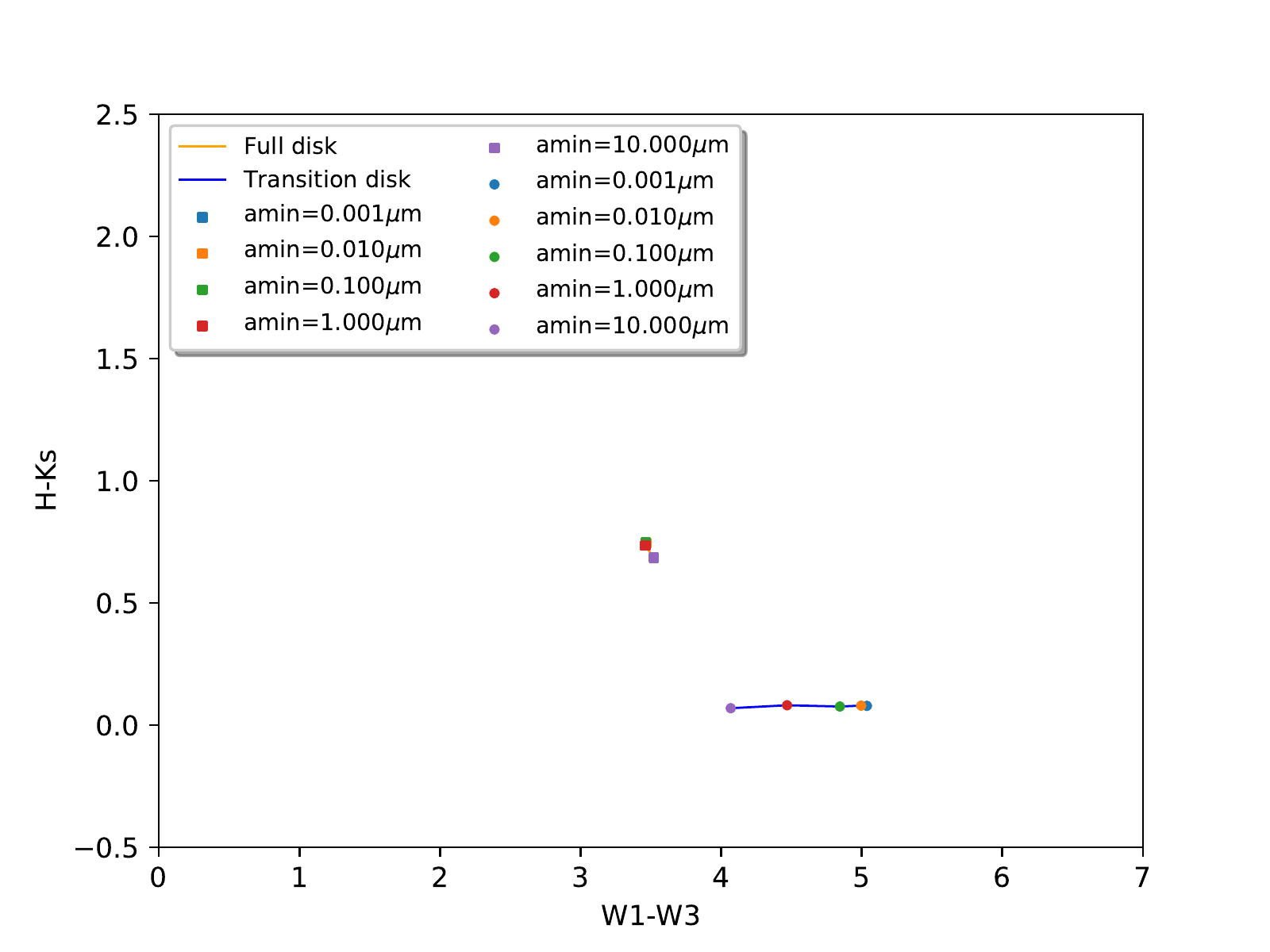}
    \caption{Effect of variation of the different parameters of the radiative transfer models on the color-color diagram. They show variations around the full disk model of IRAS\,08544-4431 (green) and the transition disk model of AC\,Her (blue).}
    \label{fig:CCeffect}
   \end{figure*}

Fig.\,\ref{fig:CCeffect} displays the effect of changing several model parameters on the position of models in the color-color diagram.

\section{Effect of the ad-hoc extended flux component on the color-color diagram.}

\begin{figure}[t]
   \centering
   \includegraphics[width=9cm]{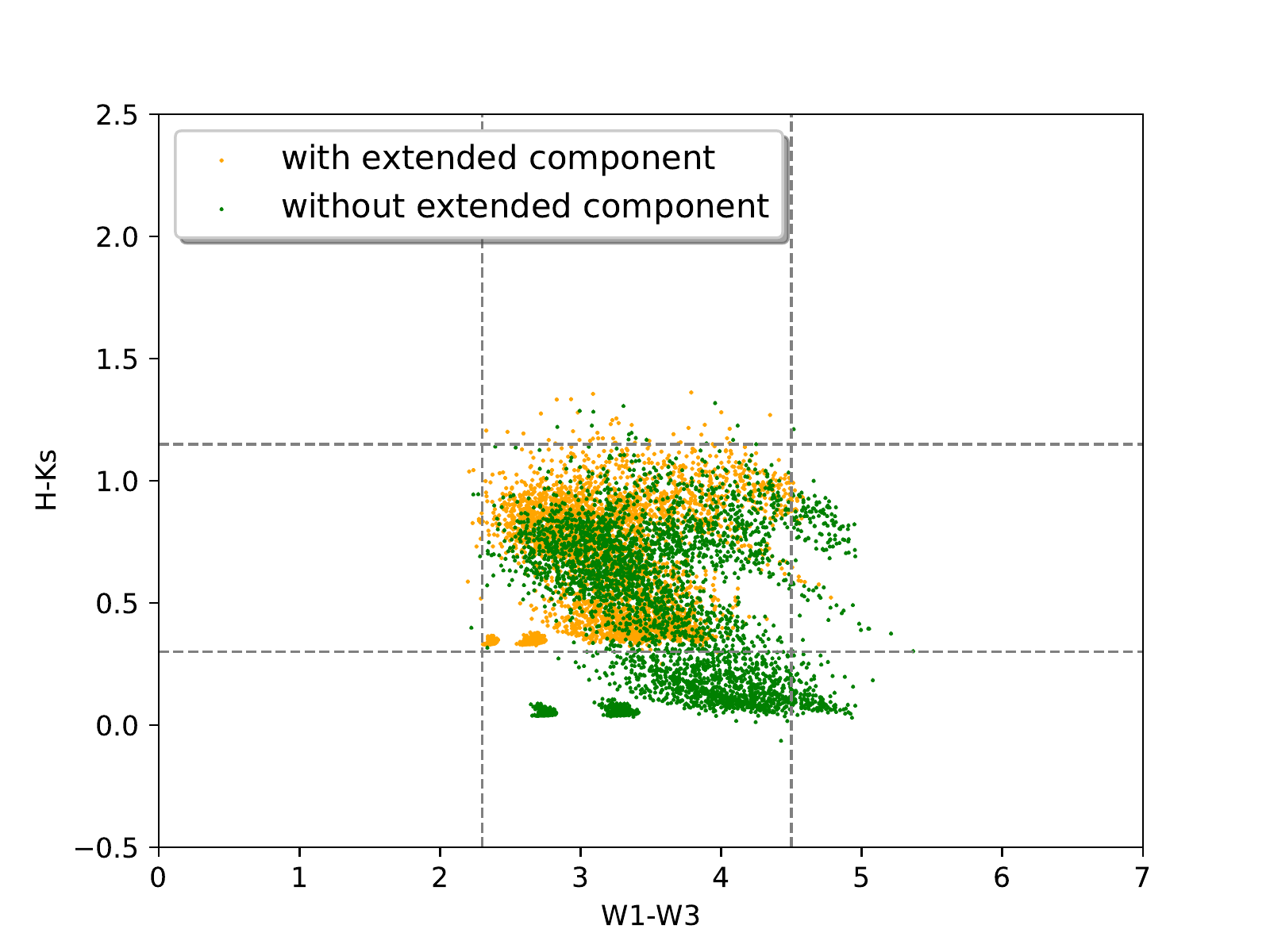}
    \caption{Effect of adding the ad-hoc extended component.}
    \label{fig:OReffect}
   \end{figure}

Fig.\,\ref{fig:OReffect} shows the influence of the extended flux component on the position of the full disk models in the color-color diagram.

\section{Distributions of effective temperatures and orbital parameters with categories.}

\begin{figure}[t]
   \centering
   \includegraphics[width=9cm]{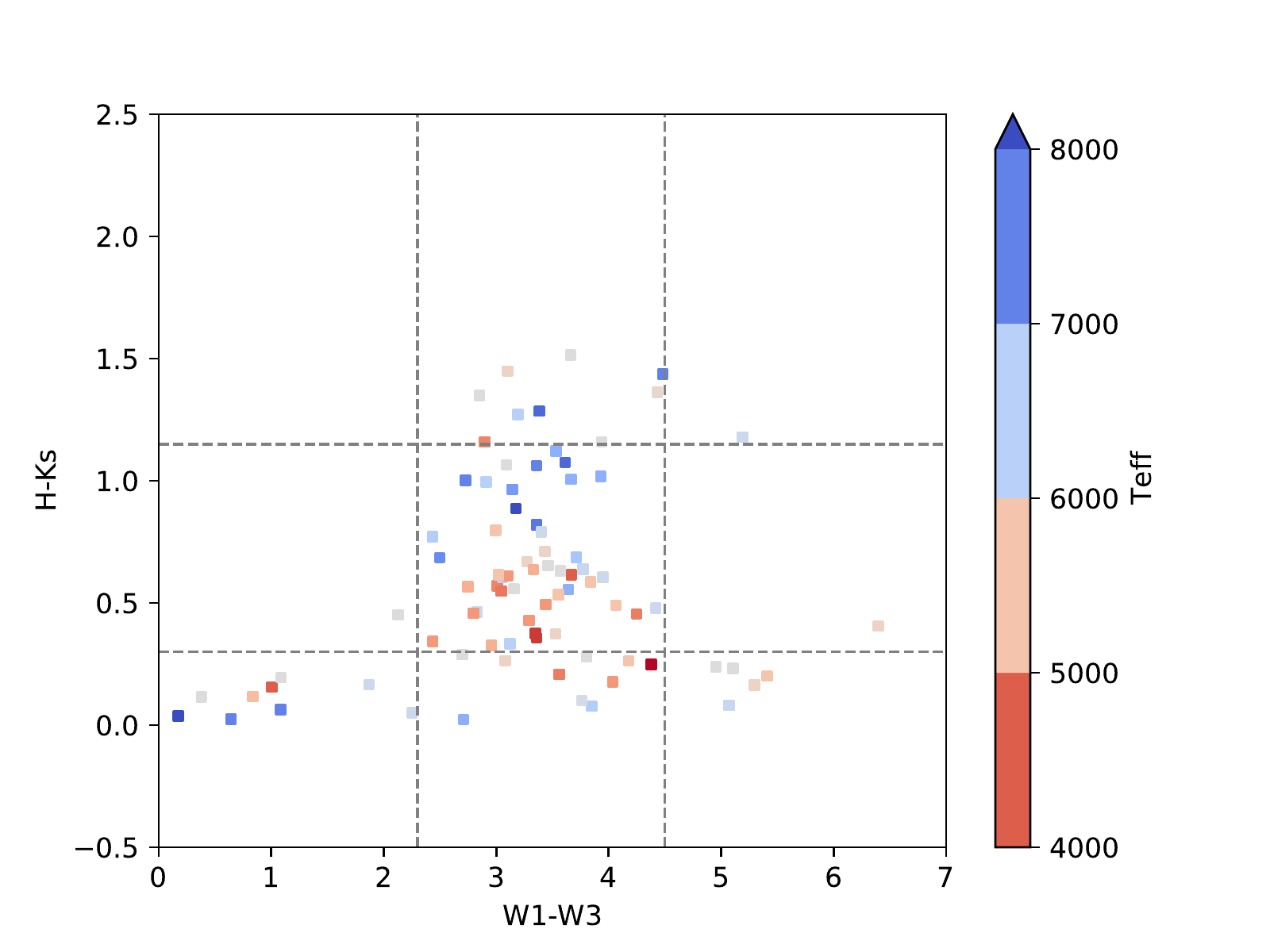}
    \caption{Distribution of effective temperature (T$_\mathrm{eff}$) in the color-color diagram}
    \label{fig:Teff}
   \end{figure}

\begin{figure}[t]
   \centering
   \includegraphics[width=9cm]{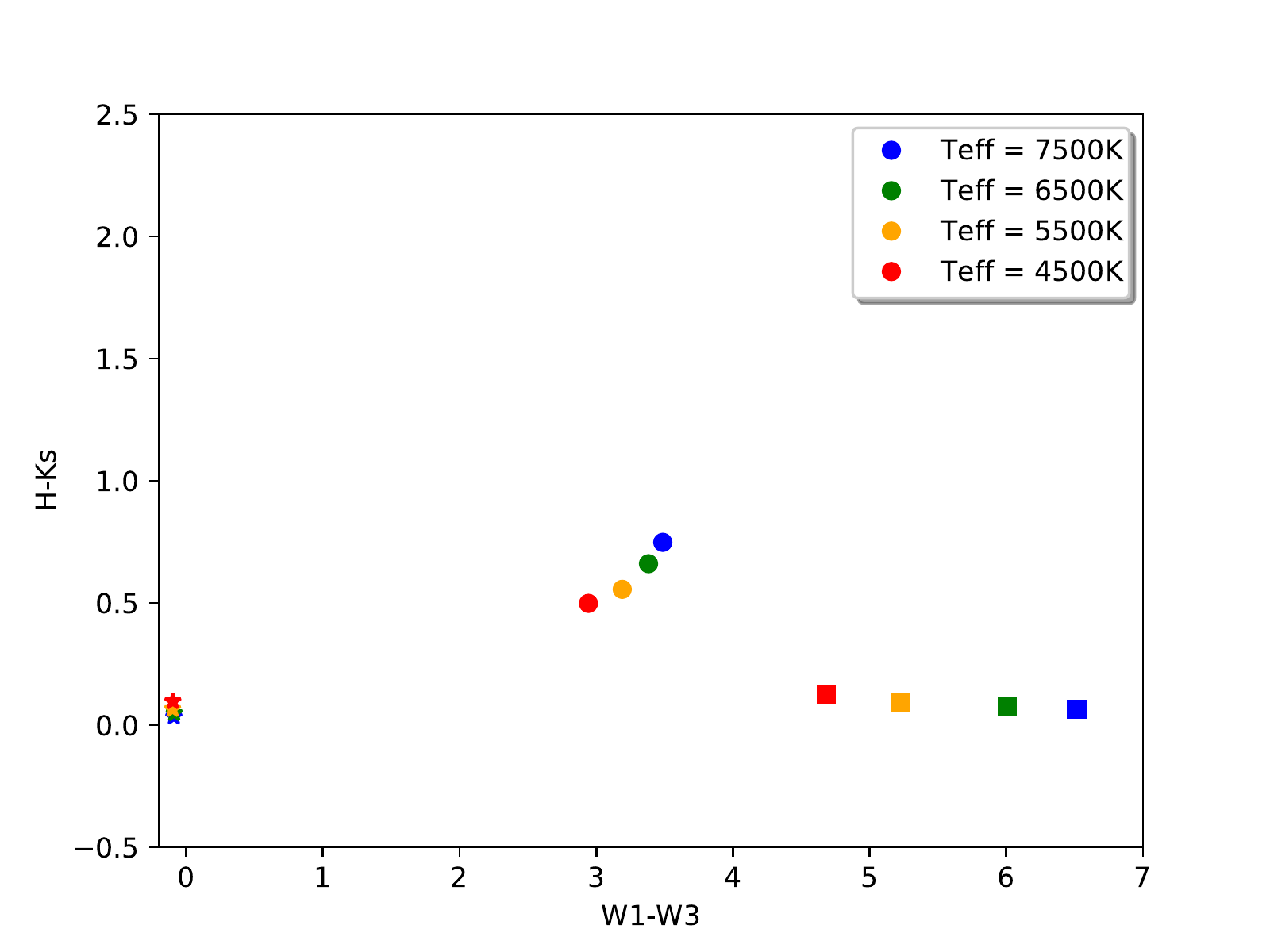}
    \caption{Change of infrared colors as a function of effective temperature (T$_\mathrm{eff}$) in the color-color diagram. Stars: no disk; circles: models of IRAS08544-4431; squares: models of AC\,Her.}
    \label{fig:TeffTest}
   \end{figure}

\begin{figure}[t]
   \centering
   \includegraphics[width=9cm]{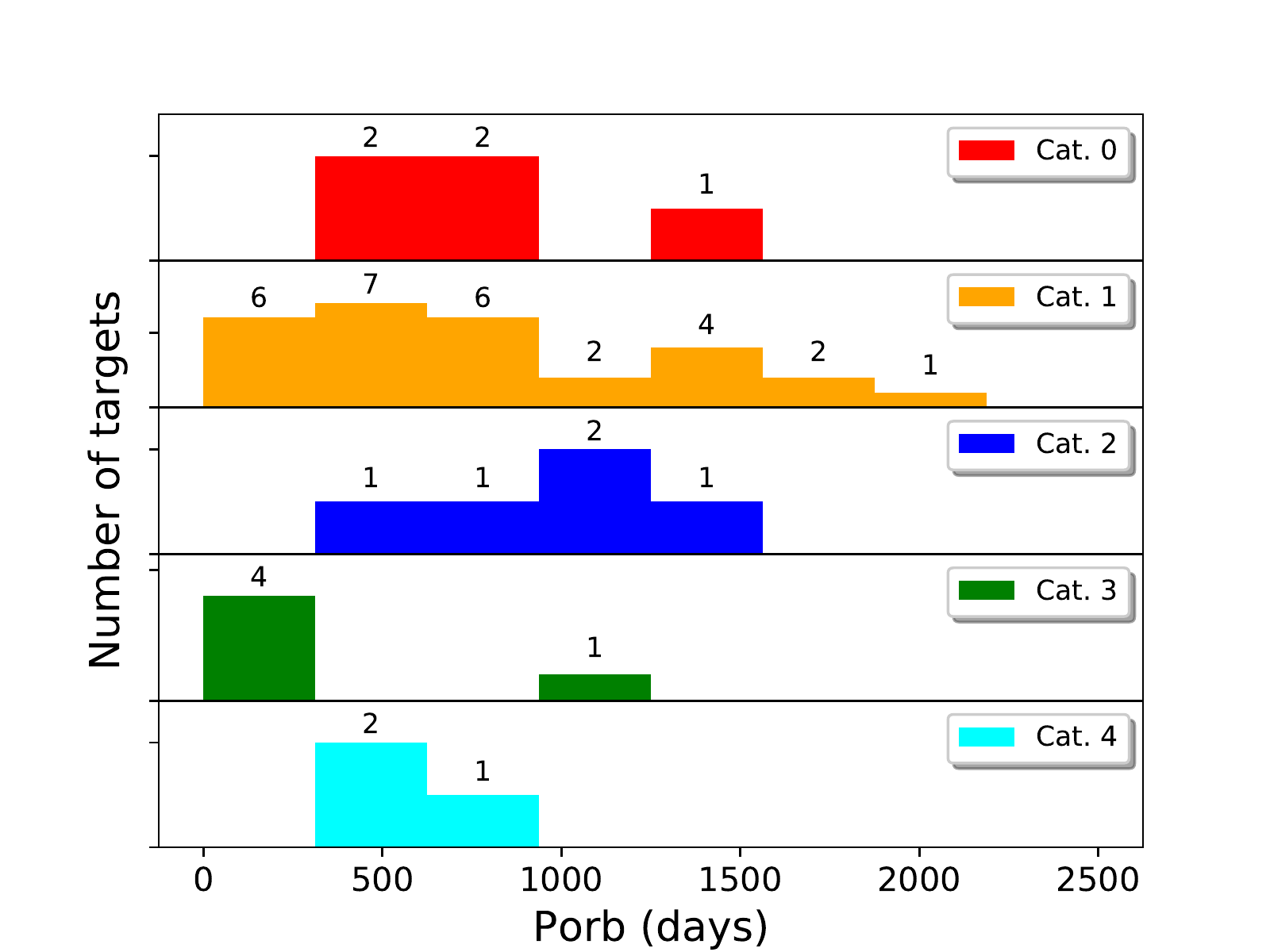}
   \includegraphics[width=9cm]{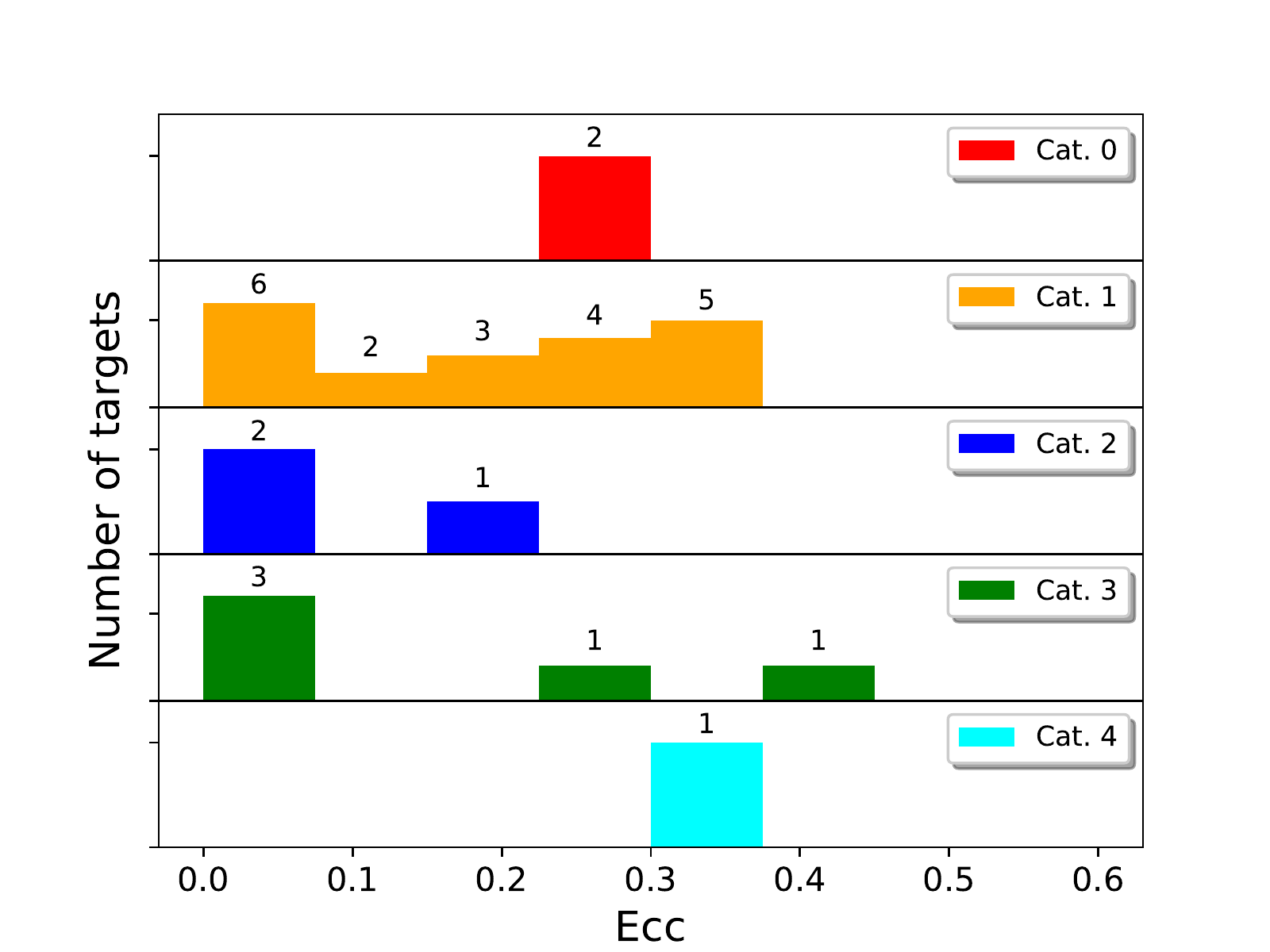}
    \caption{Histograms of periods (top) and eccentricities (bottom) for the different categories.}
    \label{fig:orbits}
   \end{figure}

Figure\,\ref{fig:Teff} shows the distribution of effective temperature within the color-color diagram. 
Figure\,\ref{fig:TeffTest} shows the influence of a changing the effective temperature of the central star for the models of IRAS08544-4431 and AC\,Her.
Figure\,\ref{fig:orbits} shows the distribution of orbital parameters (period and eccentricity) within each category.

%
%


\end{appendix}

\end{document}